# Chapter 4: From planetary exploration goals to technology requirements

Jérémie Lasue[1], Pierre Bousquet[2], Michel Blanc[1], Nicolas André[1], Pierre Beck[3], Gilles Berger[1], Scott Bolton[4], Emma Bunce[5], Baptiste Chide[1], Bernard Foing[6], Heidi Hammel[7], Emmanuel Lellouch[8], Lea Griton[1,8], Ralph Mcnutt[9], Sylvestre Maurice[1], Olivier Mousis[10,11], Merav Opher[12], Christophe Sotin[13], Dave Senske[14], Linda Spilker[14], Pierre Vernazza[10], Qiugang Zong[15]

[1]Institut de Recherche en Astrophysique et Planétologie, Observatoire Midi-Pyrénées Toulouse, France
[2]CNES, Toulouse, France
[3]Institut d'astrophysique et de planétologie de Grenoble/ISTerre, Université Grenoble Alpes, Saint-Martin-d'Hères, France
[4]Southwest Research Institute (SwRI), San Antonio, TX, USA
[5]School of Physics and Astronomy, University of Leicester, Leicester, United-Kingdom
[6]Leiden University, Leiden, The Netherlands
[7]Association of Universities for Research in Astronomy, Washington, D.C., USA
[8]LESIA-CNRS, Paris-Meudon, France
[9]Johns Hopkins University, Applied Physics Laboratory, Laurel, Maryland, USA
[10]LAM-CNRS, Marseille, France
[11]Institut universitaire de France (IUF), Paris, France
[12]Boston University, Boston, USA
[13]Laboratoire de Planétologie et Géophysique, Nantes, France
[14]NASA-JPL, Pasadena, California, USA
[15]Institute of Space Physics and Applied Technology, Peking University, Beijing, China

**Abstract:** This chapter reviews for each province and destination of the Solar System the representative space missions that will have to be designed and implemented by 2061 to address the six key science questions about the diversity, origins, workings and habitability of planetary systems (described in chapter 1) and to perform the critical observations that have been described in chapters 3 and partly 2. It derives from this set of future representative missions, some of which will have to be flown during the 2041-2061 period, the critical technologies and supporting infrastructures that will be needed to fly these challenging missions, thus laying the foundation for the description of technologies and infrastructures for the future of planetary exploration that is given in chapters 5 and 6, respectively.

## Table of Contents

# 1. Introduction: from Earth-based telescopes to sample return and human exploration

This chapter reviews for each province and destination of the Solar System the representative mission architectures that will have to be designed and implemented to move towards a more comprehensive understanding of planetary systems by the 2061 horizon. This exploration will be guided by the six key science questions, Q1 to Q6, about the diversity, origins, workings and habitability of planetary systems, identified as the starting points of the Horizon 2061 foresight exercise:

Q1- How well do we understand the diversity of planetary systems objects?

Q2- How well do we understand the diversity of planetary systems architectures?

Q3- What are the origins and formation scenarios for planetary systems?

Q4- How do planetary systems work?

Q5- Do planetary systems host potential habitats?

Q6- Where and how to search for life?

In chapter 2 (Rauer et al. 2021), these questions were addressed in the broader context of the exploration of all planetary systems. This analysis showed how synergies between exoplanet and Solar System research will play a key role in the elaboration of answers to these six questions, by placing Solar System exploration in the broader context of the study of extrasolar systems and of their objects. Then chapter 3 (Dehant et al. 2021) determined for each of the six key science questions some more specific scientific objectives that should be assigned to missions flying to the different provinces of the Solar System. Table 4.1, which is the ““science objectives vs. mission destinations” matrix of Horizon 2061, provides a concise summary of the detailed scientific objectives (and sometimes) measurement requirements described in these two previous chapters, organized with respect to the six science questions, Q1 to Q6 (along the vertical dimension) and of a list of “destinations” arranged along the horizontal dimension:

- first the solar system destinations (from the Earth-Moon system to the heliopause and ISM), directly reproduced from Table 3.3 of chapter 3);
- and then, in the right-hand side column, the symbolic “extrasolar planetary systems” destination, where additional scientific objectives derived in chapter 2 from the study of solar system-exoplanets science synergies are listed.

| Horizon 2061 science question | Earth-Moon System | Terrestrial Planets | | | Gas Giants | | Ice Giants | | Small Bodies and Dwarf planets | | | | Heliopause and beyond | Extrasolar Planetary Systems |
|---|---|---|---|---|---|---|---|---|---|---|---|---|---|---|
| | | Mars | Venus | Mercury | Planet | Moons | Moons | Planet | Asteroids & Comets | Trojans Irregular moons | TNOs | Dwarf planets | | |
| **Q1- Diversity of objects** | In-depth characterisation of the Moon as the closest model of a terrestrial planet: Internal structure, surface geology, inventory of mineral resources and volatiles, regolith and exosphere | Detailed comparative characterisation of Earth, Moon and Mars: internal structure and dynamics, crust, atmosphere, magnetic field and magnetospheres. | | In-depth characterisation of Mercury: Internal structure, surface geology, inventory of minerals and volatiles, exosphere, magnetosphere | Detailed characterisation of gas giants: structure and dynamics of interiors and atmosphere, magnetic field and magnetosphere | Comparative study of the regular and irregular moons of giant planets: bulk composition, shape and dynamics, internal layering including oceans, geology and surface properties, space environment | | Detailed characterisation of ice giants: structure and dynamics of interiors and atmosphere, magnetic field and magnetosphere | Characterise representative samples of the different populations and taxonomic types of small bodies, with emphasis on the most poorly know ones (Trojans, centaurs, TNO's): mass, density, shapes, cratering, chemical composition, degree of internal layering, refractory, organic and volatile contents | | | Characterise a few dwarf planets: bulk composition, surface properties, degree of internal layering | Explore the population of trans-Neptunian objects, possibly including very distant planets and Oort cloud. Characterise the components of the Very Local Interstellar Medium (VLISM): gas, plasma, B field, GCRs) | Characterize in situ all classes of solar system objects to use them as templates of the diversity of objects in other planetary systems. Characterize ice giants with a degree of precision comparable to other planets |
| **Q2- Diversity of architectures** | Understand the Earth-Moon system as the closest example of a secondary system | Study Phobos and Deimos as a second example of a system of moons around a terrestrial planet | | | Detailed characterisation and comparative study of the four giant planets systems: central planet, ring-moon systems, regular and irregular moon systems, populations of Trojan objects, magnetospheres and their interactions with embedded objects and plasma populations. | | | | Explore the distribution of small body populations in the solar system, particularly on in its outskirts (the Trans-Neptunian Solar System), compare to debris disks around other stars<br>Explore binary systems and secondary systems in the TNO population<br>Revisit the Pluto-Charon system with an orbiter | | | | Characterise and understand the heliosphere as the Solar System's astrosphere<br>Characterise the nature and 3-D shape of heliopause boundaries | Understand the specific architectures of the solar system and giant planet systems in the light of the diversity of other planetary systems architectures |
| **Q3- Origin of Solar System** | Discriminate between the different scenarios of Moon formation | Discriminate between the different scenarios of moons formation | | Understand how a planet forms in the hottest region of the Solar Nebula | Understand the formation scenarios of gas giants and their role in the formation of the solar system | Understand the formation scenarios of giant planets moons and where they applied, and their connection to solar system formation | | Understand the formation scenarios of ice giants and their role in the formation of the solar system | Read the messages of the Solar Nebula and of the formation of the solar system in the elementary and isotopic compositions of the different classes of small bodies.<br>Understand their connection to IDPs and meteorites<br>Remote sensing, then in situ exploration and characterisation, then sample return up to giant planets Trojans and irregular moons | | | | | Use the direct observations of protoplanetary disks around other stars to better understand the formation scenario of the solar system |
| **Q4- How does the Solar System work?** | Past and present dynamics and activity of the Moon, surface-regolith-exosphere-solar wind interactions<br>Dynamics of the Earth-Moon system | Understand the dynamics of the different layers of a terrestrial planet and their interactions.<br>Understand the processes driving their chemical evolution, loss of atmospheric species and climate evolution | | Understand the dynamics of the different layers and their interactions.<br>Understand surface-exosphere-solar wind interactions | Understand how interiors and atmospheres including magnetic field work<br>Understand the "systems within the systems":<br>- Dynamics of rings and coupling with moons<br>- Dynamics of magnetospheres, plasma populations and energetic particles, coupling to host planet and solar wind | Understand how regular moons work including their couplings to their host planet | | Understand how interiors and atmospheres including magnetic field work | Understand and monitor the dynamics of Earth-crossing objects, learn to predict and possibly mitigate their hazards | | | | Understand the transfer processes acting across the different boundaries of the heliosphere and their effect on the transfer of matter, momentum, energy, radiations and particles across the heliopause | Use the solar system as an in-situ laboratory where to study how planetary systems work.<br>Understand the heliosphere as the Sun's astrosphere, and its complex interaction with the Galaxy. |
| **Q5- Search for potential habitats** | Assessment of the role of Earth-Moon interactions in the maintenance of the habitability of the Earth<br>How to make the Moon habitable for humans? | Compare the delivery of water and organic material to the environments of Earth, Mars and Venus.<br>Understand the impact of their environments and atmospheric evolution on their habitability through time | | | | Characterise the habitability of Europa and other ocean moons | Characterise the habitability of Triton, explore Uranian moons | | Explore the water reservoirs of small bodies, assess their respective contributions to the initial water reservoirs of terrestrial planets including Earth | | | Characterise the habitability of the most promising candidates, e.g. Ceres | Understand the role played by the heliosphere in the habitability of the solar system and learn about the role of astrospheres in the habitability of extra-solar planetary systems | Study habitability in the solar system to better define the habitability conditions to be applied to other planetary systems and exoplanets |
| **Q6- Search for Life** | N/A | Active search for extinct and extant life | | | | Search for life at Enceladus, possibly Europa, Titan | | | | | | | | Use the search for life in the solar system to improve our search for biosignatures at exoplanets |

*Table 4.1. This "science objectives vs. mission destinations" matrix displays for each of the six key science questions and for each of the main destinations in the Solar System the main scientific objectives to be assigned to future space and Earth-based observations to address the six key science questions guiding the Horizon 2061 foresight exercise.*

This table illustrates the broad diversity of destinations where planetary exploration missions should go to find answers to the six key questions, and the equally broad diversity of mission architectures and instrumental techniques that will be needed to perform the critical measurements needed. To operate the adequate instruments and perform the broad diversity of measurements (see figure 3.37 of chapter 3), a large diversity of Earth-based observatories and Solar System exploration missions is needed. They can be ordered along a "scale of complexity" of planetary missions. From bottom to top of this ranking:

(1) Observations from Earth's surface;
(2) Cruise and flyby observations;
(3) Orbital surveys of a small body, planet, moon and/or secondary system;
(4) In situ probes to atmospheres and surfaces;
(5) Sample return from these objects to Earth;
(6) Human missions.

In this approach, guided by the broad science questions Q1 to Q6, one does not attempt to present a “road map” of future missions organized and prioritized in time, as space agencies have to do for their regular planning exercises (i.e., Voyage 2050 for ESA, the Decadal Surveys for NASA). Rather, the objective here is to cover as comprehensively as possible the scale of technical complexities required for future planetary missions, so as to describe in the subsequent chapters the technologies and infrastructures that will be needed to fly them.

Before diving into the future, it is useful to look back at what has been accomplished for planetary exploration, from telescope observations to space missions to Solar System destinations of increasing distances and complexities, over the past decades, in the light of this “scale of complexity” of planetary missions: one can grasp at the same time the breadth of the exploration techniques that were developed (orbiters, landers, rovers, constellations, sample returns, human exploration, etc.) and the limitations associated with distances to overcome to reach the different areas of the Solar System. Even given those restrictions, over the past 60 years or so, humankind has been extremely successful in its endeavor to better understand our universe and more specifically our Solar System. Within this relatively short time span, all the major bodies of the Solar System have been explored in situ by spacecrafts, complemented by small bodies like asteroids and comets and the first Kuiper Belt objects (Pluto, Charon and (486958) Arrokoth). Many complex technologies have been developed. Several rovers are currently exploring the surface of Mars, samples have been returned from the Moon and near-Earth asteroids, and twelve men have walked on the surface of the Moon.

As illustrated in the “mission types vs. destinations” exploration matrix presented in Table 4.2, two dynamics of limiting technologies and distances are clearly at work in the way exploration has been performed so far. The specific diagonal shape of the matrix formed by the green cells (missions already performed) illustrates horizontally that the further the bodies of interest are located, the less sophisticated the mission concept applied for exploration, up to the near interstellar medium where only telescopic observations are available. In contrast, nearly all types of space missions have been applied to the exploration of the Moon, our closest Solar System neighbor, for which even human exploration has been realized in 1969. A specific class of bodies is shown in the center part of the table by the minor bodies of the Solar System. In their case, several reservoirs remain to be explored. Small bodies with high eccentricity orbits approaching close to the Earth have often been easier to explore than giant planets.

| | | Earth-Moon system | Mars | Venus | Mercury | Planet | Moons & Rings | Planet | Moons & Rings | Asteroids | Comets | Trojans | KBO | HP Boundaries | VLISM | Oort cloud |
|---|---|---|---|---|---|---|---|---|---|---|---|---|---|---|---|---|
| Human outpost | | | | | | | | | | | | | | | | |
| Utilization of in situ resources | | | | | | | | | | | | | | | | |
| Sample Return | | Apollo, Luna , Chang'e | | | | | | | | Hayabusa | | | | | | |
| In situ exploration | Network | ILRS | | | | | | | | | | | | | | |
| | Mobile | Luna, Chang'e | MER, MSL, Mars 2020 | | | | | | | MASCOT | | | | | | |
| | Station | Luna, Chang'e | Phoenix, Insight | Vega | | | | | | | Philae | | | | | |
| | Penetrator | LCROSS | | | | | | | | ARM | Deep Impact | | | | | |
| | Atm. probe | | | Vega | | Galileo | Huygens | | | | | | | | | |
| Orbital Observation | Small satellites | | | | | | | | | | | | | | | |
| | Orbiter | LRO, Chandrayaan, SMART-1, Selene, Chang'e, etc. | Mex, MRO, Odyssey, Maven, TGO, Mangalyaan | Magellan, Vex, Akatsuki | Messenger, Bepi-Colombo | Galileo, Cassini, Juno, JUICE, EC | JUICE | | | Dawn, NEAR, Psyche | Rosetta | | | | | |
| Fly-by | | | | | | | | | | | Giotto, Stardust, Deep Impact Comet interceptor | Lucy | 1st Interstellar Heliospheric Probe | | | |
| | | | | | | | | | | | | | New Horizon | Voyager | | |
| Meteorites and cosmic dust collections | | | | ? | | | | | | | | | | | | |
| Telescope observations from Earth and its orbit | | | | | | | | | | | | | | | | |
| **MISSIONS ADDRESSING THE KEY SCIENCE TO FLY: BY 2040 (green) BY 2061 (light blue)** | | **Terrestrial planets** | | | | **Gas giant systems** | | **Ice giant systems** | | **Small bodies** | | | | **Heliopause and beyond** | | |

*Table 4.2: "Mission types vs destinations » matrix showing the "green missions" to the different Solar System destinations (missions already flown, in planning or in preparation).*

In order to push the limits of exploration forward, one will need to progressively extend the coverage of this table by new missions diagonally towards its upper right corner, developing ever more complex and ambitious missions towards more distant planetary targets: this is what the following sections will show.

The invaluable scientific return of the Apollo missions shows that human exploration dramatically enhances the scientific output that would otherwise be provided by robotic instruments. Humans may step on Mars in the time frame of Horizon 2061, and would open tremendous possibilities in terms of sample selection and collection, or infrastructures for deep drilling, for example. On the far side of the Moon, humans could build a large network of radio-telescopes where astrophysicists would benefit from a low-noise environment unrivalled on Earth. However, we will focus at the end of this chapter on identifying the technological needs of future robotic missions. In other terms, we will not elaborate the technological requirements which are essentially specific to human exploration, in areas such as, for example, life support, nuclear fission reactors, large heatshields capable of landing 20 metric tons payloads on Mars. The preparation of human exploration will be dealt with in as much as science can support it, through environmental characterization, Mars climatology, water mapping, and contribution to landing site selection.

This chapter will start with a description of the many high-value observations that can and should be done remotely using ground-based and space-based telescopes. The new generation of giant telescopes that will become progressively available in the coming decades will provide unprecedented insight into the most distant objects of the Solar System, and explore the diversity of their different families. Then the key contributions of planetary missions performing in situ measurements and enabling sample return will be presented for each province and each family of Solar System objects. Missions already in planning or in preparation (the “green missions”) and new notional missions to be flown after them (the “blue missions”) will be presented. Then the set of green and blue missions will be used to identify the new technology requirements and infrastructure support needs generated by these missions.

# 2. Exploring the Solar System with Earth or space-based telescopes

In Solar System science, Earth-based observations (conducted from the ground and space) complement *in situ* planetary missions in important ways: they provide preliminary characterizations of bodies prior to missions; they add wavelengths that complement those accessible to *in situ* missions; they generate large-scale context data for single-site *in situ* missions; they have long timelines to provide broader perspectives on time-variable phenomena; they can identify mission targets for future spacecraft exploration; and more.

In fact, some of the most important scientific discoveries in the field of planetary science have been made with Earth-based observatories, such as the first detection of a transneptunian object (after Pluto) (Jewitt and Luu 1993); the first exoplanet (Mayor and Queloz 1995) ; the first interstellar objects (Meech et al. 2017; Guzik et al. 2020); the magnetic field of Jupiter via its radio emission (Burke and Franklin 1955); and more. As a specific example, the Hubble Space Telescope has contributed to many areas of Solar System research. To name just a few: comprehensive studies of the Kuiper Belt and distant comets (e.g., Fraser and Brown 2012; Li et al. 2020b); context imaging and spectra for Mars missions (Bell and Ansty 2007; Mumma et al. 2009); assessing the stability of atmospheric features on the giant planets (e.g., Hsu et al. 2019; Hueso et al. 2020); imaging auroral activity on giant planets (Yao et al. 2019); characterizing the ocean interiors of giant planet satellites (Saur et al. 2015); and discovering moons and a follow-up target for the New Horizons missions to Pluto and beyond (Porter et al. 2018). In a similar example, a pair of recent review articles highlight science across the Solar System from the Spitzer Space Telescopes, with contributions to the science of comets, centaurs, and Kuiper Belt Objects (Lisse et al. 2020), and advances in our knowledge of asteroids, planets, and the zodiacal cloud (Trilling et al. 2020).

In this section, we briefly outline the key science questions to be addressed in Solar System science in the coming decades and how the future Earth- and space-based facilities will be used to provide elements of answer to these questions.

## 2.1. Outstanding niches for observation of Solar System objects by Earth-based telescopes

### 2.1.1. Small bodies

The huge diversity of families of small bodies, reviewed in chapter 3 (Dehant et al. 2021) makes it impossible to visit even a statistically representative subset of them by space probes. Space probes play a unique role in their close characterization, giving access to their mass (from radio science during fly-by), shape (space probe can approach them at distances where they can be resolved) and therefore density, and to the physical state, cratering record and chemical composition of their surface and tentative insights of their subsurface properties (Kofman et al. 2015). Even more importantly, they can return to Earth samples collected on their surfaces. But the number of objects thus visited will remain small.

Moreover, orbital exploration of these objects, following the Rosetta example, and even more sample return, are more and more costly and technically challenging as their distance from Earth increases. For Trans Neptunian objects, for instance, it is unlikely that a handful of them will receive the visit of orbital or sample return missions, even by the 2061 horizon.
On the same time scales, however, the giant telescopes that will come into operation either at the Earth's surface or in Earth-Moon orbit (to be described in sections 2.2 and 2.3 below) will offer new and unprecedented spectral coverage and resolution and spatial resolution capacities. They will be perfectly adapted to the spectroscopic and photometric characterization of large populations of small bodies at increasingly large distances, provided that enough observing time is given to their observation. In particular, large surveys of TNOs, a population poorly covered by past observations, will likely lead to new views of this population whose exploration is critical to improving our understanding of the formation scenario of the Solar System (see e.g., Khain et al. 2020). Deep surveys of these distant populations will also likely lead to a large number of new detections, including those of pristine objects on hyperbolic or highly eccentric orbits transiting for the first time through the inner regions of the Solar System.
Any science plan aiming at the well needed order-of-magnitude improvement in the knowledge of these classes of outer Solar System small bodies will need to combine deep surveys of large numbers of objects with a few space missions dedicated to a small number of them. Some of those missions may be decided on alert, following the example of Comet Interceptor (Snodgrass and Jones 2019).

### 2.1.2. Planets and their moons

Unlike asteroids or airless small bodies, planets and their moons are the seats of temporal changes on timescales prone to characterization by remote observations repeated to capture the relevant time scales. This is true both for objects with "dense" atmospheres (Venus, Mars, the Giant Planets, Titan) where atmospheric dynamics undergo seasonal or unpredictable changes, with timescales from hours to tens of years, and for moons and other planet-sized bodies (such as Pluto) characterized by interactions between their sub-surfaces, surfaces and atmospheres. Earth-based observations also provide information about the seasonality of the exosphere of Mercury.

### 2.1.3. Planetary atmospheres

Space missions to Venus, Mars, and the Jupiter and Saturn systems have amply shown that planetary atmospheres are strongly coupled objects. They are the place of complex interplay between gas composition, neutral and ion chemistry, haze and cloud formation, thermal field, vertical mixing, atmospheric dynamics and escape, all themselves strongly dependent on the seasonal variations of insolation whenever is the case (Mars, Saturn and Titan). As an example, Cassini has secured detailed 3-D maps of composition, hazes, and temperatures of Titan during its 13-year exploration (see review in Hörst 2017). Space probes have also enabled unique measurements on elemental and isotopic abundances not accessible remotely, including the long-awaited measurement of the O/H elemental ratio in Jupiter (Li et al. 2020a) as well as evidence for an unexpected horizontal/vertical variability of condensable species below the clouds. Most remarkably, they have also determined the vertical extent of the deep winds in Jupiter and Saturn and their internal structure (Kaspi et al. 2018, Iess et al. 2019) based on gravity field measurements. All of these measurements bear fundamental constraints both on formation models and on atmospheric meteorology below cloud tops, including the role of small-scale convection and lightning (Guillot et al. 2020). Obviously, the role of space exploration for these increasingly well-characterized bodies is being overwhelming. However, there are still areas in which remote observations from Earth are invaluable and will remain so for some time in the future:

- Some measurements are achievable by techniques not yet carried out on space probes. This includes for example wind measurements from Doppler spectroscopy, or probing atmospheres above the clouds where no wind tracers are available (e.g., Lellouch et al. 2019 for Titan). The JUICE probe will carry a submillimeter instrument capable of such measurements, for the first time on a planetary mission.

- The time basis, time resolution and rapid response offered by ground-based observations is suited to the study of fundamentally changing phenomena, often in an unpredictable way. Even the 13-year baseline offered by Cassini represents a little less than half a Saturn/Titan year. Large meteorological disturbances such as Saturn's superstorms, that normally occur once a Saturn year, do not repeat in a regular fashion (the 2011 storm observed by Saturn happened 10 years « in avance »), and characterizing those is key to understand how moist convection occurs and how episodic storms may carry heat flux from the interior. Externally-driven events such as cometary impacts occur essentially randomly. Yet, by modifying abruptly the atmospheric composition and thermal state, they provide unique opportunities to constrain chemical and dynamical processes on timescales from minutes to tens of years.

- Besides short flybys by Voyager 2 in 1986 and 1989, Uranus and Neptune haven't been explored by spaceborne missions. Yet, as much as the Gas Giants, these Ice Giants (that also serve as benchmarks for the numerous sub-Neptune exoplanets discovered) display complex meteorology (e.g. strong equator-to-pole gradients in volatiles) and show surprisingly large disparities in terms of observable atmospheric composition both in stratospheres and tropospheres (see e.g. reviews by Moses et al. 2020 and Cavalié et al. 2020), that might ultimately be related to the vast differences in their internal heat flux and axial tilt. Ground-based observations are needed to monitor the observable meteorology and winds on timescales relevant to their orbital periods (84 and 165 years, respectively). Thirty-meter class telescopes (ELT, TMT) will have diffraction-limited resolution of 8 mas at 1 µm, i.e. 200 km at Neptune's distance, almost rivalling the Voyager-achieved resolution of ~80 km.

### 2.1.4. Moons, ocean worlds, dwarf planets

A number of satellite / dwarf planets show evidence for the presence of sub-surface oceans. Confirmed ocean worlds include Europa, Ganymede, Titan and Enceladus, and bodies such as Pluto, Triton, Dione and Ceres are considered as candidates (Hendrix et al. 2019). Among those, Enceladus (e.g. Porco et al. 2006) and Europa (e.g. Roth et al. 2014) stand out as objects with direct evidence of interaction between the sub-surface, surface, and a tenuous, localized and time-variable, atmosphere sourced by plume activity (episodic only at Europa). Io is another example of a (volcanic, in this case) interior ultimately determining the object's atmospheric and surface composition, injecting sulfur species ($SO_2$, SO, $S_2$) and salts (NaCl, KCl) to form a permanent but time- and space- variable atmosphere, that ultimately affects the Io plasma torus and the surface chemistry of both Io and the other Galilean satellites. Cryo-volcanism is also thought to be powering Triton's plumes.

Given the inherently time-variable character of (sub)-surface activity, monitoring observables (atmospheric density and composition, surface composition, surface morphology) on these bodies with high time sampling is needed, and achievable only from ground-based monitoring.

The atmospheres of Pluto and Triton result more directly from sublimation-condensation exchanges with volatile-rich, spatially heterogeneous, surfaces. Such exchanges vary along the orbit with the interplay of seasonal and (in Pluto's case) heliocentric distance effects, and are thought to lead to the redistribution of the volatile ices ($N_2$, $CH_4$, CO) with time, and to attendant evolution of the atmospheric composition and structure. While the New Horizons encounter has provided an exquisite view of Pluto's system, it was limited to a single point in time in 2015. Temporal monitoring over orbital timescales is needed, to address for example the question of the fate of Pluto's atmosphere when it will reach its 49.3 au perihelion by the year ~ 2110. Stellar occultations (e.g. Sicardy et al. 2016), which do not require large instrumentation, but multi-telescope campaigns, are ideally suited to monitor the evolving surface pressure. On the other hand, large ground-based facilities such as ALMA, the VLT, and in the near future the Webb and the ELTs (that do/will provide spatially-resolved information on Pluto's surface) are needed to follow the detailed aspects of Pluto's complex climatic cycle. The same considerations holds for other KBOs susceptible to harbor atmospheres (e.g. Eris, Makemake).

## 2.2. Facilities with expected implementation by 2035

Several giant ground-based telescopes and new space telescopes are currently under development and planned to start operations before 2035. This section gives a short review of their capacities, timeline and relevance to Solar System observations.

The **James Webb Space Telescope (Webb)** is a joint NASA/ESA/CSA mission to launch an infrared-optimized 6.5-meter space telescope (Gardner et al. 2006). WEBB will launch to the Sun-Earth Lagrange 2 point on an Ariane 5. Its early 2021 launch date has been deferred due to schedule slow-downs introduced by the covid-19 crisis. WEBB's portfolio has included Solar System observations since its formal inception. A comprehensive overview of the planned planetary observations, ranging from near-Earth asteroids out to the distant Kuiper Belt, can be found in a series of papers in a special issue of the Publications of the Astronomical Society of the Pacific (Milam et al. 2016 and references therein). Also, all of the planned Solar System proposals from the Guaranteed Time observations and Early Release Science programs are publicly accessible (see https://www.stsci.edu/jwst/observing-programs/approved-gto-programs).

In a nutshell, the high sensitivity and broad wavelength coverage afforded by WEBB with NIRSpec (0.6-5 μm) and MIRI (5-28 μm) will enable, for the first time, spectroscopy of faint outer Solar System small bodies (>5 au) in spectral regions where the absorption and emission bands diagnostic of ices, silicates, and organic materials can be detected and analyzed. The WEBB will also provide new constraints regarding the surface composition (especially for the volatiles, the hydrated minerals and the organics) of inner Solar System small bodies (<5 au) as former generation telescopes could not allow properly investigating the 2.5-5 micron spectral region for these bodies. The study of most bright planets and their satellites will benefit from the spectroscopic capabilities of the WEBB in the near to mid infrared (Norwood et al. 2016).

**Roman Space Telescope (Roman)** (formerly WFIRST) is a wide-field 2.4-m telescope planned for launch after 2024. This mission, the top-ranked priority in the previous US Astrophysics Decadal Survey (CDSAA 2011), will also host a robust Solar System program for observations that take advantage of its wide field. The Wide-Field Instrument is a 288-megapixel multi-band near-infrared camera covering a field of view of 0.28 square degree, 100 times the one of HST while keeping a similar resolution (Figure 4.1). This capability will be especially attractive to small-body science such as near-Earth objects, main-belt asteroids, comets, and Kuiper Belt objects, where statistical surveys provide important complements to in situ missions to a very limited number of individual bodies. Potential groundbreaking discoveries from Roman could include detection of the first minor bodies orbiting in the inner Oort Cloud, identification of additional Earth Trojan asteroids, and the discovery and characterization of asteroid binary systems similar to Ida/Dactyl (Holler et al. 2018).

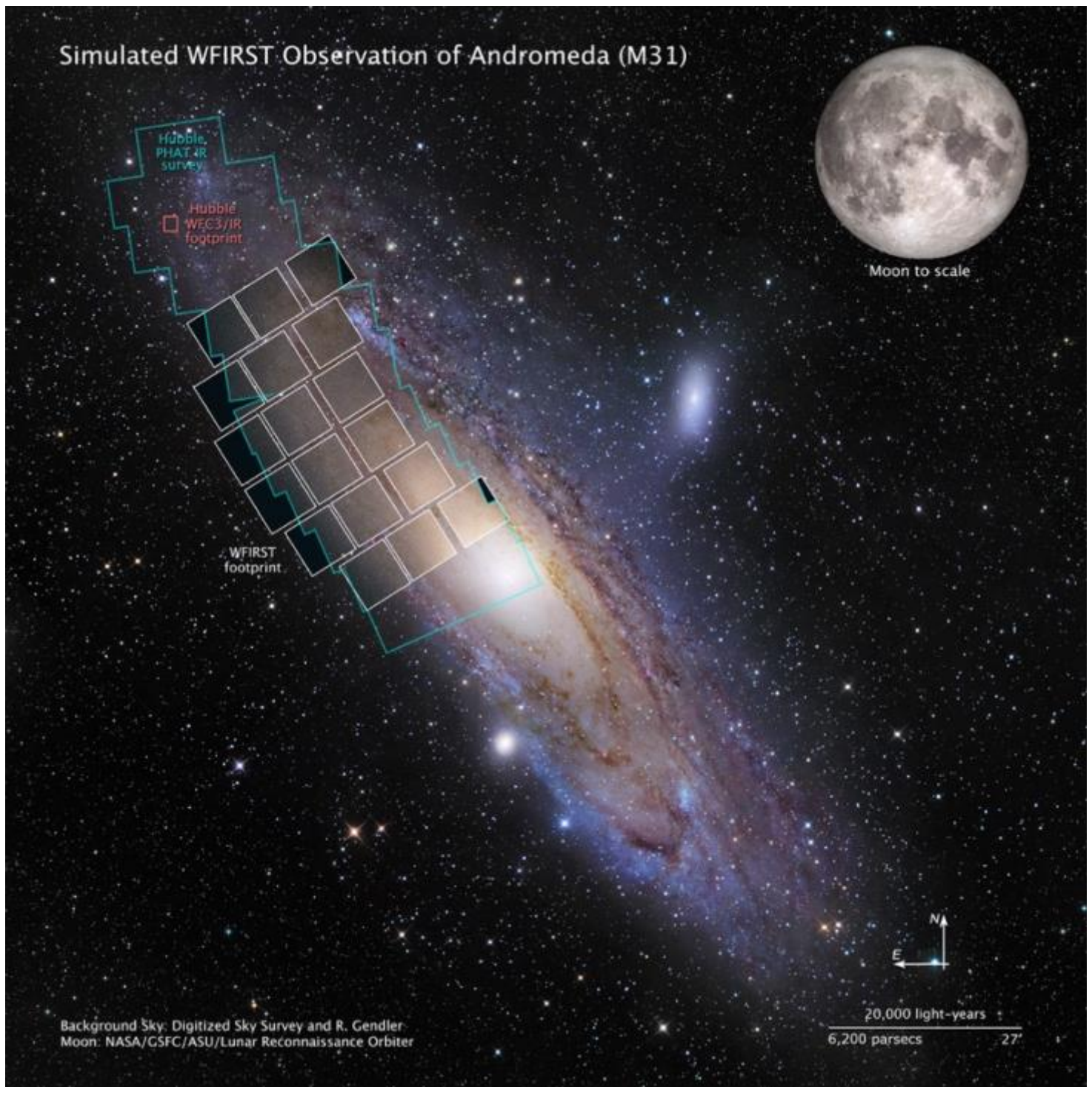

*Figure 4.1. A comparison of the Roman Space Telescope and HST field of views views (credit NASA/GSFC/ASU/Lunar Reconnaissance Orbiter)*

**European Extremely Large Telescope (E-ELT)** is a 39-m telescope under construction in Chile, with first light anticipated in 2025 (McPherson et al. 2012, Ramsay et al. 2020). Future ELT adaptive-optics imaging observations of main belt asteroids will allow to resolve craters down to ~2-5 km in size implying that we will be able to characterize their geological history from the ground in addition to the global 3D shape of D>30km asteroids. Resolving more distant small bodies will become possible with the ELT implying that the shape of the largest Jupiter Trojans and TNOs will be constrained. In terms of surface composition, it will be possible to collect spectra for TNOs as small as New Horizons' target Arrokoth. In a different register, ELT observations of Jupiter with the near-infrared integral field spectrograph HARMONI will have a higher spatial resolution (at least a factor of 3) than those performed in-situ by the ESA JUICE mission with the MAJIS near-infrared imaging spectrometer. An illustration of the ELT's resolution is given in Figure 4.2.

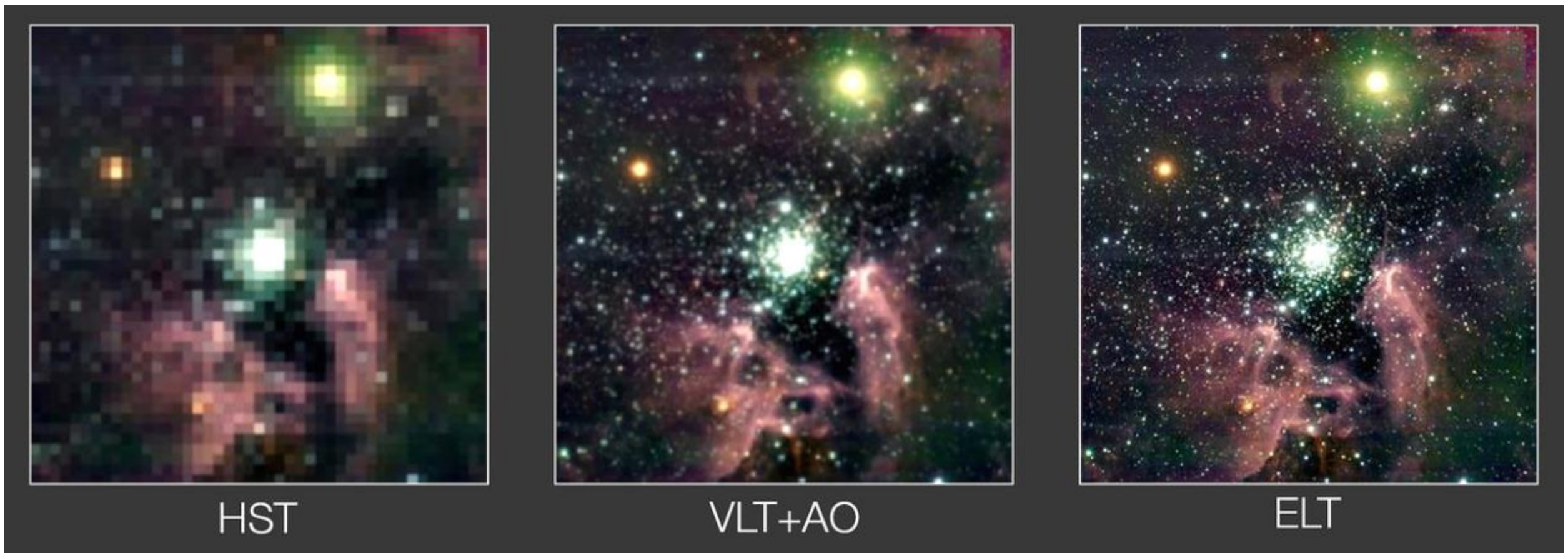

*Figure 4.2. This illustration shows how the nebula NGC 3603 as observed by three different telescopes: the NASA/ESA Hubble Space Telescope, ESO's Very Large Telescope with the help of its adaptive optics modules, and the future Extremely Large Telescope. NGC 3603 is a star-forming region in the Carina spiral arm of the Milky Way, 20 000 light-years away from Earth. Note that this illustration is approximate. (credit ESO)*

**Thirty Meter Telescope** (**TMT**, first light 2027) and the **Giant Magellan Telescope** (**GMT**, first light 2029) are working collaboratively with the US-Extremely Large Telescope Program (US-ELTP) to create two telescopes of similar class to the E-ELT, with one telescope in each hemisphere for full-sky coverage. Both facilities offer the promise of exquisite Solar System science with their expected high spatial resolution and high sensitivity. They have included teams of planetary scientists in their science planning (Chisholm et al. 2020, Fanson et al. 2020, Wong et al. 2021).

Figure 4.3 illustrates a comparison between the size of the primary mirrors of all the future giant telescopes discussed.

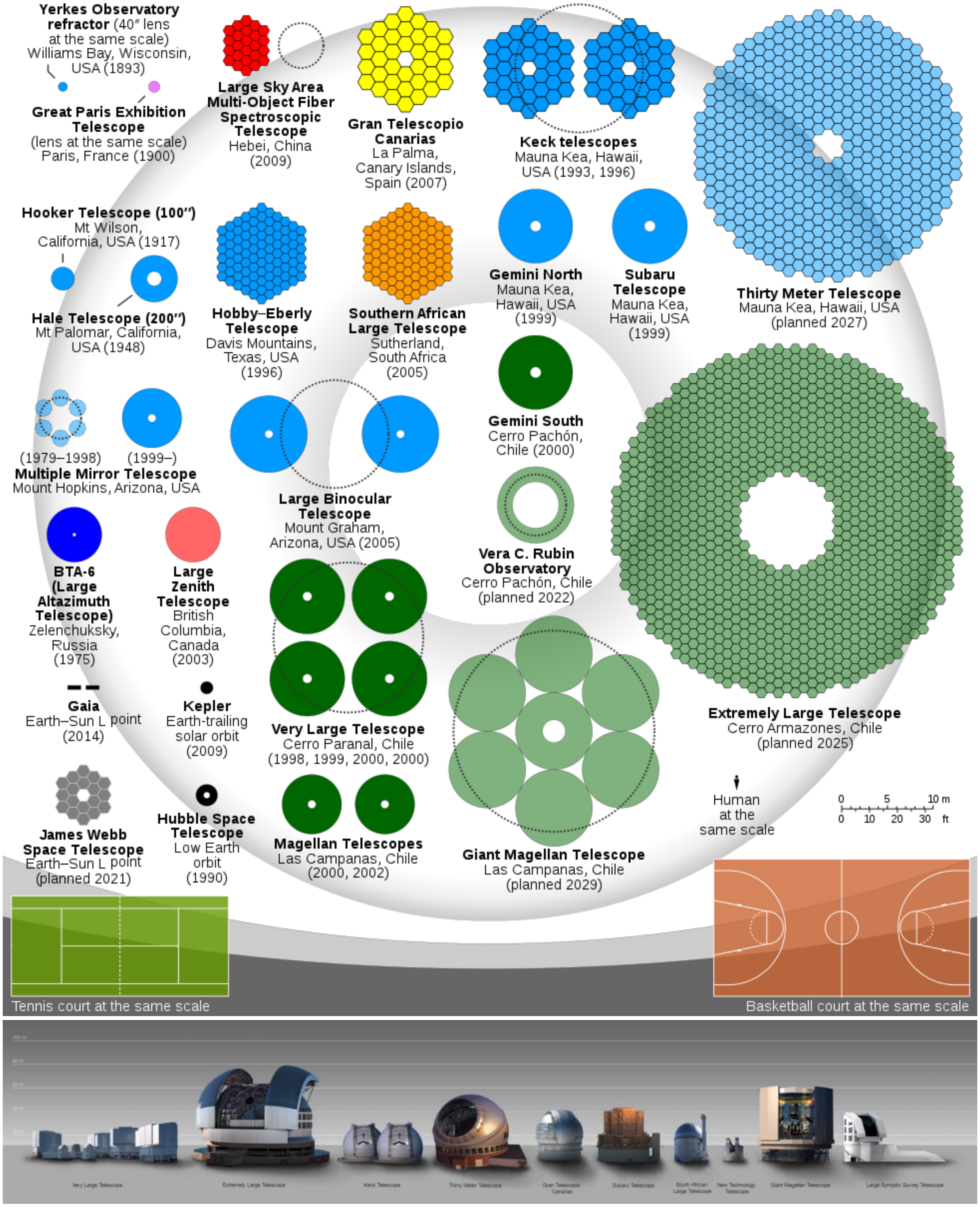

*Figure 4.3. A comparison of the size of primary mirrors and domes for ground-based astronomical telescopes (credit Wikimedia commons)*

**Next generation Very Large Array (ngVLA)** is a planned significant upgrade to the existing VLA. **ngVLA** will be ~10 times more sensitive and provide ~10 times higher spatial resolution than the current VLA and the Atacama Large (sub)Millimeter Array (ALMA). VLA has produced significant understanding of giant planet atmospheres (e.g. de Pater et al. 2019). ngVLA will produce unique scientific observations for giant planets (e.g., full maps over a broad wavelength range, and mm-wavelength spectroscopy), providing invaluable complementary information to in situ space missions. It also is a potential ground station for spacecraft telemetry (Bolatto et al. 2017).

## 2.3. Facilities and capabilities under study for operations beyond 2035

Building on new and emerging technologies, a new generation of even more ambitious telescopes is currently under study and should become operational after 2035. While they are designed to respond to the extreme measurement requirements such as the characterization of biosignatures in the atmospheres of Earth-like exoplanets (see chapter 2, Rauer et al. 2021), their unprecedented capacities will open totally new possibilities for the observation of Solar System objects, such as for instance imaging of TNOs and some of their dwarf planets. This section gives a short review of their capacities, expected development timelines and applications to Solar System observations.

**NASA's "New Great Observatories"** refer to four flagship concept studies conducted for the US Astro2020 decadal survey activity. All four space telescope—nicknamed LUVOIR (illustrated in Figure 4.4), Lynx, Origins, HabEx—included Solar System observations in their science plans. As was true for Hubble and Spitzer, and will be true for Webb and Roman, there will be unique ground-breaking science results from these facilities to complement future in situ space missions (see *greatobservatories.org* for the complete study reports). Like WEBB, these facilities would likely support international partnerships. The first gate is a top ranking in Astro2020; thereafter, discussions with interested parties will begin in earnest.

On the ground-based side, astronomers are already contemplating what the future may hold. Whether the traditional path is followed to the increased aperture of an "over-whelmingly large telescope," or a new trail is blazed into optical interferometry, there will undoubtedly be use cases for Solar System observations. Whether they map the surface spectroscopy of Kuiper Belt objects and outer Solar System moons, find evidence for activity of cometary nuclei at extreme distances from the Sun, or uncover the source of the time variability of Ice Giant ring systems, these advanced facilities offer the promise of new vistas in planetary exploration.

## Solar System Science with LUVOIR

The bodies of the solar system, from smallest Kuiper Belt objects to the giant planets, represent a treasure trove of information on the formation of the solar system, atmospheric processes, and dynamical evolution. LUVOIR's observational capabilities in the solar system extend from the orbit of Venus outward. Here, we highlight a few additional solar system science cases LUVOIR could address beyond the Signature Science Cases described in this report. For some bodies, LUVOIR's imaging is comparable to flyby and orbiter quality.

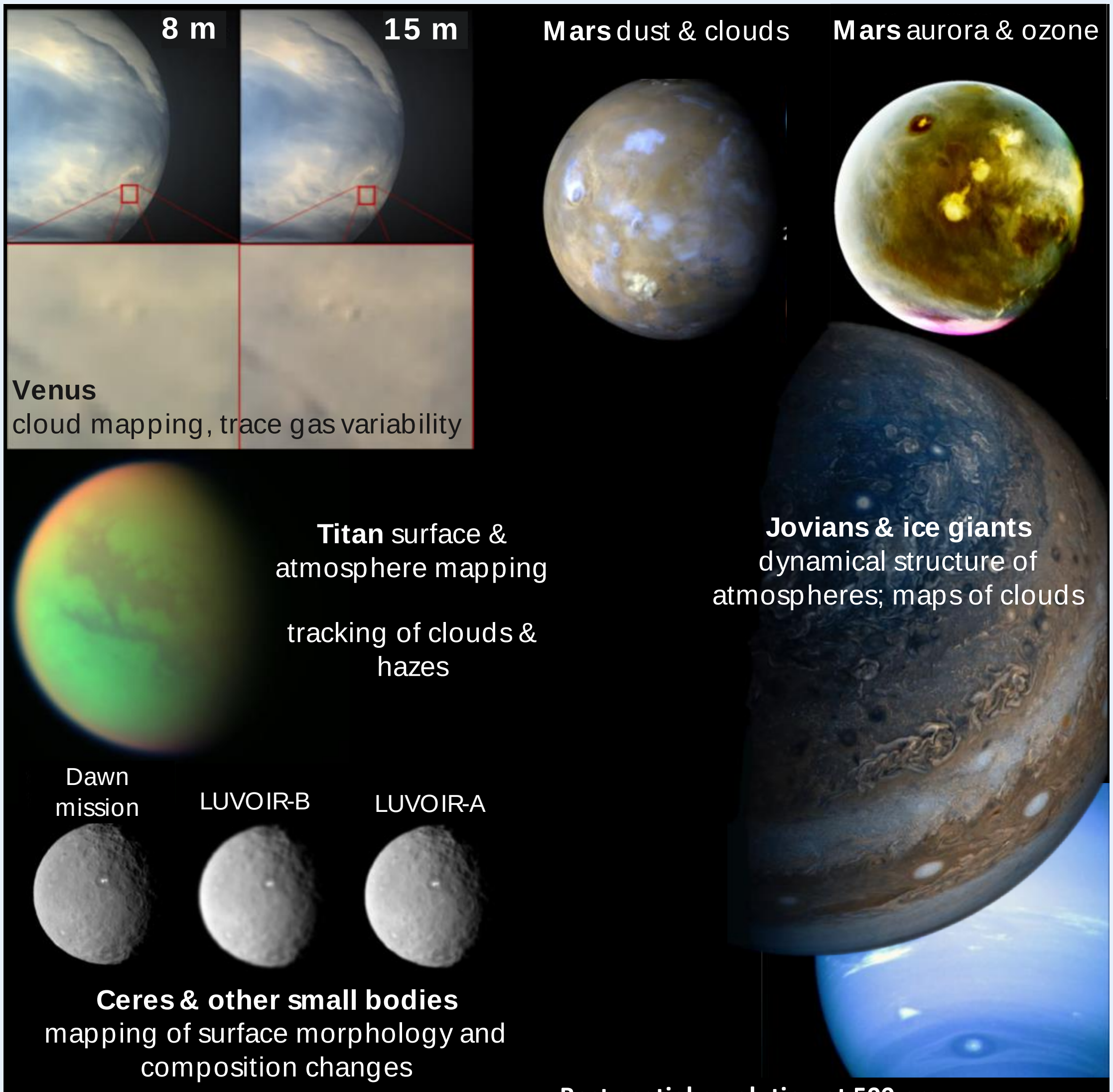

**Best spatial resolution at 500 nm**

| | LUVOIR-A | LUVOIR-B |
|---|---|---|
| **Venus, Mars, Ceres, Jupiter** | 7 km, 3 km, 11 km, 25 km | 14 km, 5 km, 20 km, 47 km |
| **Saturn, Uranus, Neptune, 40 AU** | 51 km, 108 km, 173 km, 232 km | 96 km, 204 km, 327 km, 438 km |

*Figure 4.4. Solar System science enabled by LUVOIR (LUVOIR Team, 2019)*

## 2.4. Summary

Astronomical facilities have for generations aided in our quest to understanding the formation, evolution and workings of objects within our own Solar System. With the emergence of more and more challenging space missions which will likely focus only on a small subset of these objects, the expected contributions of Earth-based telescopes to Solar System exploration will remain extremely important: only they will be able to provide broad and deep spectroscopic surveys of the largest and most distant populations of objects such as comets and trans-neptunian objects, and to capture both the short-term and multi-decennial variability of planets and moons. Earth-based observations will also continue to be the dominant sources of discovery of new objects, and their observations will be instrumental in the preparation of future space missions. Looking to the future, these continued synergies between Earth-based observations and in situ exploration will be one of the main drivers towards satisfactory answers to the most burning scientific questions about the Solar System and its objects.

# 3. In situ Space Missions to the different provinces of the Solar System

Addressing the six science questions of the Horizon 2061 exercise, we will divide the different destinations of Solar System exploration into six provinces, from the nearest to the farthest:

- The Earth-Moon system,
- Venus,
- Mars,
- Mercury,
- Giant planets and their systems,
- Small bodies: asteroids, comets, Trojans, Trans-Neptunian Objects,
- The "frontier regions" of the Solar System, extending from the Trans-Neptunian Solar System to the interstellar medium

For each of these provinces, this section will first summarize the expected contributions of their exploration to an improved understanding of the Solar System. It will then successively describe the missions that will fly in the 2021-2040 time scale, most of which are already defined and in preparation, and the new set of notional missions that should fly in the 2041-2061 period to address the six science questions of the Horizon 2061 exercise.

## 3.1. The Earth-Moon system

### 3.1.1. Main scientific objectives of Moon exploration.

**The Earth's Moon** is of special interest as a direct witness to the Earth-Moon system origin and as a track record of the Solar System evolutionary history in the near vicinity of Earth. A large fleet of space missions have continued lunar exploration even after the major achievement reached by the human exploration from the Apollo program. Moon samples in our laboratories can be investigated via a suite of modern instruments. In the special context of the Horizon 2061 exercise, lunar science can be addressed as the study of the smallest and most accessible of secondary systems: the Earth-moon system.

Studies of this system cover a broad spectrum of science questions, including the six science questions of Horizon 2061 (see also Table 4.1):

- Q1- Diversity of objects : the Moon occupies a special place among Solar System objects, as the closest model of a terrestrial planet;
- Q2- Diversity of systems: indeed, the Earth-Moon system is the closest secondary system in the Solar System, and the most accessible one;
- Q3- Origin of planetary systems: well-chosen space measurements may provide the capacity of discriminating between the different current models of formation of the Earth-Moon system, including the giant impact model, a particularly interesting scenario among other moon formation scenarios which can best be studied at Earth's Moon; the Moon also preserves a unique record of the history of collisions in the early Solar System, which is far from having been fully interpreted.
- Q4- How does the Solar System work? The Moon offers a unique laboratory to study and understand the dynamics of the different layers of a terrestrial planet and their interactions.
- Q5- Search for potential habitats: What is the role of Earth-Moon interactions in the maintenance of the habitability of the Earth? How to make the Moon habitable for humans in the future?

Major progress in lunar science are expected to be achieved by acquiring an ever larger collection of diverse lunar samples, and especially by exploring the South Pole-Aitken Basin, the oldest and deepest observed impact structure on the Moon, where lower crust and possible lunar mantle samples may be found (NRC 2007). This section reviews the current, planned and future space missions that will address this broad range of questions.

### 3.1.2. Current and future missions up to 2035: preparing human missions to the Moon

The number of current and planned near-future missions to the Moon is steadily increasing, showing the rising scientific interest, as well as the interest to bring back humans to the Moon with the Artemis program, and even to use the Moon as a base for further manned Solar System exploration.

The latest technical achievements and results of recent missions (SMART-1, Kaguya, Chang'e-1, Chandrayaan-1 &2 orbiters, LCROSS, LRO, GRAIL, LADEE) were discussed at a plenary panel and technical sessions, with the Lunar Reconnaissance Orbiter (LRO) still in operation. All these missions made particular progress towards a better understanding of the lunar processes (impact processes, interior and surface properties, as well as dust environment and volatile cycles, Denevi et al. 2018). In particular, the set of Chang'e 3 to 5 missions results highlighted the importance of these new lander missions and sample returns for understanding the Moon, with Chang'e 4 being the first landed mission to explore the far side of the Moon (Li et al. 2021a).

Recent and future robotic missions to the Moon participate from several different dynamics. The state-led robotic exploration remains a major leader, with the Chinese Chang'e program which successfully landed a rover on the far side of the Moon for the first time with Chang'e 4 in 2019 (Jia et al. 2018; Li et al. 2021a) and successfully returned the first lunar samples in 45 years with Chang'e 5 in 2020 (see Figure 4.5; Qian et al. 2021; Xiao et al. 2021). Further Chinese missions to the moon include Chang'e 6 in 2023-2024 to investigate the South Pole area and return samples and Chang'e 7-8 are planned to explore resources and demonstrate utilization in this decade (Li et al. 2019). India also successfully inserted its Chandrayaan-2 orbiter around the Moon in 2019, but its attempt at landing a rover failed (Mathavaraj et al. 2020). Other robotic lunar missions prepared for this decade include the Russian Luna program from Luna 25 in 2021, Luna 26 orbiter in 2024, Luna 27 lander to characterize lunar ice 2025, and Luna 28 to 31 to study technology for a future lunar base (Zelenyi 2018); the plans from South Korea to launch a Pathfinder Lunar Orbiter (KPLO) in 2022-23 (Ju 2017) and the Korean Lander Explorer (KLE) with orbiter, lander and rover in 2025 (Kim et al. 2016) and Europe's preparation for a Large Logistic Lander (Gollins et al. 2020). Commercial initiatives are also starting with the failed attempt by Space IL Beresheet to land on the Moon in 2019 (Gibney 2019) and the NASA CLPS Commercial Lunar Payload Services initiative (Voosen 2018), which include the CLPS Masten to South Pole, Intuitive Machines mission with Drill for 2022 and the CLPS Astrobotic VIPER to South Pole and the Firefly Aerospace to non-polar region missions at the 2023 horizon. NASA also prepares the robotic lander MoonRise mission among its potential candidates for the next New Frontiers mission call, targeting sample return from the South Pole–Aitken basin.

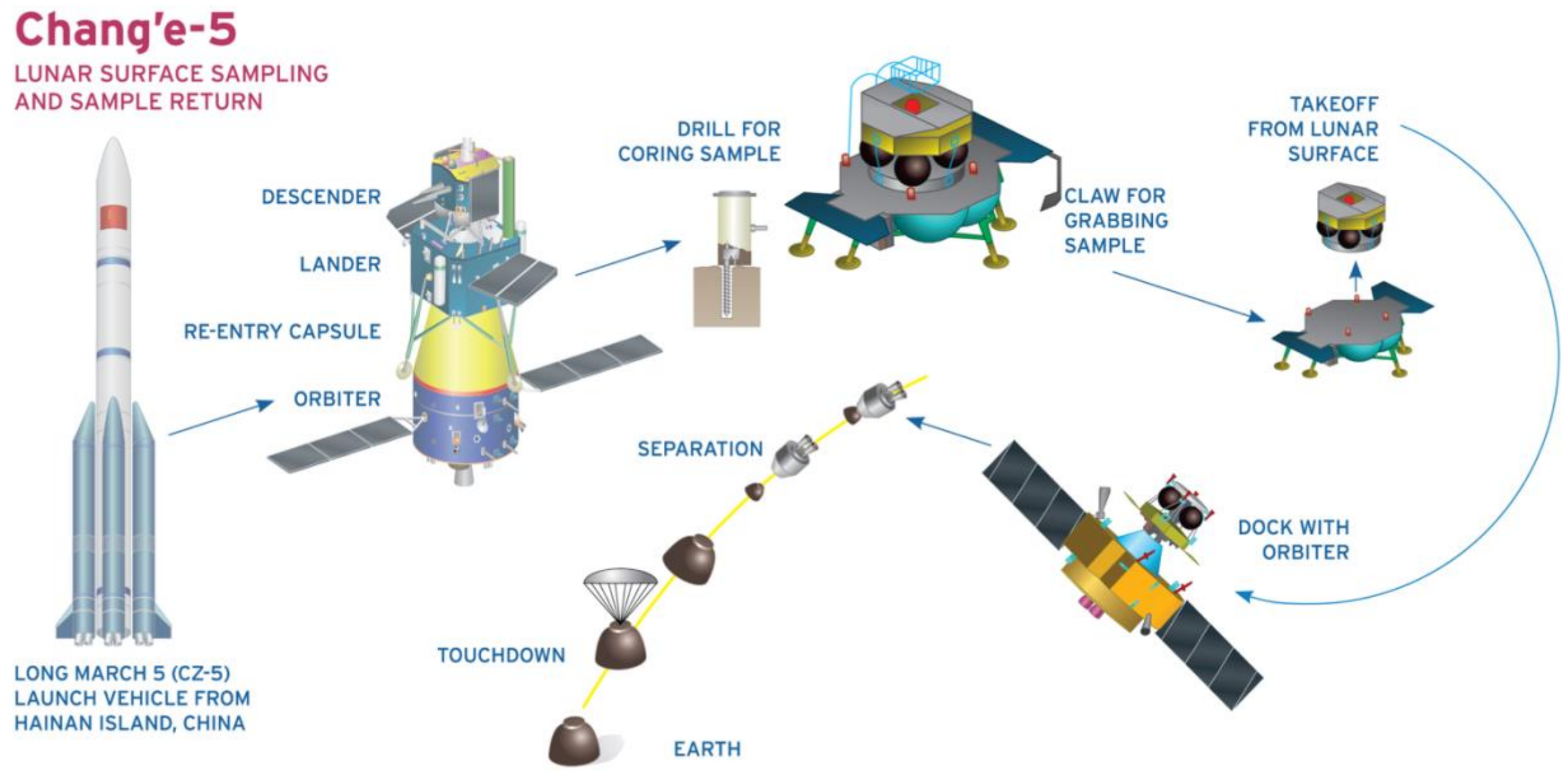

*Fig 4.5. Architecture for the Chang'e 5 Lunar sample return mission from CNSA (Wikimedia commons)*

Finally, a major endeavor from NASA and its partners on this program, including ESA, is the planned human return to the Moon with the Artemis program (see Figure 4.6), which will also allow significant embarked science In this framework, the planned missions so far correspond to the Artemis 1 missions (NASA/ESA) with Orion/ESA service module in cis-lunar orbit in automatic mode (no crew) with 13 secondary payload cubesats by 2021 (McIntosh et al. 2020), then Artemis 2 bringing 4 crew in lunar orbit for 10 days in 2023; Artemis 3 with 4 crew in orbit, and 2 on surface in 2025 and yearly Artemis missions, with surface human landing and operations expected thereafter. This architecture also makes use of a cislunar orbital space station named the Gateway, which is expected to be operational around 2024. Visions going beyond the mid-2020s include manned missions and the set-up of structures suitable for humans, like villages or gateways for future exploration of the Solar System (Arvidson et al. 2010). For an updated list of current and future lunar space missions, see e.g. https://nssdc.gsfc.nasa.gov/planetary/planets/moonpage.html

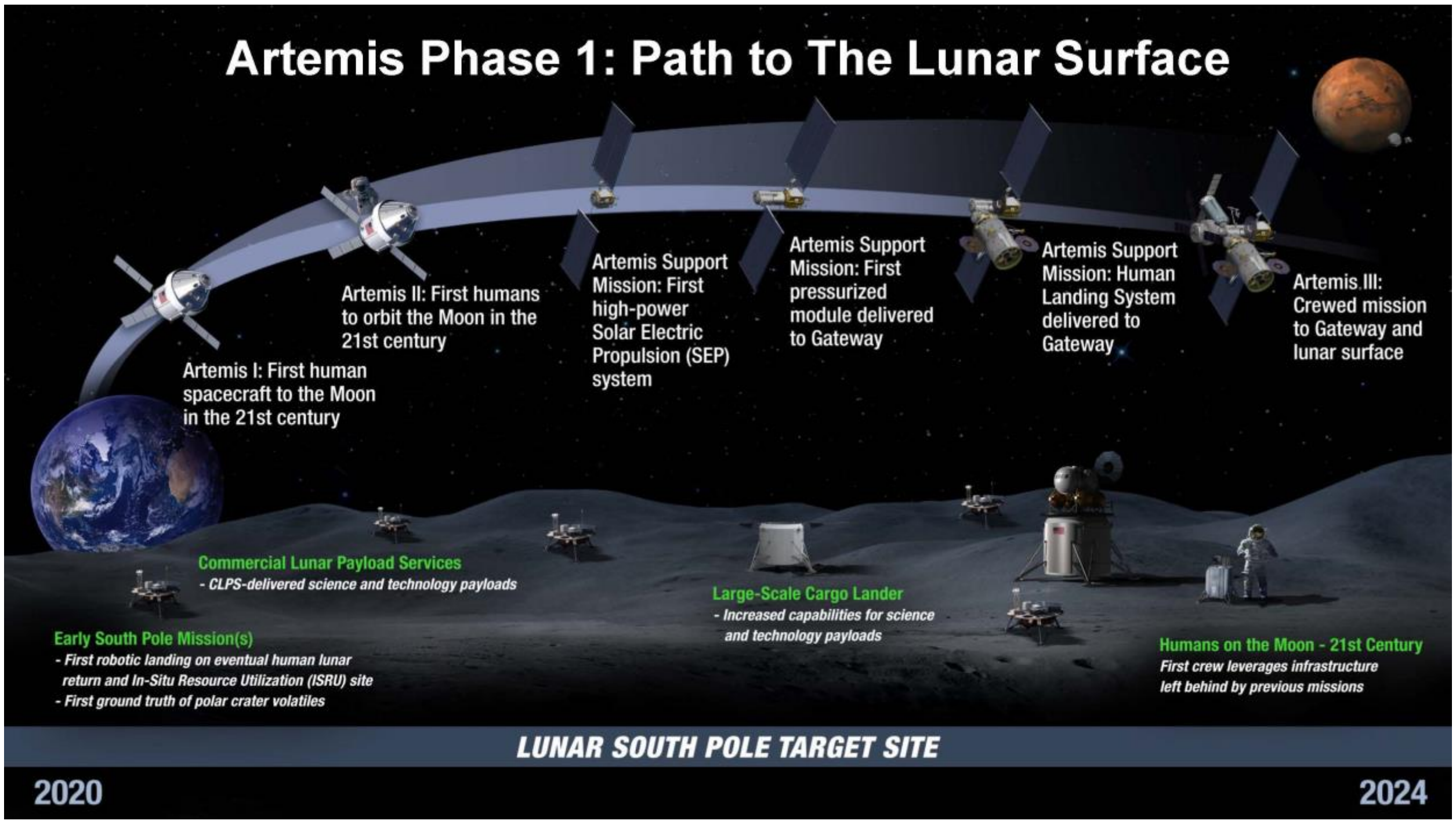

*Fig 4.6. Artemis NASA calendar of programmed missions for a human return to the Moon (credit NASA)*

A number of recommendations were given by the community for future lunar missions and their scientific scopes (see e.g. ICEUM 4,5,6,8,9, Foing 2008; Arvidson et al. 2010; Cohen et al. 2020). Most of the important outstanding questions about the Moon highlighted in previous reports still remain to be explored: by understanding the structure and composition of the crust, mantle, and core of the Moon, by better exploring the origin and evolution of the Earth-Moon system. Understanding the evolution of the Earth-Moon system with time is strongly dependent on the timing, origin, and consequences of the late heavy bombardment; and the related impact processes, regolith and volatile content evolution (Jawin et al. 2021). Finally, technological exploration should also be pursued with the developments required in in situ resources utilization (Anand et al. 2012).

Specific topics highlighted in the ICEUM reports address the following issues:
- the ground truth information on the lunar far side is still lacunar and is needed to address many important scientific questions, e.g., with a sample return from South Pole-Aitken Basin (Jolliff et al. 2021).
- the knowledge of the interior remains poor relative to the knowledge of the surface, and is needed to address a number of key questions, e.g., with International Lunar Network for seismometry and other geophysical measurements.
Future lunar missions will be driven by exploration, resource utilization, and science so we should consider relevant science payloads for every mission, for example, landers and rovers should be systematically equipped to determine surface composition and mineralogy (e.g. Lasue et al. 2012; Schröder et al. 2019; He et al. 2019). Furthermore, new research topics in such areas as in life sciences, partial gravity processes on the Moon should be pursued to support future missions (Jones 2019). Answering all these questions in a cost-effective manner will require international cooperation and sharing of data and results.

Currently advocated exploration missions to 2035 in the ILEWG roadmap phase 3 "Lunar resources" & 4 "Human outposts" cover a range of science objectives and instruments (see Fig. 4.7; Foing 2016):

- Global multi-messenger cartography of Moon surface, water cycle, shallow interior combining surface stations and orbiters;
- Network of geophysical stations including seismometers;
- Campaign of sample returns from a comprehensive set of representative terrains (South Pole Aiken basin, etc…);
- Exploration of the permanent shadowed poles of the Moon + cryogenic sample return;
- Low Frequency radio-interferometer in orbit;
- Geochemical study of subsurface with drilling;
- Radiation exposure astrobiology facilities;
- First demo of ISRU in-situ resource utilization and of life support;
- Precursor Lunar Radio telescope at poles and on farside;
- Science and exploration enabled by short duration human outpost missions at poles, lava tubes and non-polar regions with Artemis program and International Lunar Research Station (China, Russia & partners).

### 3.1.3. Representative missions for 2035-2061: A gateway for deep space exploration

The envisioned missions for this longer-term timeframe of the Horizon 2061 perspective will benefit from the progressive development of robotic and human infrastructures associated to the operations and services. All these developments will contribute to the establishment of a permanent sustainable Moon Village (Foing 1996; Casini et al. 2020).
A number of instruments concepts already proposed, after having been studied and prototyped on Earth, will find ideal locations and support for their deployment in this particularly favorable context. The instruments and laboratories that have been envisioned include:

- Geoscience observatories network on the Moon with a goal to probe the interior using seismometers (Yamada et al. 2011);

- Astronomical observatories on the Moon: these may include liquid mirror giant telescope; optical and infrared interferometers and hyper-telescopes; radio-telescope network on the far side of the Moon, including Lunar Crater Radio Telescope (Clery 2019);
- Moon-based Earth observatories for studying global scale phenomena on Earth (Guo et al. 2016);
- Science and exploration enabled by permanent human research bases at poles, lava tubes and non-polar regions (Heinicke and Foing 2021);
- In-situ manufacturing facilities for research and economical exploitation;
- Curation facilities for extraterrestrial sample analysis ;
- Astrobiology and life science laboratories ;
- Research infrastructures benefitting from habitation, business, tourism, entertainment, and citizen science exploration.

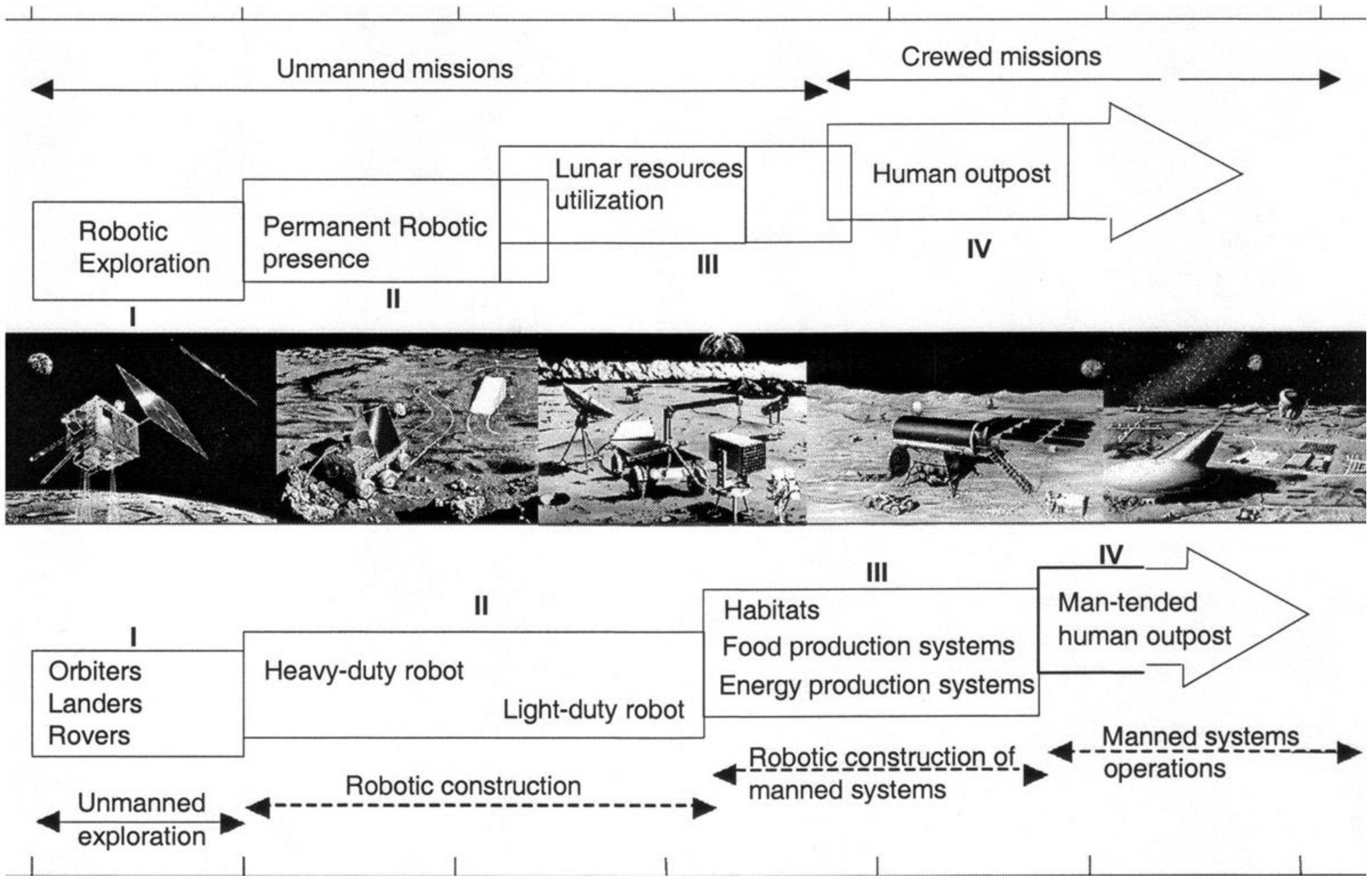

*Fig 4.7. ILEWG roadmap for lunar science, exploration and utilization with phased approach: fleet of orbiters, surface robotic village, human outpost, permanent and sustainable human/robotic presence (MoonVillage)*

## 3.2. Venus: towards sample return

### 3.2.1. Main scientific questions

Venus is a potential next target for in-situ robotic exploration. Since the Venera missions in the 70-80s, which provided in-situ analyses of few surface sites, space missions to Venus focused on its atmospheric composition (Venus Express, Akatsuki). Nevertheless, and despite its dense atmosphere, radar and infrared spectroscopy provided some hints about its surface composition at global scale, suggesting unidentified chemical reactions (highlands radar anomalies, Garvin et al. 1985) and possible differentiated rocks, Gilmore et al. 2017). In addition to comprehensive morphologic observations (e.g., Pioneer Venus Orbiter, Venera 15 and 16, Magellan) suggesting a past active tectonics (tesserae, late resurfacing), the recent observations raise questions on the exact surface composition. Thermodynamic (Fegley et al. 1997; Semprich et al. 2020) and recent experimental studies depict the possible mineral reaction affecting, or having affected, the surface composition (Berger et al. 2019; Filiberto et al. 2020). These reactions are driven by redox reactions and are linked to the sulfur and carbon cycle. They also suggest the possible persistence of alteration features acquired under past wetter conditions if liquid water was ever on Venus.

Concerning the atmosphere, its complex stratification, ranging from a dense and immobile low atmosphere to a super-rotating high and light atmosphere also raises unsolved questions such as the nature of the UV absorber (Marcq et al. 2020), missing reservoirs in atmospheric cycles and the conditions of coupling between topography and gravity waves breaking at cloud top level or giant structure (Kashimura et al. 2019). In addition, the intermediate clouds may meet conditions propitious for the emergence and sustainability of life (Morowitz and Sagan 1967).

As of 2021, several projects of space missions to Venus orbit and Venus lower atmosphere have been selected by the space agencies: Veritas and DAVINCI+ (NASA), EnVision (ESA), Shukrayaan (ISRO), or are still under evaluation such as VeneraD (IKI/RosCosmos). These selections highlight a well-deserved revival of the interest of the scientific community for Venus: from origin to (lack of) habitability, this often called Earth's twin sister, which followed a different evolution path from Earth leading to a radically different environment today, challenges our understanding.

### 3.2.2. Measurement requirements and mission types:

The key measurements to be performed to address these issues encompassing the Venus geodynamics, atmospheric processes and biological potential are available in Earth's laboratories: microscopic observations and chemical analyses of rock samples; chromatography, mass spectrometry, isotopic measurement, NMR for gas/aerosols. For rock samples, the comparison of analytic data from low lands to high lands will also better constrain the possible elemental transfer at large scale through the atmosphere (radar anomalies).

Most of the required techniques have been space-qualified for in-situ measurement, mainly on Mars. However, for Venus surface or deep atmosphere, analyses by remote sensing are strongly limited by the composition and density of the atmosphere (clouds, IR absorber, etc.) and the extreme conditions of the Venus surface constitute a serious limitation for any in-situ analyses (soil or deep atmosphere). Running a complex scientific rover like Discovery on Mars, for example, is not possible at 470°C and an alternative is a sample return mission for both surface rocks and atmosphere.

### 3.2.3. Technology challenges and synergies with existing or planned space missions:

Return sample mission is a challenge in space exploration even if it is sometimes envisaged in alternative to human spaceflight program. With the exception of cosmic dust (Stardust), asteroidal (Hayabusa, Osiris-Rex) and lunar samples (Apollo), no robotic or human return mission from distant terrestrial planet was carried out. The major obstacle in the case of a terrestrial planet is the energy required for a rocket ascent vehicle and the return fly and supposes advanced and breakthrough technologies. For Mars, a return mission is under evaluation by NASA and the next Mars2020 mission will prepare a selection of samples for a future sample return mission. It is clear that a Venus sample return mission will benefit from the Mars program. But in the case of Venus, the higher gravity and the surface temperature imply that still more several critical technical issues have to be overcome such as the electronic accommodation to in-situ high temperatures or the earth return vehicle and trajectory (Several concepts were studied earlier, Scoon and Lebreton 1998; Sweetser et al. 2003). In addition, for a successful scientific mission, multisite sampling at both the surface and the atmosphere would be envisaged. By contrast with Mars, a sequence of several launches (Mars2020 in 2020 and Sample Retrieval lander in 2026) and the use of mobile rovers for the rock collection (Perseverance and later a fetch rover) cannot be envisaged. The rock collection will probably be done randomly. By contrast, retrieving atmospheric sample is less difficult and scenarios are discussed in Shibata et al. (2017) (see Figure 4.8). In addition to these governmental agencies programs, private initiative are to be considered in the future, such as the Venus atmosphere sampling project of Rocket Lab (USA-New Zealand) for 2023 (Limaye et al. 2021).
Concerning the feasibility, technological advances allow us to be optimistic. We may cite in particular the recent progress in high temperature electronics (Neudeck et al. 2016), the concept of aero-platforms (Cutts 2018) to be pursued and the development project of gateways in the next future allowing the launch of vehicle at low gravity, as well as the development of propelled spacecraft (pulsed or thermal nuclear propulsion?).

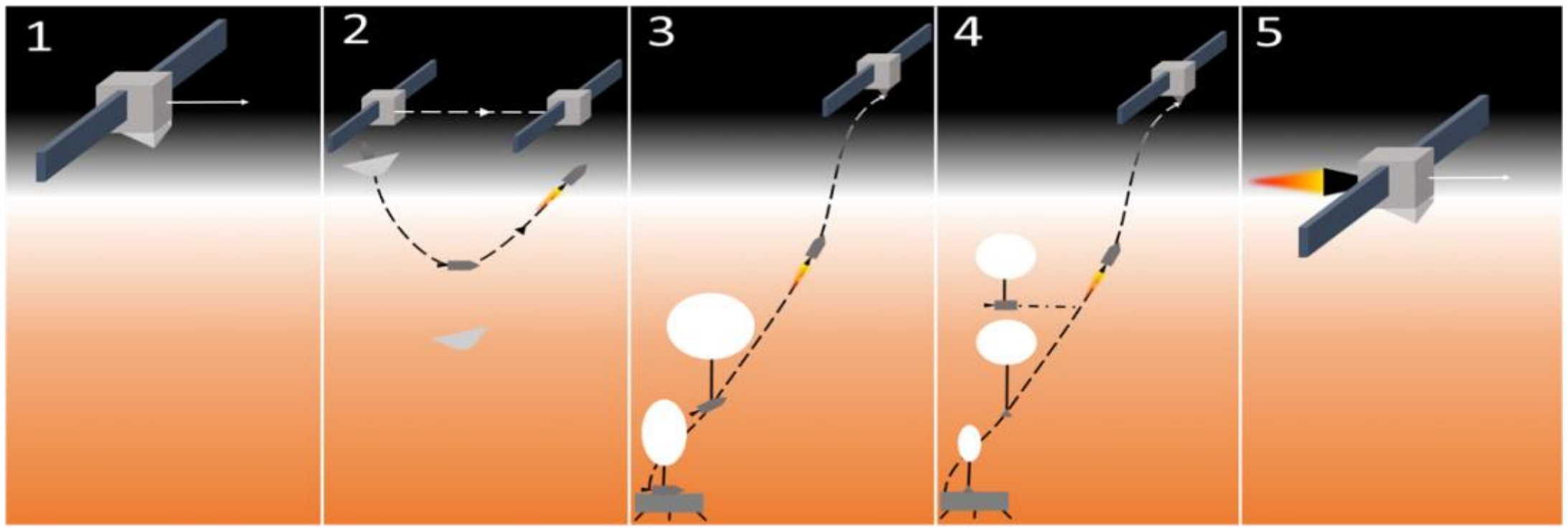

*Figure 4.8: Five possible architectures of Venus sample return missions that have been analyzed. Architecture 1 is the atmospheric skimmer, using a flyby spacecraft. Architecture 2 uses a low-altitude probe that collects samples at low velocities. Architecture 3 and 4 use a lander to collect*

*surface samples, with a balloon that brings the VAV (in 3) or just the sample (in 4) to the VAV launch height. Once in orbit, it rendezvous with an orbiting tug. Architecture 5 has a high-altitude spacecraft burn at the flyby periapsis while collecting samples (Shibata et al. 2017).*

## 3.3. Mars: Sample return and beyond

### 3.3.1. Main scientific objectives

These are great times for Mars planetary scientists! There have been so many missions since the end of the 1990s: nine orbiters[1], two landers[2], six rovers[3], despite a few failures[4]. Eight orbiters were still operational in 2021, as well as three rovers and one lander; most of these spacecraft have exceeded their design lifetime, except (yet…) for those ones which have arrived in 2021. That's a lot for a relatively small fraction of the international astronomical community, but discoveries are piling up and Mars science is making great strides.

The study of Mars geology and geophysics (keywords: composition, internal structure, dynamics, evolution, crust, and interior), of its atmosphere (present/ancient climates, climate processes, orbital configurations, climate change), and the search for life (habitability, life today, traces of life, carbon cycle, prebiotic chemistry) are the themes that backup a series of specific Mars science objectives which comprehensively address the six science questions of Horizon 2061 (see Table 4.1):

Q1: Diversity of objects. Mars occupies a special place in the family of terrestrial planets, due in part to its small mass and to its position on the outer edge of the habitable zone of the Solar System. Were these peculiar characteristics the main drivers of its internal, geological and climate evolution? How does Mars geology compare to Earth or Venus? Is Martian hot-spot volcanism similar to that of the Earth? How does the atmosphere of Mars compare to the other terrestrial atmospheres? Similar questions exist for the internal structure of Mars, which we are only starting to discover.

Q2: Diversity of systems. With Phobos and Deimos, Mars also offers a second example of a system of moons around a terrestrial planet, undoubtedly very different from the Earth-Moon system, which deserves an in-depth characterization.

Q3: Origins. Unravelling the formation scenario of Mars, of its moons (capture, giant impact, in situ formation…?) is not only interesting in itself. In a more global perspective, it helps us to explore and understand the diversity of formation scenarios of planet/moon systems in the solar system… and beyond.

---

[1] Orbiters by year of launch: NASA/Mars Global Surveyor (1996), NASA/Mars Odyssey (2001), ESA/Mars Express (2003), NASA/Mars Reconnaissance Orbiter (2005), ISRO/Mars Orbiter Mission (2013), NASA/MAVEN (2013), ESA-RSA/Trace Gas Orbiter (2016), UAE/HOPE orbiter (2020), CNSA/Tianwen-1 orbiter (2020)

[2] Landers: NASA/Phoenix (2007), NASA/Insight (2018)

[3] Rovers: NASA/Pathfinder (1996), NASA/Spirit and Opportunity (2003), NASA/Curiosity (2011), NASA/Perseverance rover (2020), CNSA/Tianwen Zhurong rover (2020)

[4] Failures: NASA/Mars Observer (1992), ISAS/Nozomi (1992), RSA/Mars 96 (1996), NASA/Mars Climate Orbiter (1999), NASA/Polar Lander (1999), ESA/Beagle 2 lander (2003), RSA/Phobos-Grunt (2011), ESA/Schiaparelli lander (2016)

Q4: How do terrestrial planets work? Mars is a natural laboratory to test our understanding of the dynamics and energetics of the different layers of a terrestrial planet, their interactions and the processes driving their chemical evolution, the loss of atmospheric species, decay of the magnetic field and climate evolution through time. Studying these processes at Mars helps us to better understand how they work on our own planet.

Finally, the questions of the habitability of Mars (Q5) and the search for a Martian life (Q6) have received by far the largest attention in the Mars exploration program, which has been designed to successively accomplish three major quests:

1/ Search for liquid water. It is now established that water flowed on the surface of Mars in quantity, more than 3.5 billion years ago (Ehlmann et al. 2013; Lasue et al. 2013). There are many morphological traces of it (e.g. valley networks, open basins, outflow channels, Carr 2012). The quantity of liquid water is not well known, but one can imagine a Global Equivalent Layer (GEL; Alsaeed and Jakosky 2019) up to ~200 meters deep. For this liquid water to be stable at the surface, one must imagine that Mars had a dense atmosphere, which generated a greenhouse effect that no longer exists today. There was also a magnetic field (Acuna et al. 1999), which did not last, and whose absence precipitated the loss of the atmosphere ripped off by the solar wind (Kass and Young 1995). Today, there is no dense atmosphere, and therefore no stable liquid water at the surface. The stagnant water over the Noachian and Hesperian eras transformed the original igneous rocks into various alteration products such as clays, sulfates, carbonates, etc. (Bibring et al. 2006; Ehlmann et al. 2008a).

2/ Characterization of Mars habitability. There is enough evidence now to think that Mars was habitable at that time (Grotzinger et al. 2013): the pH of the liquid water was about neutral. Water was probably not very salty, fresh water you might say. The elements of organic chemistry were present, especially the basic atoms, C, H, N, O, P, and S. There were also reserves of energy. How long did this period of habitability last? At what time? There are still many open questions, particularly about the end of this period (transition to a more acidic era, loss of the atmosphere, etc.). The products of alteration created during this era (e.g. smectites, carbonates) can also preserve the organic memory of this period. The evidence of the past habitability of Mars makes it possible to consider the next phase.

3/ Search for traces of life. This theme is the most exciting and most complex. The period of habitability having been short, it is likely that evidences of life are thin, akin to what is found on Earth for the same time period. The observables will not be fossils stricto sensu but, from isotopic evidence, more or less degraded organic molecular chains, biominerals, morphological and chemical structures. Terrestrial stromatolites serve as a reference for establishing robust observational processes. They allow us to say, for example, that a site rich in carbonates has better chances of preserving such traces (Horgan et al. 2020).

We now first describe the missions that will continue to focus on these major questions in the coming two decades with the objective of returning samples from Mars, and then discuss the future of the science of Mars beyond the perspective of this sample return.

### 3.3.2. Missions for the coming decade and the Mars sample return.

The “search for liquid water” was mostly conducted in the 2000s at orbital distances (Odyssey, Mars Express, Reconnaissance Orbiter), and confirmed on the ground by all landers and rovers. Then “the study of habitability” really took off with Curiosity (Figure 4.9) and the loss of atmosphere is currently studied by Maven and Trace Gas orbiter. HOPE and Tianwen-1 will follow along the same line. The “search for life” was prematurely addressed by the Viking missions in the 1970s (Klein et al. 1972) but it led to inconclusive results (Guaita 2017). More than four decades later, the objective remains the same and it will be the focus of the new generation of rovers, Perseverance (currently operating on Martian surface) and ESA/ExoMars rover (launch expected 2022). Needless to say, all missions address “comparative planetology” with a special mention for Insight and its seismic sounding of Mars interior (Banerdt et al. 2020). For the “preparation of human exploration”, the strategy is to do nothing in most cases, and occasionally to test the in situ feasibility of some key technologies (e.g. oxygen production on Perseverance by MOXIE; Hecht et al. 2021), to measure important environmental parameters for humans (radiation, dust, toxic elements, …) and to characterize potentially usable resources for the future (ISRU: In situ Resources Utilization, see chapter 5, Grande et al. 2022).

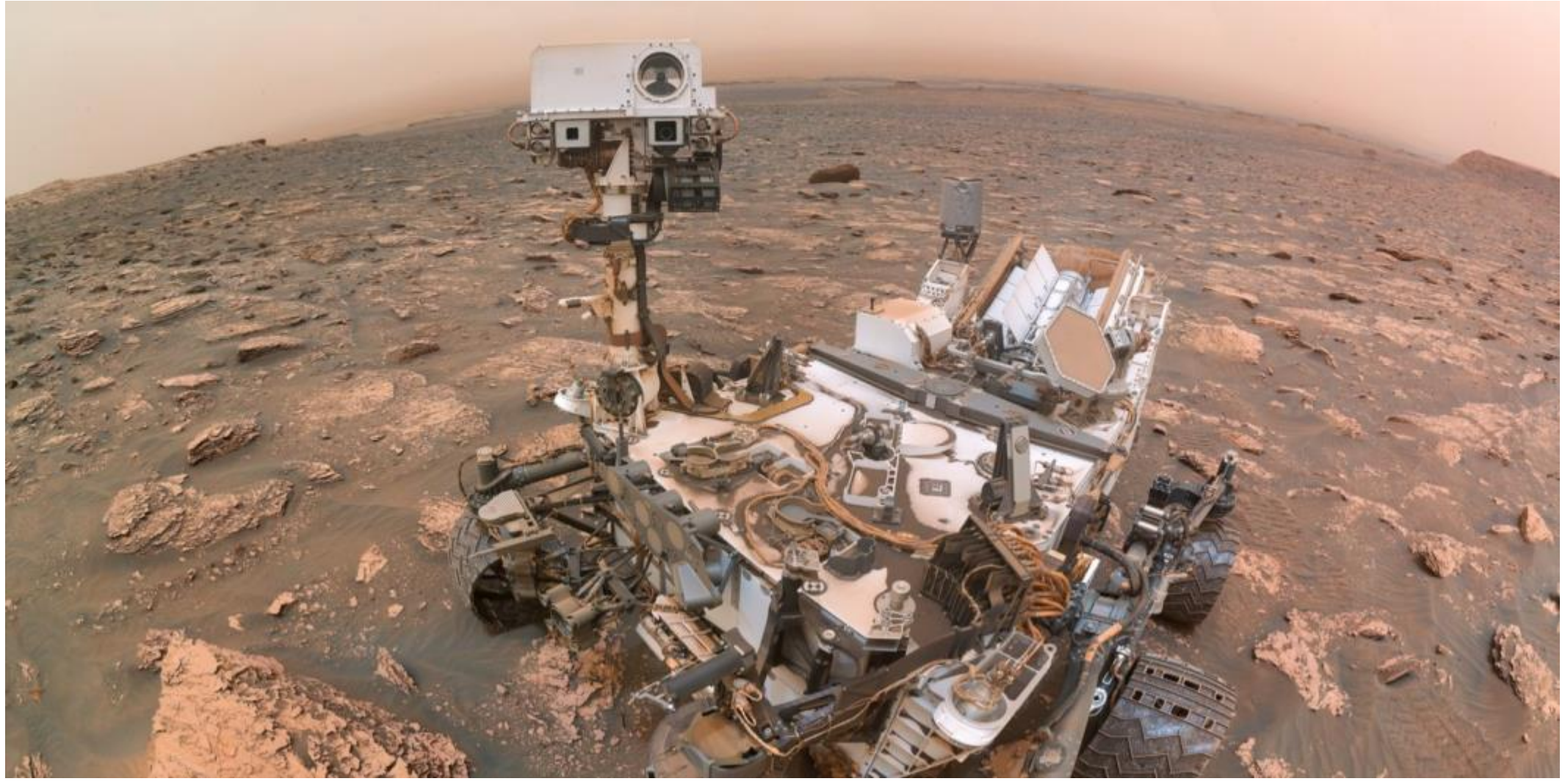

*Figure 4.9. A selfie by NASA's Mars Curiosity rover during a Martian dust storm (dec. 2016). © NASA/JPL-Caltech/MSSS*

So the goal of the 2020s is the search for life, at last. The approach is twofold: in situ exploration and sample studies on Earth. The return of samples is recognized as the grail in our quest to understand the Solar System (Drake et al. 1987), following the example of what has been returned from the Moon (Apollo and Luna missions), from the solar wind (Genesis), and from small bodies (Hayabusa-1, Stardust, Hayabusa-2 and Osiris-Rex to come). Analytical performance in our laboratories is 10, 100, 1000x better than in situ, sample preparation is well controlled and repeatable. Therefore, it allows critical measurements such as precise radiometric dating. This does not diminish the interest of what is done in situ, but when it comes to traces as tenuous as those of ancient life, only laboratory studies will be able to provide a definitive answer. For this to happen, the samples in question must be well chosen. Of course, Martian meteorites constitute a library of Mars samples available on Earth (about ~120 samples). Although they provide highly valuable information (Sautter et al. 2016), they give only a limited view of the Martian processes due to their lack of diversity and their absence of geological context. This is where in situ exploration brings its greatest added value: the choice of the sampling site and of the samples themselves. The landing site must contain sediments, which themselves contain mineral phases resulting from aqueous weathering that are most likely to retain traces of life. As for the samples, the keyword is diversity: suites of sedimentary rocks, suites of rocks formed and/or altered by hydrothermal fluids, of rocks/veins representing water-rock interaction in the subsurface, fine-grain dust and regolith, suites of ancient igneous rocks, and if possible, current atmospheric gas and ancient atmospheric gas trapped in older rocks. A powerful payload is required to select those samples and also to characterize their close geological environment and anything else that may later help in the interpretation of the data.

Perseverance landed on Feb. 2021 in Jezero crater (Farley 2018). This 49 km diameter crater was filled by water and drained away on at least two occasions. More than 3.5 billion years ago, river channels spilled over the crater walls and created a lake with a deltaic deposit (Goudge et al. 2015). We see evidence that water carried clay minerals from the surrounding Noachian-aged Nili Fossae area into the crater after the lake dried up (Ehlmann et al. 2008b). Conceivably, microbial life could have existed in Jezero during one or more of these wet times. If so, signs of their remains might be found in lakebed sediments. Perseverance has a unique payload for in situ studies and for the selection and characterization of samples, and above all a whole apparatus to encapsulate and seal the samples, a device that is called “cache system”.

The Mars 2020 mission is the first segment in a series of three missions to implement the Mars Sample Return (MSR, Figure 4.10; Grady 2020). The Sampling and Cache System of Perseverance was designed to collect and seal the Mars samples according to the most stringent planetary protection class V protocols (Moeller et al. 2021). Sample tubes, not less than 20, up to 35, will be « dropped » by Perseverance in groups at several locations during its nominal mission 2021-2023. As it stands for now, then NASA will launch the Sample Return Lander mission to land a platform within Jezero. From here, a small ESA rover, the Sample Fetch Rover, will collect the cached samples. Once it has collected them, it will return to the lander platform and load them into a single large canister at the top of the Mars Ascent Vehicle. This vehicle will perform the first liftoff from Mars and carry the container into Mars orbit. ESA’s Earth Return Orbiter will be the next mission. Its role is to capture the basketball-size sample container, which are in orbit around Mars. The samples will be sealed in a biocontainment system to prevent contaminating Earth with unsterilized material before being moved into an Earth entry capsule. The spacecraft will then return to Earth, where it will release the entry capsule for the samples to end up in a specialized handling facility. The dynamics are set in motion, if all goes well, the samples could come back to Earth in 2031, not earlier.

The MSR program is the climax of the Martian program and a pivot: nothing afterwards will be the same. First of all, we have to succeed in this return even though the technological challenges are numerous, as we just described. Let's not neglect the challenge of sample management on Earth either (Chan et al. 2020). Do we need a P4 laboratory to protect us from malignant Martian life? Or two independent laboratories to secure science? If early investigations for the search for life have to be done in these laboratories, how can they be sterilized before being released into the community to search for traces of life without altering them?

The future of Mars exploration is set for the next decade along these lines, and failure is not an option.

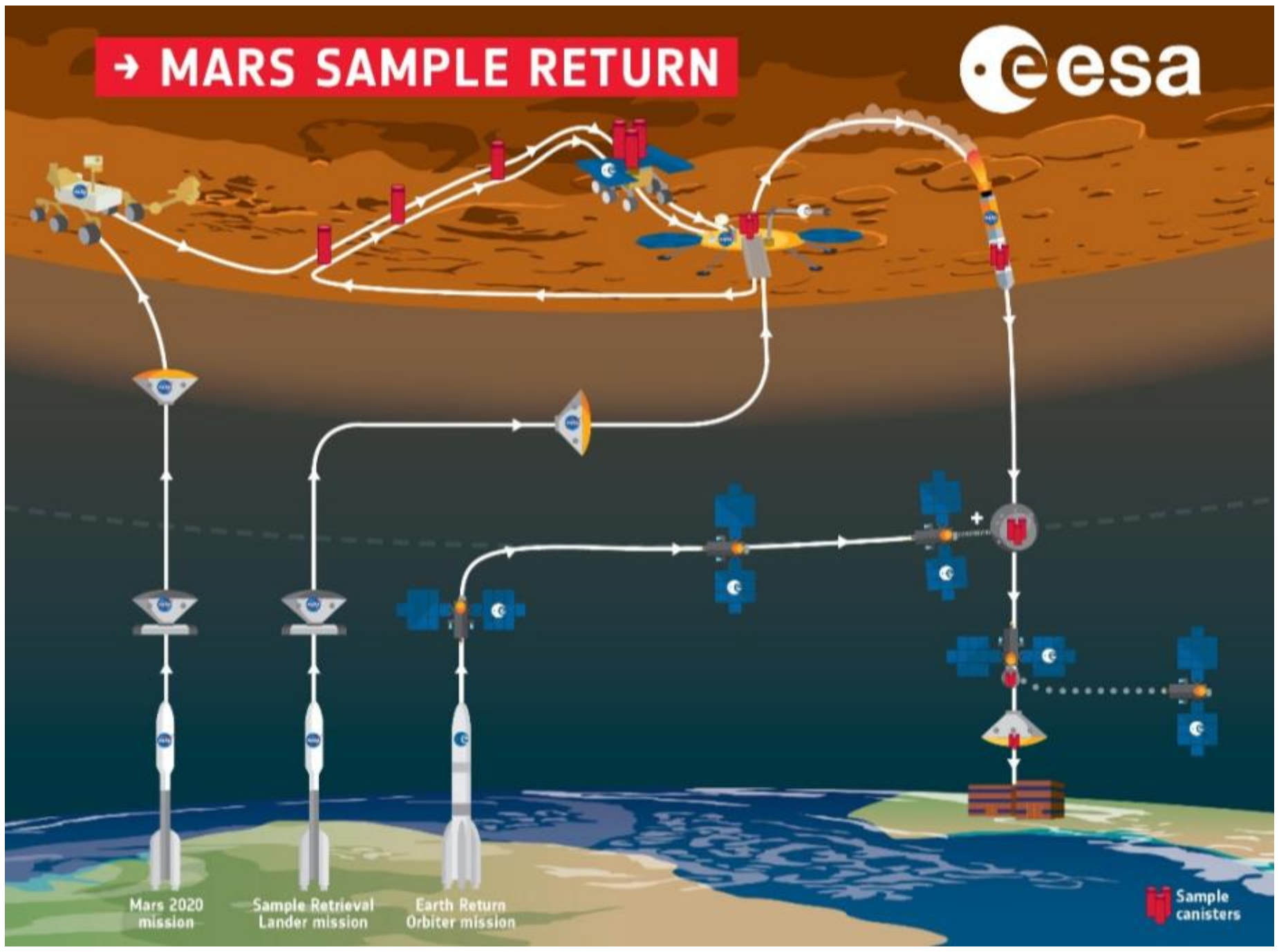

*Figure 4.10. Mars Sample Return overview infographic by ESA, as of June 2020.*

It is also worth mentioning that the Japanese Space Agency JAXA is also developing a sample return mission for the lunar Moon Phobos named Mars Moon Explorer (MMX). To be launched in 2024 with a sample return expected by 2029, it will allow to decipher the origin of the lunar moons, while possibly returning fragments of Mars that could have exchanged between the surface of the planet and Phobos (Kuramoto et al. 2021).

### 3.3.3. Toward human exploration of Mars

In parallel to the MSR campaign and beyond, how will the exploration of Mars address the scientific objectives of Horizon 2061, described in section 3.3.1, in the future? It will likely continue to develop with a combination of missions belonging to the three major mission types described vertically in Table 4.2:

- robotic exploration, combining in situ and orbital observations, which we will call Path 1;
- follow-on sample return missions (Path 2);
- and finally human exploration (Path 3), which occupies a unique place in our perspective as Mars is the only Solar System planet where it can be foreseen and initiated by 2061.

The relative importance of these three exploration paths and of the different scientific questions they address will dramatically change depending on the astrobiology findings of the MSR campaign.

#### 3.3.3.1 Case 1: MSR finds traces of life

In this first scenario, the discovery of a second genesis of life in the Solar System will trigger a new scientific revolution by itself (see chapter 3, Dehant et al. 2061). Mars should and likely will become our major astrobiology field site laboratory! This first discovery will inevitably raise new questions, such as : did life emerge at only one site, or independently at several sites? How, and how far, did life spread to the planet? These burning questions, which are right now impossible to address for our own Earth life, will drive irresistible motivations for new robotic and sample return missions from different regions and sites of the Red Planet, mainly oriented towards Questions 5 and 6: where and in which way was Mars habitable, and how did life emerge?

In this context, however, human exploration will likely be stopped or severely restricted, at least for the long period needed to perform all necessary astrobiology investigations and possibly secure some areas where human presence could be accommodated with due respect of the stringent rules of planetary protection – maybe not before 2061.

#### 3.3.3.2 Case 2: MSR results are not fully conclusive

In that case, from a scientific point of view, Mars exploration will have to continue along paths 1 (robotic) and 2 (more sample returns) while cautiously exploring to what extent this intermediate scenario can accommodate human exploration. This is likely the least favorable scenario for the future of Mars exploration: the MSR program is so promising that it is success oriented. As such, a non-conclusive answer may slow down exploration for some time, as happened already in the past, when high expectation missions led to disappointments regarding life: the first orbital images from Mariner 4 did not show the canals imagined by Percival Lowell, the soil samples tested by Vikings were proven to be sterile, and the exploration program paused for years. Even though, new approaches, new considerations and new ideas later gave a new impulse and the Mars exploration program started all over again a couple of decades later. This should teach us that, for science, a non-conclusive answer is also a new element in the building of more efficient strategies to successfully answer a scientific question later.

#### 3.3.3.3 Case 3: MSR convincingly concludes that life is and has been absent from Mars

In this third scenario, Questions 5 and 6 will no longer be relevant. The search for a second genesis of life in the Solar System will focus on the ocean moons of the outer Solar System, which have recently been identified in ESA's Voyage 2050 Senior Committee report (2021) as the next Solar System destination for an L-class mission beyond the Cosmic Vision program. Mars science will re-focus on Questions 1 to 4: understand the special place occupied by Mars in the family of terrestrial planets, and of its moons system in the family of secondary systems; retrieve their formation scenarios; study how a small planet located on the outskirts of the habitable zone, with faster cooling times and atmosphere loss times, works; retrieve its internal structure and understand its internal heat transfer engine, including the history of its magnetic field, the formation history of its surface and sub-surface, and its weather and climate.

Mars is a unique place to find detailed answers to this long list of planetary science questions on a planet which has hosted a hydrosphere, witnessed erosion and developed a rich sedimentary record. There are still so many geological settings to explore that remote exploration can last for years with orbiters, landers, rovers and drones, to study Mars geology. These future missions could be driven by recent discoveries, such as recurring slope lineae (RSL), vents and caves for instance. New technologies may be developed (e.g. Stamenković et al. 2019): geophysical network, deep drill, long traverse, heavy drones, etc. Missions could be cheaper if we tend to reuse the vehicles and adapt “only” the payload.

The Martian surface and subsurface host a precious water reservoir as well as unexplored mineral resources whose inventory needs to be done, not only to understand the planet’s history, but also to know which resources will be available for future long-term human settlements. Science of Mars as a planetary object will go on, riding along with science in support to human exploration!

As one can anticipate, in this perspective of an abiotic Mars, Question 5 will take a different but equally exciting meaning: is Mars habitable … by Humans? Or rather, how will the space program develop to create the conditions for long-term human settlements on the Red Planet?

#### 3.3.3.4 The perspective of human exploration

Human exploration of Mars, if not colonization, has been announced in the Media for decades (Figure 4.11). It receives quite some support in the public, the same audience that doesn't appreciate the sum of technologies that don't exist yet for such a journey. The astronaut is the true embodiment of the human dream.

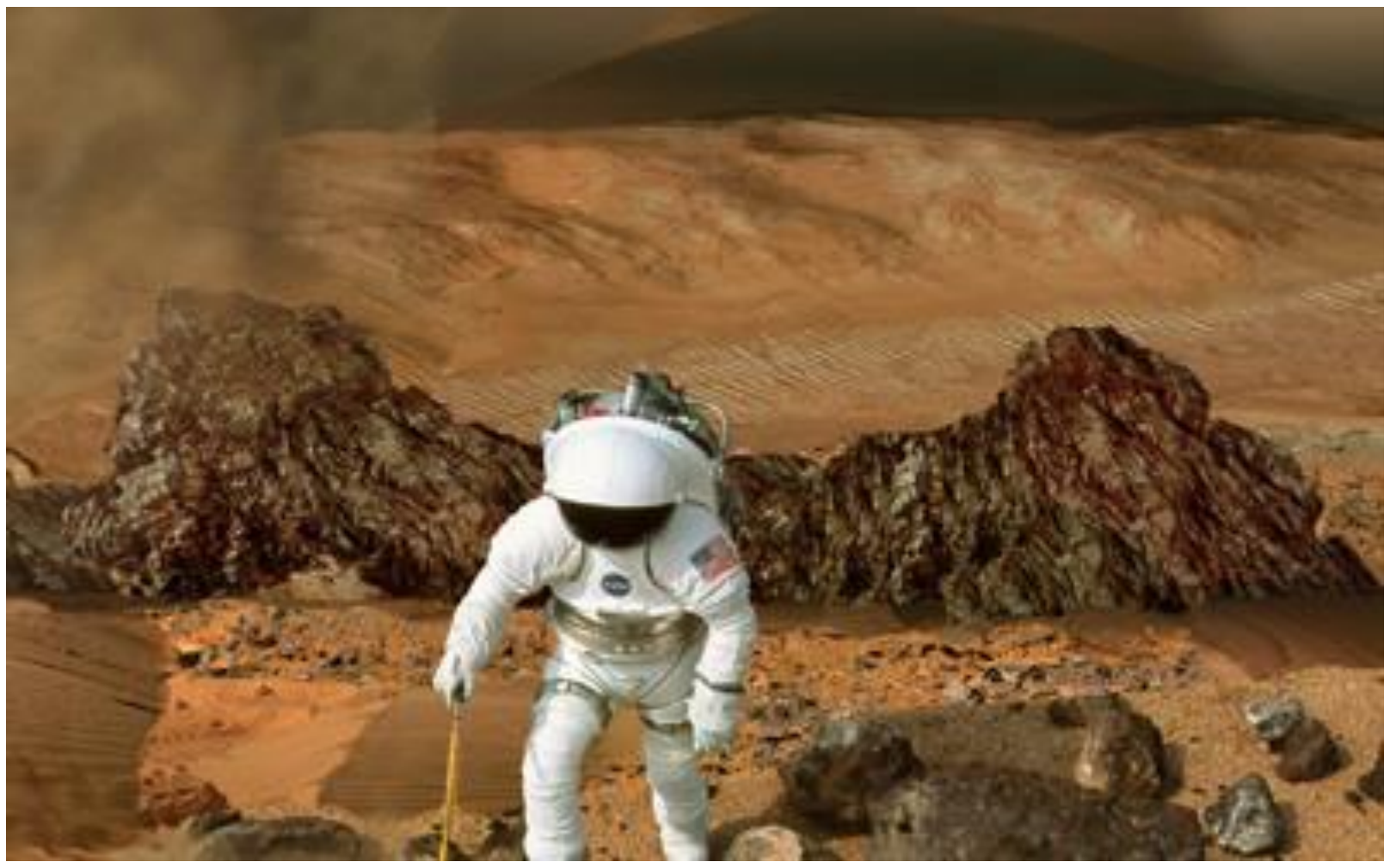

*Figure 4.11. Preparing for Human exploration on Mars. © NASA*

As a prerequisite, is MSR required prior to sending humans to Mars? The answer is clearly yes. These samples will enhance our understanding of the Martian environment: properties of Martian materials relevant for ISRU, toxicity of Martian materials with respect to human health and performance, or information related to engineering surface hazard. The sample return is obviously a ‘proof of concept’ for a potential round-trip human mission to the planet, and a potential model for international cooperation.

Science can hardly be the driver of human exploration, given the level of investment that such an endeavor represents, well beyond the cost of any current large-scale basic research facility. But, like for all sectors of human activities, science must and will take its part in the treat, just like it did for Moon exploration in the 1960s. As mentioned before for case 1 (discovery of life), since astronauts carry life with them, the question of life on Mars must be answered before their arrival, otherwise the problems of planetary protection will make it impossible.

But are humans an advantage for Mars exploration? Garvin (2004) has tried to evaluate the relative advantages of humans vs. robots for Mars exploration, with regards to various skills: strength, endurance, precision, cognition, perception, speed, reliability, agility, dexterity, fragility, etc. In most cases humans have a clear advantage. But with the endurance of recent rovers, Spirit, Opportunity, Curiosity and their increasing sophistication, robots are regaining ground. So, for most science objectives, humans are not necessary on Mars, but there is no opposition between robotic and human exploration. Robots lead the way, and they will always be needed when humans are in orbit, or even at their side on the surface. When will that happen? Reasonable projections are not earlier than the 2040s. Possibly in 2061! A return at the surface of the Moon with the Artemis program seems to be a prerequisite before Mars.

The Martian community has been well served in recent decades and has made good use of the means given to it. To go on, Mars exploration must be ever more innovative, more exciting, perhaps more dramatic. Path 1 “robotic exploration” will go on in all the cases foreseen here. Scientists are creative and new questions arise every day. Path 2 “Return sample” is expected to bring a clear answer to the question of life on Mars. Path 3 “human exploration” will bring a new momentum to Mars exploration, and no doubt more opportunities for more science. In return, it is essential that science takes all its part in the perspective and implementation of Human exploration of the Red Planet.

## 3.4. Mercury

### 3.4.1. Main scientific objectives of Mercury’s exploration

Out of the four rocky planets in our Solar System, the innermost planet Mercury has been far less explored than the Earth’s two nearest neighbors, Venus and Mars. Three flybys by Mariner 10 (NASA) in 1974 and 1975 - by then, Mars and Venus had both benefited from several flybys and one orbiter each - revealed that Mercury is a very specific planet, with a rich geological history, a unique interior, a global magnetic field which results in its own intrinsic magnetosphere and a unique configuration of orbital parameters due to its proximity to the Sun. Three decades later, MESSENGER (NASA) became the first spacecraft to orbit Mercury, from 18 March 2011 to 30 April 2015 (Solomon et al. 2018). Thus it acquired the first global observations and measurements of the Hermean surface, gravitation field, magnetic fields, exospheric and magnetospheric populations of molecules, atoms and ions.

The ESA/JAXA BepiColombo mission (Benkhoff et al. 2010) is now en route to Mercury, after its first flyby of the innermost planet in October 2021. Two orbiters will be deployed simultaneously around Mercury (a first in the history of space exploration) at the end of 2025. Those two complementary spacecrafts are fully equipped to provide new insights concerning the six following regions: Mercury’s surface, interior and intrinsic magnetic field, exosphere, magnetosphere and its interaction with the Hermean plasma environment, Mercury as the closest planet to its star (for comparison with exoplanets in the same situation) and Mercury as a test for Einstein's General Relativity theory, due to the precession of its perihelion.

The scientific objectives of future exploration of Mercury are thus multifold. Mercury's uncompressed density is markedly higher than that of all other terrestrial planets, Moon included. This fact raises many questions about the nature and the properties of Mercury's core. In particular, current unanswered questions are why does such a small planet posess an intrinsic magnetic field, and how does the smallest planetary intrinsic magnetosphere adjust and resist in a harsh and permanently changing solar wind. Indeed, the strong pressure from the solar wind, which is denser at Mercury's orbit than at the Earth and carries a stronger interplanetary magnetic field, leads to the fact that external sources of magnetic field can dominate the magnetic field at Mercury, which is thus difficult to determine and model. The magnetosphere of Mercury is a unique laboratory for plasma physics, as the magnetosphere is small enough to be covered from one edge to the other in a 10-hour orbit by one or two spacecrafts, bringing new insights on magnetospheric dynamics and temporal evolution. In terms of surface and geology, spectro-morphological analysis of MESSENGER's data reveals a complex history of the Hermean soils, combining tectonic deformation, past volcanism and all kinds of impact cratering events. Moreover, the permanently shadowed craters of the polar regions contain radar-reflective materials, which could be sulphur or water ice. Another important question concerns the sources and sinks of volatile species, and the production mechanisms of the exosphere. At Mercury more than anywhere else, the exosphere is bounded to the surface, and its composition and behavior are controlled by interactions with both the surface and the magnetosphere (especially in the absence of an ionosphere). The interaction of the surface and the magnetosphere is also interesting to study the impact of space weather on the surface properties.

Studying such a unique planet as Mercury will bring new clues about terrestrial planets formation, evolution, and how it depends on the planet size, but also about its initial conditions such as the composition of the primordial solar nebula.

Finally, considering fundamental physics, and considering that the advance in Mercury's perihelion was explained in terms of relativistic space-time curvature, sending orbiters around Mercury and tracking their position with high accuracy can allow to further test general relativity, taking advantage of the proximity to the Sun.

### 3.4.2. Challenges for Mercury's exploration from 2030 to 2061

Mercury is the least explored terrestrial planet so far, but there are many reasons for that. First of all, the planet never separates from the Sun by more than 28° of arc when observed from the Earth. It is thus much too close to the Sun for most of the ground-based facilities, because their optical systems would be damaged by a direct exposure to sunlight. This is also the case for space telescopes such as the Hubble Space Telescope for instance.

However, sending an orbiter is not easy either: the planet is so close to the Sun and its huge mass, that mission design is challenging. The combination of the proximity of the Sun and of the high temperature of the planet's surface itself also makes missions to Mercury extremely challenging in terms of keeping the solar panels safe and maintaining the thermal balance of the spacecraft .

The only ongoing mission to Mercury is BepiColombo, a joint project between the European Space Agency (ESA) and the Japanese Aerospace Exploration Agency (JAXA). The Mission consists of two orbiters, the Mercury Planetary Orbiter (MPO) and the Mercury Magnetospheric Orbiter (MMO, also called Mio). BepiColombo is in the middle of its cruise to Mercury, where it will deploy two spacecraft at the end of the year 2025. In addition to carrying to the innermost planet a set of complementary instrumental suites with unprecedented capabilities, it will provide a rare opportunity to collect multi-point measurements in a planetary environment. This will be particularly important at Mercury because of short temporal and spatial scales in Mercury’s environment. The foreseen orbits of the MPO and MMO will allow close encounters of the two spacecraft throughout the mission.

Beyond BepiColombo, NASA’s Planetary Mission Concept Studies (PMCS) program recently awarded a study to a multidisciplinary team led by Dr. Carolyn Ernst of the Johns Hopkins University Applied Physics Laboratory (APL), to evaluate the feasibility of a landed mission to Mercury in the next decade. The resulting mission concept (Ernst et al, 2020) achieves one full Mercury year (~88 Earth days) of surface operations, based on a launch in 2035 with a landing on Mercury’s surface in 2045. All technical requirements are described in the report and are now being discussed in conferences (Kubota et al. 2021).

Finally, sending a lander and planning sample return from Mercury — the final dream of many planetary scientists (Vander Kaaden et al. 2019), which could be foreseen in the future with the development of more powerful launchers and more efficient energy suppliers — would require to face the unique configuration of a planet that rotates so slowly that its surface temperature varies by more than 600° C over the course of a solar day. Also, a given region remains in the night side or in the day side for a very long time with respect to the typical orbital period of an orbiter around Mercury (about 10 hours), which brings tedious but challenging questions of telemetry, energy and heat control for the lander.

## 3.5. Small bodies: asteroids, comets, TNOs

### 3.5.1. Key scientific objectives of future missions to small bodies

As reviewed in chapter 3 (Dehant et al. 2021) and summarized in table 4.1, space missions to small bodies play a unique role in characterizing their in situ properties, such as their mass, shape and therefore density, and also their physical state, cratering record, surface chemical composition and possible internal layering. Such measurements are critical to produce a statistical picture of the small bodies populations in the Solar System, using in situ detected properties to possibly generalize them from ground based surveys to other objects of the same class. Constraining the distribution and orbital properties of the small bodies reservoirs will open the door to possible comparisons with extrasolar debris disks. Besides, understanding the origin and dynamics of binary systems in the small bodies populations remains a challenge to this day.

A major goal of current missions is to return asteroidal and cometary samples to Earth: these returned samples will provide context to the collections of cosmic dust and meteorites that are curated in the laboratories and hints to the processes that occurred in the formation of the Solar System from the protosolar nebula.

The material that makes up small bodies (minerals, refractory material, organics, volatiles such as water) is critical to understand which objects may have provided water to the terrestrial planets after their formation, and whether organic matter such as amino-acids that have been found in asteroids and comets could have played a role in the emergence of life on Earth.
Finally, a better understanding of the properties of such objects will also help mitigate the potential threat that Earth-crossing asteroids and comets may present in the future.

## 3.5.2. Currently planned exploration missions to 2040

In order to improve our understanding of Solar System small bodies by 2040, a number of ambitious missions have been selected or are considered for implementation by the space agencies. They typically tackle technological challenges such as operations around and sample collection from an object without gravity, or in situ measurements to better understand the interior make up of these objects.

### 3.5.2.1 AIDA mission

Very little is known about the internal structure of asteroids. Density measurements performed either in situ or from the ground indicate that a significant fraction (>95%) of the smaller asteroids (D<~60km) are under-dense (Carry 2012): their density is lower than that of their surface composition. These under-dense bodies have been interpreted as being pervaded by large cracks and voids in their interiors, resulting from cataclysmic impacts and subsequent uneven re-accumulation of material. The fraction of volume occupied by these voids is called macroporosity. Our current census of density and macroporosity for about 300 asteroids indicates that some asteroids may have macroporosities up to 50%.

Most if not all near-Earth objects may have large macro-porosities. Understanding how such bodies hold together and how these objects would respond to impacts is of primary importance for humankind in case one of these bodies would be on a trajectory that would intersect that of the Earth.

The ESA-NASA Asteroid Impact and Deflection Assessment (AIDA) mission consists in a pair of space probes (NASA DART and ESA Hera; Cheng et al. 2018; Michel et al. 2018) that will study and demonstrate the kinetic effects of crashing an impactor spacecraft into a near-Earth asteroid moon. The architecture of this dual mission is illustrated in Figure 4.12. The mission is intended to test and validate impact models of whether a spacecraft could successfully deflect an asteroid on a collision course with Earth.

AIDA will target binary near-Earth asteroid 65803 Didymos. The primary asteroid is about 800 km in diameter whereas its moon Dimorphos is about 150 meters in diameter. It is expected that the impact of the 300 kilograms DART spacecraft at 6.25 km/s will produce a velocity change on the order of 0.4 mm/s. Such impact will only induce a minimal change in the heliocentric orbit of the system. The NASA DART mission is due to collide with the Didymos system in 2022: ground based optical and radar observations will allow to monitor changes in the binary system orbit. Then, from the end of 2026, HERA will characterize in detail the physical properties of Didymos and Dimorphos and of the crater: these measurements will allow a better quantification of the DART momentum transfer efficiency (Rivkin, 2021;Michel, 2022).

The NASA DART mission is due to collide with the Didymos system in 2022 whereas the ESA Hera mission will be launched in 2024 and reach Didymos in 2026.

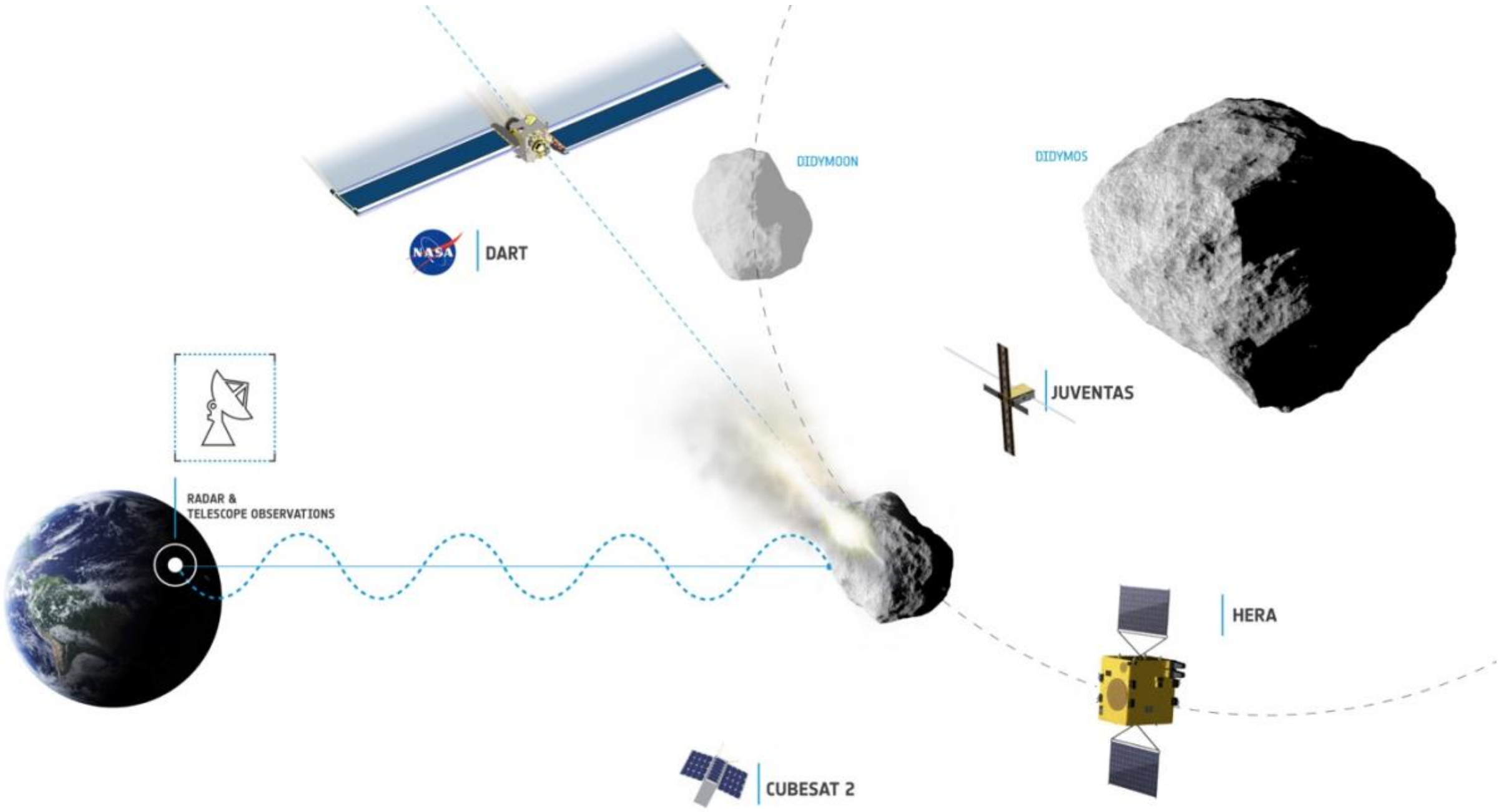

*Figure 4.12. HERA and DART mission architecture; the moon is called Didymoon or Dimorphos (credit ESA-NASA)*

### 3.5.2.2 NEOSM mission

The **Near-Earth Object Surveillance Mission** (**NEOSM**), will set up in space an infrared telescope designed to survey the Solar System for potentially hazardous asteroids. Its launch is planned by NASA in 2025 within the frame of its planetary defense program. The NEO Surveillance Mission will be carried out by the **NEO Surveyor** spacecraft, which will survey from the Sun–Earth $L_1$ (inner) Lagrange point, allowing it to look close to the Sun and see objects inside Earth's orbit, that would otherwise be difficult to detect from Earth or from low Earth Orbit.

The main objective of the mission is to find 65% of undiscovered Potentially Hazardous Asteroids large than140 m in 5 years (goal: 90% in 10 years), and to characterize their orbit. The field of view of NEOSM will be large enough to allow the mission to discover tens of thousands of new NEOs with sizes as small as 30 m in diameter. Secondary science goals include detection and characterization of approximately one million asteroids in the asteroid belt and thousands of comets, as well as identification of potential NEO targets for human and robotic exploration.

### 3.5.2.3 Psyche mission

Generally speaking, metal-rich asteroids and their meteoritic counterparts (iron, stony-iron meteorites and possibly iron-rich chondrites such as CB and CH chondrites) are among the most perplexing and mysterious objects among Solar System small bodies. It is still not understood how these asteroids with diameters up to 220 km formed (e.g. the asteroid Psyche; Ferrais et al. 2020) and what they actually represent. Notably, are they remnant cores of primordial differentiated protoplanets or did they form as iron-rich chondritic bodies? In the first scenario, Psyche for instance would require that the mantle of a parent body of the size of Vesta (since Psyche's size is similar to Vesta's metallic core) has been totally blown off. If so, where has such a huge mantle gone? Another perplexing feature of metal-rich asteroids is their apparent low density <4.5 g/cm$^3$ that appears more compatible with that of stony iron meteorites (pallasites, mesosiderites; density in the 4-5 g/cm$^3$ range) than that of iron meteorites (~7.8 g/cm$^3$). Yet, iron meteorites represent 4.6% of meteorite falls whereas stony-iron meteorites represent only 1.1% of the falls. In essence, where are the parent bodies of iron meteorites if the currently known metal-rich bodies are related to stony-iron meteorites?

In situ observations of the metal-rich asteroid Psyche (Elkins-Tanton et al. 2020) will improve our understanding of the formation of metallic worlds. The Psyche mission will be launched in 2022 and reach Psyche in early 2026. It will characterize 16 Psyche's geology, shape, elemental composition, magnetic field, and mass distribution.

#### 3.5.2.4 Comet interceptor

Comets have been studied from the ground and through a few space missions. Nowadays, they are believed to consist of the most primitive Solar System materials given their richness in very volatile ices (Bieler et al. 2015), organic compounds (Hérique et al. 2016; Bardyn et al. 2017) and salts (Altwegg et al. 2019; Poch et al. 2020). Because of flight dynamics constraints, space missions to comets have been targeting short period comets, which have experienced devolatilization and geological modification in response to solar irradiation in earlier orbits. Therefore, they cannot be considered as fully fresh pristine material. Comet Interceptor will be the first type of F-class mission by ESA (Snodgrass and Jones 2019), and will be launched together with the ARIEL M-class mission. The spacecraft will be delivered to the L2 Earth Lagrange point and rest for possibly a few years until an appropriate target is identified. The main objective of Comet Interceptor is to perform a flyby of a Dynamically New Comet (DNC) in order to study its comae, shape and surface geology, before the surface is modified by Sun. Comet interceptor offers thus the opportunity to investigate from close distance fresh Oort Cloud material, from a target that has not been identified at the moment. The mission design is based on three distinct spacecrafts, one of them staying at a relatively large distance to avoid damage by cometary dust, and two other smaller spacecraft (including one by JAXA) that will perform close approach and high-risk, high-gain science.

One exciting perspective of Comet Interceptor is the possibility to eventually target an Interstellar Objects (ISO). This would be a grand premiere and may offer unique insights on the composition of such objects. For the first time, we may have constraints on the nature of ices, their isotopic composition and possibly the mineralogy of a small body from another planetary system.

The launch of Comet Interceptor is planned in 2028, and the mission should visit its target sometimes within about 5 years after that. The target is not identified yet, and may not be known before launch as it will depend on the opportunities for new comets during the time of the mission. The mission will be a true test of a new style of fast and “cheaper” missions, that may pave the way for future exploration of very dynamic objects.

### 3.5.2.5 Lucy mission

Jupiter Trojan asteroids are small primitive bodies located beyond the ‘snow line’, around respectively the $L_4$ and $L_5$ Lagrange points of Jupiter at ~5.2 AU. Their origin – still unknown – remains a major challenge to current theories of the formation of the Solar System (Marzari et al. 2002). There are two current scenarios, each one having distinct implications for the origin and early dynamical evolution of the Solar System.

The first model proposes that the Trojans originally formed where they are seen today – with the Trojans being captured onto their current orbits while Jupiter was growing. An alternative model (the so-called Nice model) suggests that a large fraction (if not all of them) of the Trojans has formed in more distant regions - typically in the primordial trans-Neptunian disk, which is also the precursor of the Kuiper belt - and subsequently chaotically migrated towards the inner Solar System during the time when all four giant planets migrated, before being trapped at their current location. The radial migration of the giant planets is believed to be the last major event that sculpted the structure of the Solar System.

Up to now, no mission has gone through the regions in space where Jupiter Trojans are located while all major populations of objects in our Solar System have been visited by spacecraft. The Lucy mission to these objects will thus allow the exploration of the last major – yet unexplored - population of primitive bodies and contribute to answering the following key questions about the early history of the Solar System:

- Did the Trojans originate near Jupiter’s orbit or farther out in the Solar System? Are they the building blocks of the giant planet cores or instead more accessible TNOs?
- What do the compositions of these primitive bodies tell us about the region(s) of the solar nebula in which they formed? How much compositional variability is there among them?
- Is there any evidence of life precursors in the form of simple organics and water ice on these bodies? Are they volatile-rich or volatile-poor?
- What is the mineralogy of the silicates in these objects? To which extent did radial mixing of protoplanetary dust occur during the early phase of the Solar System formation?
- How do the geological processes that have occurred on Jupiter Trojans compare to those that have affected other small bodies? Are they homogeneous or differentiated?
- How are the properties of Jupiter Trojan surfaces modified over time by the space environment?
- How diverse are these objects?
- How do they compare to comets, TNOs, outer planet satellites, and Main Belt asteroids?

The Lucy mission was selected at the same time as the Psyche mission in the framework of the NASA Discovery program. It will flyby six Trojans from both the L4 and L5 Lagrange swarms and provide information on density, geology, and surface composition (See Figure 4.13). The Lucy mission should be launched in 2021 and reach the Jupiter Trojans after a 7 years journey.

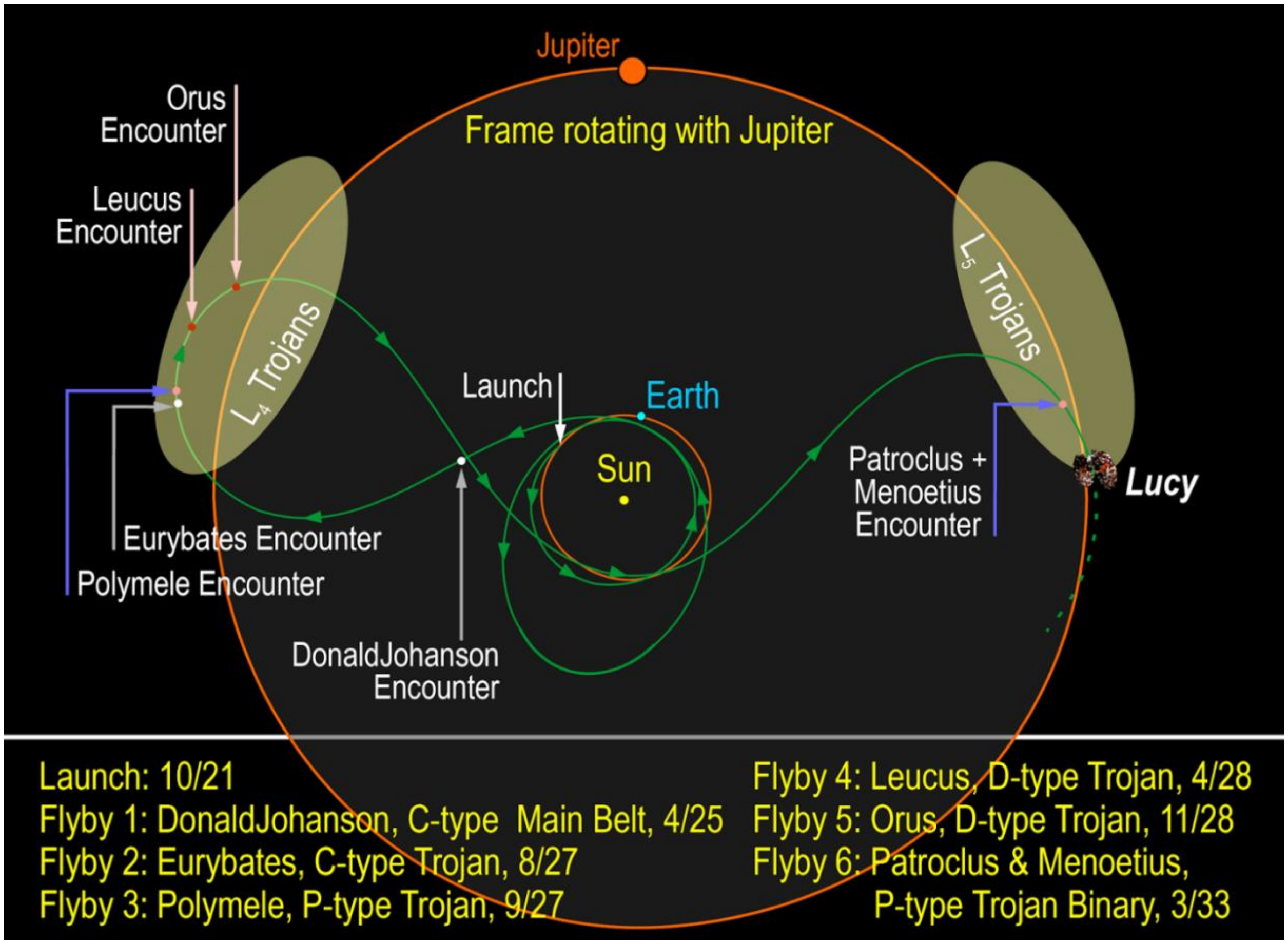

*Figure 4.13. Planned trajectory and flybys of the Lucy mission (credit SwRI)*

### 3.5.3. Representative missions for 2041-2061

#### Sample return of primitive matter from the outer Solar System

The last thirty years of cosmochemistry and planetary science have shown that one major Solar System reservoir is vastly under-sampled in the available suite of extra-terrestrial materials, namely small bodies that formed in the outer Solar System (>10 au). Because various dynamical evolutionary processes have modified their initial orbits (e.g., giant planet migration, resonances), these objects can be found today across the entire Solar System as P/D near-Earth and main-belt asteroids, Jupiter and Neptune Trojans, comets, Centaurs, and small (diameter <200km) trans-Neptunian objects. This reservoir is of tremendous interest, as it is recognized as the least processed since the dawn of the Solar System and thus the closest to the starting materials from which the Solar System formed. This is underlined by the extremely interesting results obtained by in-situ studies of isotopic compositions of matter from comet 67P/Churyumov-Gerasimenko by ESA’s Rosetta mission (see Hoppe et al. 2018 for a review), and from laboratory studies of anhydrous chondritic porous interplanetary dust particles (CP-IDPs) (Ishii et al. 2008), ultra-carbonaceous Antarctic micrometeorites (UCCAMs) (Duprat et al. 2010), and matter from comet 81P/Wild 2 returned to Earth in 2006 by NASA's Stardust mission (Brownlee et al. 2006).

The next major breakthroughs in planetary science will come from studying outer Solar System samples in the laboratory, but this can only be achieved by an L-class mission that directly collects and returns to Earth materials from this reservoir. The proposed strategy consists in 1) a direct trajectory to the rendezvous target, 2) a reconnaissance of the terrain with an orbiter payload including at least an optical camera, a near-infrared spectrometer and a thermal infrared camera, 3) collection of surface/subsurface samples (at least two locations) that are volatile and dust rich and 4) return of the samples to Earth. The re-entry capsule must be able to preserve the samples at cryogenic temperature. The selected target should be as primitive as possible which might exclude near-Earth objects from the candidate list. Comets and P/D main belt asteroids including main belt comets would then appear as the most accessible and scientifically valuable targets, with comets being the preferred targets because of their activity that can be used to characterize the volatiles and also because their surface should be more “primitive” (Vernazza et al. 2021b).

## 3.6. Giant planets and their systems

### 3.6.1. Main scientific objectives of giant planet systems exploration.

Giant planet systems are four different occurrences of secondary systems in the solar system. As Table 4.1 shows, their exploration addresses the six key science questions of this foresight exercise:

(Q1) The diversity of objects that compose them, from planet through ring particles, gas and plasma tori to moons, is unique in the Solar System, and these objects are probably fair analogues of objects populating other planetary systems. A detailed characterization of their different objects is therefore of the utmost importance: characterisation of the structures and dynamics of their interiors, atmospheres and magnetic fields; comparative characterization of their regular and irregular moons, including their bulk composition, shape and dynamics, internal layering including oceans, geology and surface properties, and their space environment; physical, chemical and dynamical properties of ring particles.
(Q2) The diversity of the four giant planets systems architectures remains to be understood: ring-inner moons systems, regular and irregular moon systems, their Trojan object populations, their giant and fast rotating magnetospheres and their interactions with moons and plasma populations.
(Q3) This diversity also challenges the current understanding of Solar System formation scenarios at all scales: what has been the role of gas and ice giants in the formation of the Solar System as a whole and in the radial redistribution of volatiles in its early ages? What were the formation scenarios of gas and ice giants and their different moon and ring-moon systems?
(Q4) How well does one understand how giant planets interiors and atmospheres work and maintain a planetary magnetic field? How do their regular moons work and what are their coupling processes to their host planet and to the other objects in the system? What are the main ring-moon coupling processes? How and how much is the dynamics of their magnetospheres, plasma populations, energetic particles, driven by their coupling to the central planet, to the different moons and to the solar wind?

Giant planet systems are clearly a unique in-situ laboratory to study this outstanding diversity of working mechanisms.

(Q5) The different ocean moons hosted by giant planets systems are potential habitable words: it is urgent to characterise the habitability of Europa and Ganymede in the Jupiter system, of Titan in the Saturn system. Ice giant systems also offer promising candidates: characterising the habitability of Triton and possibly of the active moons of Uranus are high priorities for the exploration of ice giants.

(Q6) Missions to search for life should be sent to the best candidates among these moons: Enceladus, whose habitability has been assessed by the Cassini mission, Europa, and possibly Titan and Ganymede.

### 3.6.2. Future missions to giant planet systems

A plan to explore giant planets systems in the Horizon 2061 perspective must take into account previous investigations and current knowledge of gas giants, ice giants, and their moons.

Both gas giants have been the subject of comprehensive orbital exploration missions: Galileo for Jupiter, currently followed by Juno and its extended mission, Cassini-Huygens for Saturn. The next missions should no longer focus on exploring the diversity of their systems, but rather on their origins, workings and habitability during the 2021-2040 period. In complement to the JUICE mission (Grasset et al. 2013) which will study the Jupiter system and Ganymede's habitability, and to Europa Clipper's investigation of the habitability of Europa (Howell and Pappalardo 2020), new missions addressing the formation scenarios of these systems would be particularly desirable. They could use combinations of in situ atmospheric probes, planet and moon orbiters and moon landers to achieve this objective.

In contrast, ice giants have been visited only once by the fly-by of Voyager 2 and remain largely unknown (see the review by Blanc et al. 2021b and the references therein). The next logical step for the 2021-2040 time frame is to fly ambitious orbiter missions that will provide a first comprehensive survey of their objects, architectures and workings. Combined with the delivery of an atmospheric probe, they could also explore their formation and migration scenarios. As these first orbiter missions will reach the ice giants in the 2040-2061 time frame, they should also be designed to characterize the habitability of their active moons: Triton at Neptune and several of the poorly known Uranian regular satellites.

A systematic search for habitability and life at ocean moons can be designed along the same lines. In the gas giant systems, several moons (Europa and Ganymede at Jupiter, Enceladus and Titan at Saturn) have already been the subject of close-in observations by the previous generation of orbiters, and the Huygens probe even landed on Titan. Time is ripe, in the 2021-2040 time frame, for missions aiming at an in-depth characterization of their environments, exospheres/atmospheres, internal oceans and magnetospheric interactions in view of assessing their habitability. They will prepare astrobiology-oriented missions searching for traces of life in their surfaces, sub-surfaces and immediate environments in the 2040-2061 time frame. In the ice giants systems, on the other hand, a similar task can be accomplished only by the first generation of orbiters reaching these planets in the 2040-2061 time frame, via a detailed exploration of Triton and of the regular moons of Uranus. These missions should return the data needed to design a mission to the most promising habitable moon of the ice giants: while Triton is the current best candidate, a first orbiter mission to Uranus may well bear many surprises and reveal equally compelling targets for the search for life around that planet.

Table 4.3 summarizes the timeline of missions already flown, in preparation, under study and notional missions addressing the six key science objectives. In the remainder of this section, we describe in more detail these missions as a function of their destination: giant planets systems as a whole, atmospheres, moons, rings and magnetospheres.

| Time | Gas giant systems | | | | Ice giant systems | |
|---|---|---|---|---|---|---|
| | Planet | Moons | Rings | Magnetosphere | System | Moons |
| 2021-2040 | Workings: JUICE<br><br>Origins:<br><br>Atmospheric Probes | Workings:<br>Titan – DragonFly<br>Io<br>Habitability:<br>Europa - EC<br>Ganymede - JUICE<br>Origins:<br>Callisto<br>Irregular moons | Origins & workings:<br><br>Saturn Ring skimmer<br><br>Saturn Ring Observer | | Diversity + Origins<br><br>Initiation of orbiter + probe missions to Uranus & Neptune | Habitability:<br><br>Triton<br><br>Active moons of Uranus |
| 2041-2061 | | Workings:<br>Io<br>Search for life:<br>Europa, Enceladus, Titan…<br>Moon orbiter + lander | | Workings<br>+<br>Role in habitability<br>Multi-point investigation of Jovian m'sphere | Implementation | Search for life:<br>Design of a mission to the best candidate habitable moon |

*Table 4.3: A summary of the missions already decided and of future notional missions to giant planets and their systems to be flown during the two (approximate) periods considered in this study: 2021-2040 and 2040-2061.*

### 3.6.2.1. Comprehensive missions to giant planets systems

In the 2021-2040 time frame, one mission in preparation aims at a comprehensive study of the Jupiter system: ESA's JUICE mission (Grasset et al. 2013). JUICE, to be launched in 2022, should be in Jupiter orbit by 2031 to execute a four-year tour of the system, studying its atmosphere, magnetosphere and three of the Galilean moons (Europa, Callisto and Ganymede) before going into Ganymede orbit for a detailed characterization of its geophysics, internal ocean, surface geology and magnetosphere, leading to a global assessment of this moon's habitability. The choice of the Jovian system, the largest and in some respect the most complex of giant planets systems, offers a template within reach of other giant planets systems, including probably the still-to-be-discovered giant exoplanet systems.

In the same time frame, the absolute priority in giant planet systems exploration is to conduct two exploration missions to each of the two ice giants Uranus and Neptune. Flagship missions to these destinations have been proposed by the U.S. community to NASA for the last two NASA decadal surveys (see Hofstadter et al. 2017) with a strong support of European scientists. The international planetary science community also met at the Royal Astronomical Society in London in January 2020, united in the goal of realizing the first dedicated robotic mission to Uranus and Neptune (Fletcher et al. 2020a). A white paper was also presented at ESA following the call for the Voyage 2050 program (Fletcher et al. 2020b).A joint orbiter + probe mission to either of these planets (Mousis et al. 2019) would perform the first detailed exploration of these systems while addressing the question of their origin. Blanc et al. (2021b) reviewed the science case and mission scenarios for these missions. For all missions to the ice giants, a fly-by of Jupiter on the way is the only way to reach these systems in a reasonable amount of time with the current propulsion capabilities. Based on an analysis of Jupiter flyby opportunities, Blanc et al. (2021b) came up with a timeline for the four notional missions, I to IV, presented in Figure 4.14.

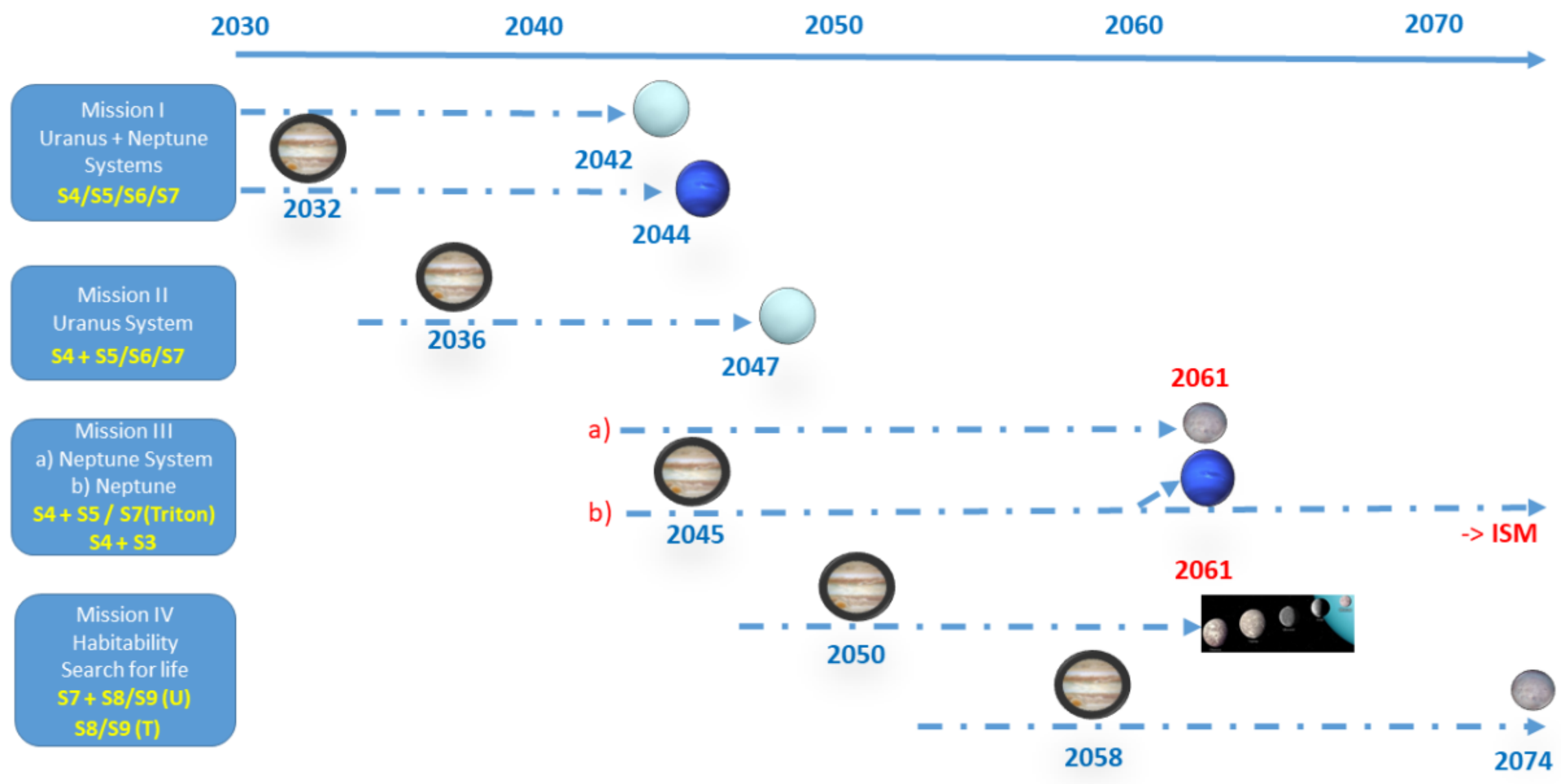

*Figure 4.14: An illustration of the different mission opportunities offered by the fly-bys of Jupiter en route to ice giants up to 2061, together with the preferred flight sequences (S4 to S9) for each of these windows and destinations. From Blanc et al. (2021b).*

Mission I uses a 2032 Jupiter flyby, which makes it possible to send two missions to the two ice giant systems simultaneously. If this opportunity cannot be met, one should plan to take advantage of the next two windows, one to Uranus with Jupiter fly-by in 2036 (mission II), one to Neptune with Jupiter fly-by in 2045 (mission IIIa), to fly a mission to each of these planets. If only one of the two missions can be afforded, this scenario proposes to fly mission II to Uranus and take the opportunity of a fly-by of Neptune by a different mission to the outer Solar System, for instance a mission to the TNOs, the heliosphere boundaries and/or the local interstellar medium, to deliver a probe into Neptune's atmosphere (mission IIIb).

Finally, a follow-on mission dedicated to the search for life, either at one of the active moons of Uranus (mission IVa), or at Triton (mission IVb), could be designed and launched before 2061 once missions II and III have determined the best candidate for habitability. Mission IVa would fly by Jupiter in 2050, while mission IVb would fly by Jupiter in 2058: a very appealing window for an astrobiology mission to Triton.

### 3.6.2.2. In situ exploration of giant planets atmospheres

Much of our understanding of the origin and evolution of the outer planets comes from remote sensing by necessity. However, the efficiency of this technique has limitations when used to study the bulk atmospheric composition that is crucial to the understanding of planetary origin, primarily due to degeneracies between the effects of temperatures, clouds and abundances on the emergent spectra, but also due to the limited vertical resolution of orbital observations. In addition, many of the most abundant elements are locked away in a condensed phase in the upper troposphere, hiding the main volatile reservoir from the reaches of remote sensing. It is only by penetrating below the “visible” weather layer that we can sample the deeper troposphere where those elements are well mixed.

A remarkable example of the unique contributions of in situ probe measurements to resolving the puzzle of giant planets formation is illustrated by the exploration of Jupiter, where key measurements such as the determination of the abundances of noble gases and the precise measurement of the helium mixing ratio have only been possible through in situ measurements by the Galileo probe (Owen et al. 1999; Gautier et al. 2001; Mousis et al. 2009, 2012). Similar results from a Saturn, Uranus or Neptune entry probe, and particularly at least from one ice giant, would be the indication as to whether the enhancement of the heavier noble gases found by the Galileo probe at Jupiter is a feature common to all the giant planets, or is limited only to the largest gas giant (Mousis et al. 2014, 2016, 2018). This could have broad implications for the properties of known exoplanets of both giant and ice types, especially in planetary systems sharing both types of exoplanets.

The primary goal of a giant planet entry probe mission is to measure the well-mixed abundances of the noble gases He, Ne, Ar, Kr, Xe and their isotopes, the heavier elements C, N, S, and P, key isotope ratios $^{15}N/^{14}N$, $^{13}C/^{12}C$, $^{17}O/^{16}O$ and $^{18}O/^{16}O$, and D/H, and disequilibrium species CO and $PH_3$, which act as tracers of internal processes, and can be achieved by a probe reaching at least 10 bars (Mousis et al. 2014, 2018). In addition to measurements of the noble gases, chemical, and isotopic abundances in the atmosphere, a probe would measure many of the chemical and dynamical processes within the upper atmosphere, providing an improved context for understanding the chemistries, processes, origin, and evolution of all the atmospheres in the Solar System. A descent probe should sample atmospheric regions far below those accessible to remote sensing, well into the cloud forming regions of the troposphere to depths where many cosmogenically important and abundant species are expected to be well-mixed. Along the descent, the probe would provide direct tracking of the planet's atmospheric dynamics including zonal winds, waves, convection and turbulence, measurements of the thermal profile and stability of the atmosphere, and the location, density, and composition of the upper cloud layers.

By extending the legacy of the Galileo probe mission, a Saturn probe and at least one ice giant probe will further discriminate competing theories addressing the formation and chemical, dynamical, and thermal evolution of gas and ice giants, and assess the role of these classes of planets in the formation of the Solar System and of other planetary systems.

### 3.6.2.3. Missions to giant planet moons

Giant planets host a large family of regular and irregular moons that interact strongly with the central planet magnetosphere (Saur 2021). While most regular moons (with the exception of Neptune's Triton) are believed to have formed in situ as by-products of the formation and evolution of their host planet, irregular moons are likely captured objects, i.e. former planetesimals formed in the Solar Nebula that fell later into the potential well of their parent planet. All inform us on the formation history of giant planet systems, which led to a very broad diversity of objects (see Chapter 3, Figure 3.15). A fraction of the largest moons are massive enough to have experienced differentiation and to display a complex geological history and geophysical activity, and an even smaller number of them host internal oceans which are candidate habitats for life: these are the Ocean Worlds, one of the priority targets of future exploration. Future missions to the Galilean moons and their connections to the six key science questions are presented first, before focusing on the exploration of Ocean Worlds.

**Missions to the Galilean moons**

The Galilean moons, the first discovered among the family (Galileo 1610) have kept until today a central place in the exploration of giant planets systems. There are several reasons to it, illustrated in Figure 4.15 (from bottom to top): (1) two of the moons, Io and Europa, display an active and young surface, dominated by volcanism in the case of Io and by intense tectonic activity in the case of Europa; (2) they offer a particularly interesting template for a planetary system, because of the role played by the 1:2:4 mean motion resonance (the so-called Laplace resonance) in the dynamics and evolution of three of the moons, similar for instance to the Trappist-n system of p planets in resonance (see chapter 2, figure 2.4); (3) and (4) their chemical composition and internal structure are very diverse, with two moons (Io and Europa) dominated by a rocky interior and three moons likely bearing an internal ocean. All these characteristics, combined with the fact that they are the closest destinations among giant planets moons, make them a destination of choice for future missions.

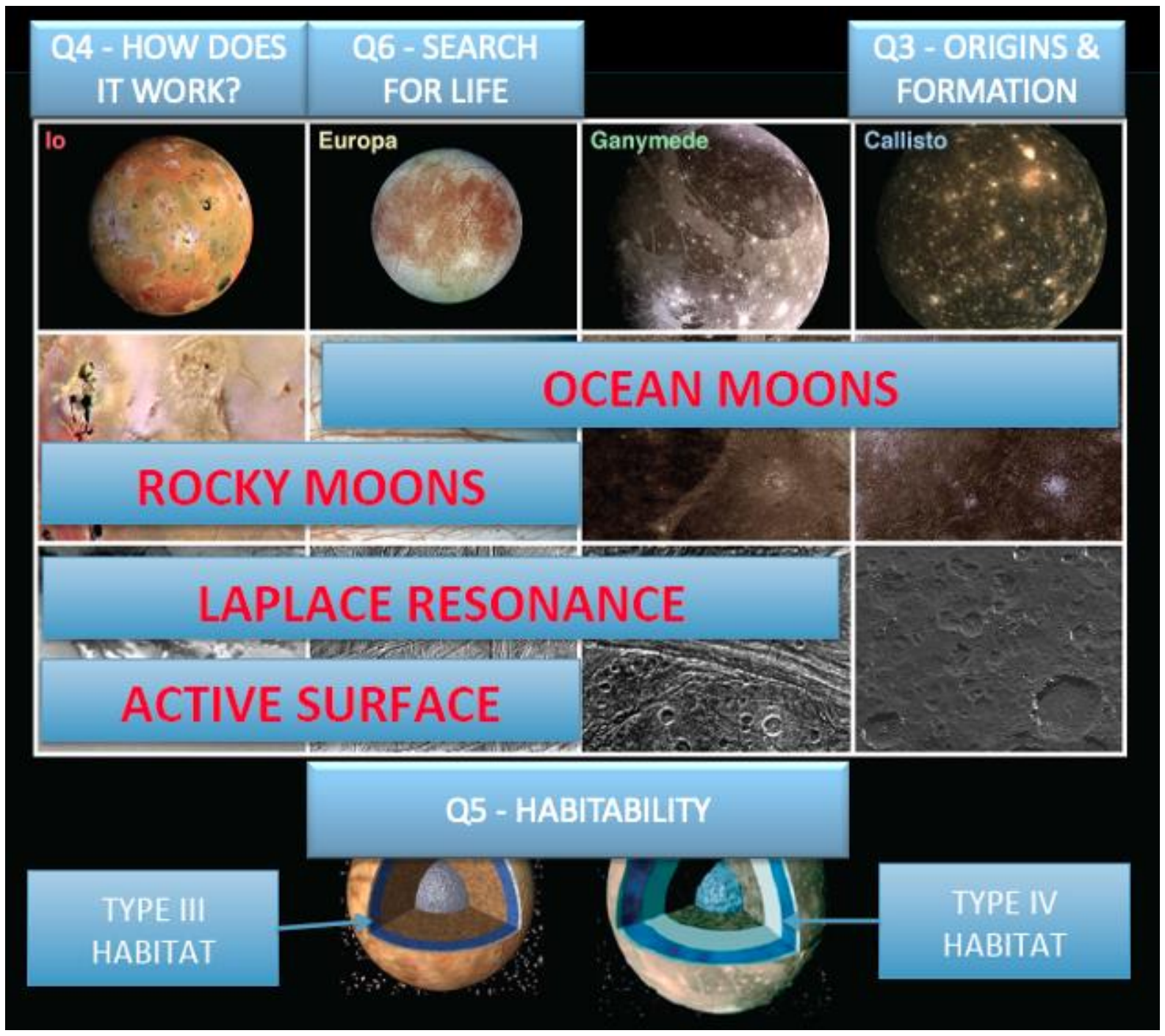

*Figure 4.15: This summary of the main properties of the four Galilean satellites shows the diversity of their internal structure and composition, of their surface activity, and of their resonant tidal coupling to each other and to Jupiter (Io, Europa, Ganymede on the Laplace resonance) or lack of it (Callisto). Three of the moons are "ocean worlds" and potential habitats, though of different types: type III for Europa (liquid ocean extending between an ice crust and silicate floor) and type IV for Ganymede (liquid ocean extending between two ice layers) according to the classification of habitats by Lammer et al. (2009). Space exploration of these moons specifically address different key science questions, as shown (see text).*

A plan for their comprehensive exploration up to the 2061 horizon will address altogether four of the six different key science questions (Figure 4.15 again).

**Question 3 – origins of planetary systems.** Callisto, which is outside of the Laplace resonance, seems to be the least differentiated of the four moons, and its heavy cratering record probably keeps the memory of the early ages of the Jupiter system. Studying its cratering, bulk and surface chemical composition and internal differentiation appears to be one of the best ways of discriminating between the different formation scenarios of the system (see Canup and Ward 2006). A mission combining flybys of several of the irregular satellites with an orbital characterization of Callisto, such as JUICE and the Gan De mission currently considered for flight by the Chinese National Space Agency (CNSA) (Blanc et al. 2019; Li et al. 2021b) will be in the best position, following on the heritage of Juno's study of the origins of Jupiter itself, to unravel the formation scenario of the Jupiter system.

**Question 4 – how do planetary systems work?** Io is the target of choice to address this question. A mission focusing on Io and its coupling with the Io torus, such as the Io Volcano Observer (IVO) (McEwen et al. 2014) will not only make it possible to understand how the "internal engine” of Io works. It will also make it possible to study how the Io engine is fed by its tidal interactions with Jupiter, and how in return the dense Io torus continuously maintained by Io’s volcanic exhausts is the main driver of the dynamics of its magnetosphere and of part of the Jovian aurorae. A mission on an eccentric Jupiter orbit experiencing a series of flybys of Io will be in an ideal position to describe the chain of mass, momentum and energy transfer processes which links the interior of Jupiter to the one of Io, and then, via its volcanism, Io to its torus and to Jupiter’s magnetospheric and auroral activity. In complement to this approach focusing on Io, a mission monitoring the whole system with adequate multi-wavelength instruments, for instance placed in orbit around the L1 Lagrangian point of the Sun-Jupiter system, would establish a global observatory monitoring energy transfer processes within the system (Li et al. 2021b).

**<u>Question 5: searching for habitable worlds</u>.** This will be a well addressed question in the coming two decades with NASA’s Europa Clipper and ESA’s JUICE missions sent to Europa and Ganymede, respectively, to characterize their habitability. Both missions will fully characterize the ocean and associated habitability of their main target, using two different mission scenarios: an orbital exploration of Ganymede for JUICE, and multiple fly-bys of Europa for Europa Clipper, a strategy illustrated in Figure 4.16 which reduces the effect of accumulated doses from Jupiter’s radiations belt particles on the spacecraft equipment while offering an excellent coverage of the surface.

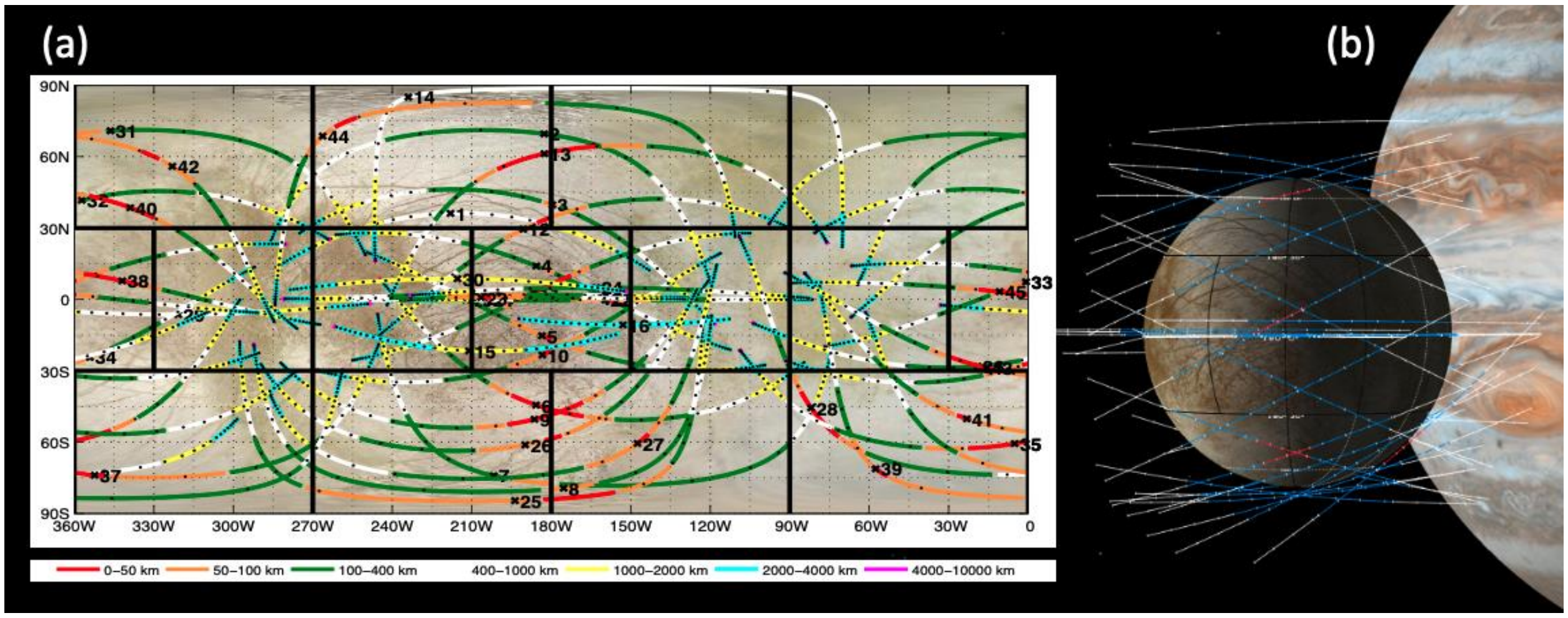

*Figure 4.16: an illustration of the exploration strategy used by NASA’s Europa Clipper mission to characterize the habitability of Europa. To avoid the extreme and lethal radiation doses that the spacecraft would experience if in Europan orbit, the mission stays in Jupiter orbit and executes a well-designed series of 45 or so flybys of Europa with different inclinations and phases, so that taken together the orbital arcs around Europa (b) constitute a bird’s cage which, projected onto Europa’s surface, provides a good coverage of its surface (a).*

**<u>Question 6: Searching for life</u>.** Among all Galilean moons, Europa appears as the best choice to search for life. Europa belongs to a broader family of moons and small bodies where life may have developed, which should remain one of the top priorities of planetary exploration during the next decades: Ocean Worlds.

**Missions to Ocean Worlds**

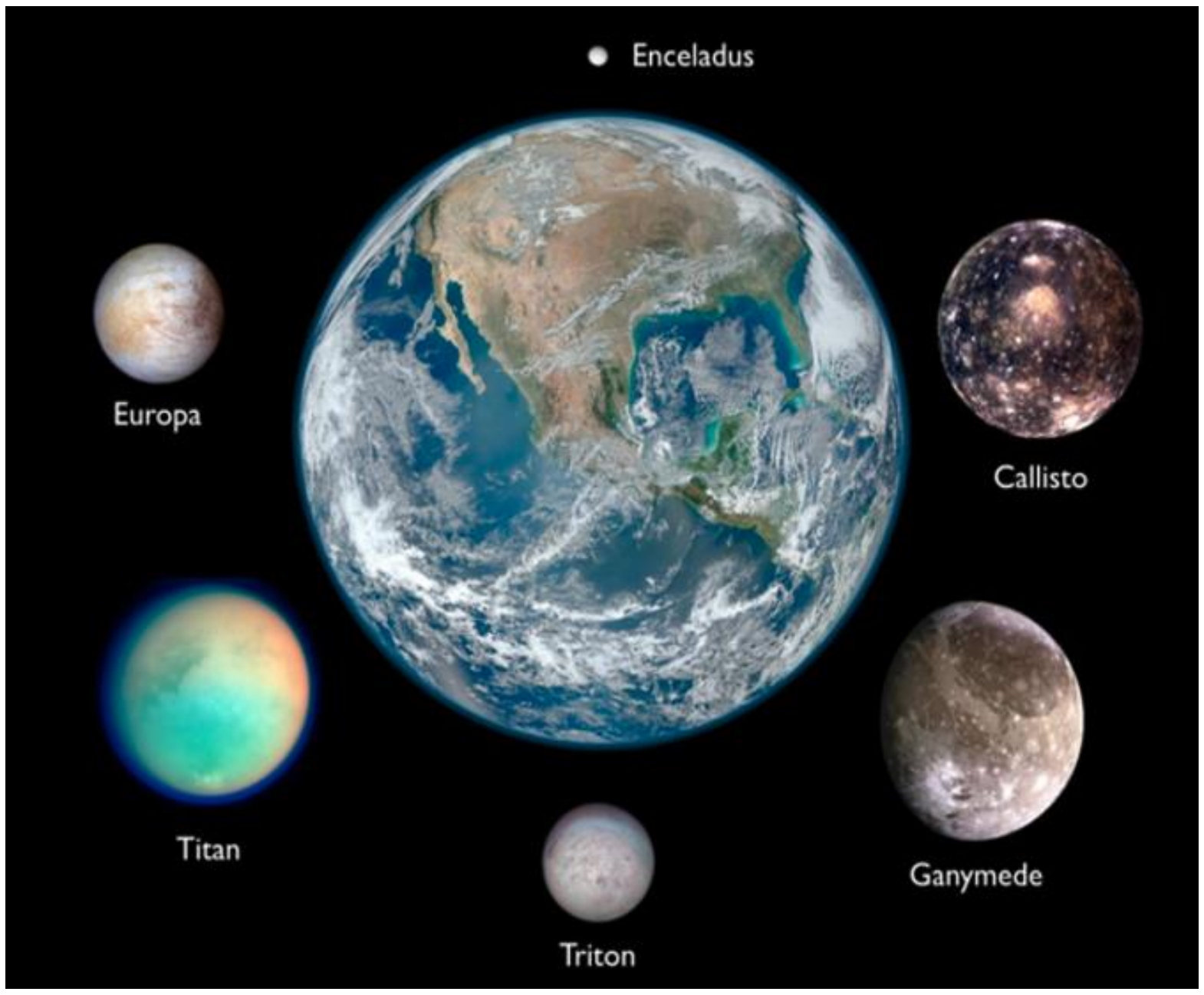

*Figure 4.17: "family portrait" of "Ocean worlds" orbiting giant planets, scaled to Earth for comparison. Credit: K.P.Hand, NASA-JPL. From Ocean Sciences Across the Solar System NASA report.*

Figure 4.17 shows a family portrait of ocean worlds orbiting giant planets discovered by Voyager, Galileo and Cassini. While for five of them the presence of an ocean is regarded as confirmed, Triton's ocean is still to be detected, and the relation of this possible ocean with the moon's surface activity and plumes needs to be investigated. The search for life at these moons should be one of the highest priorities of planetary exploration up to the 2061 horizon. NASA's Outer Planets Assessment Group (OPAG) was tasked to draft a "Roadmap to Ocean Worlds (ROW)" document which provides a detailed review of the different ocean worlds, or our state of knowledge concerning their habitability, and of the main science objectives and technology challenges that will drive the design of these missions. We refer to this comprehensive document (Hendrix et al. 2019) for a comprehensive presentation of these missions. Figure 4.18 shows the state of the art in progress towards search for life at each of these moons, using a set of logical steps leading to it: (1) identify oceans; (2) characterize oceans; (3) assess habitability; (4) search for life. The vertical borders between the colored and white areas correspond to the state of the art for each objects, and thus indicates what the next missions after JUICE (for Ganymede) and Europa Clipper (for Europa).

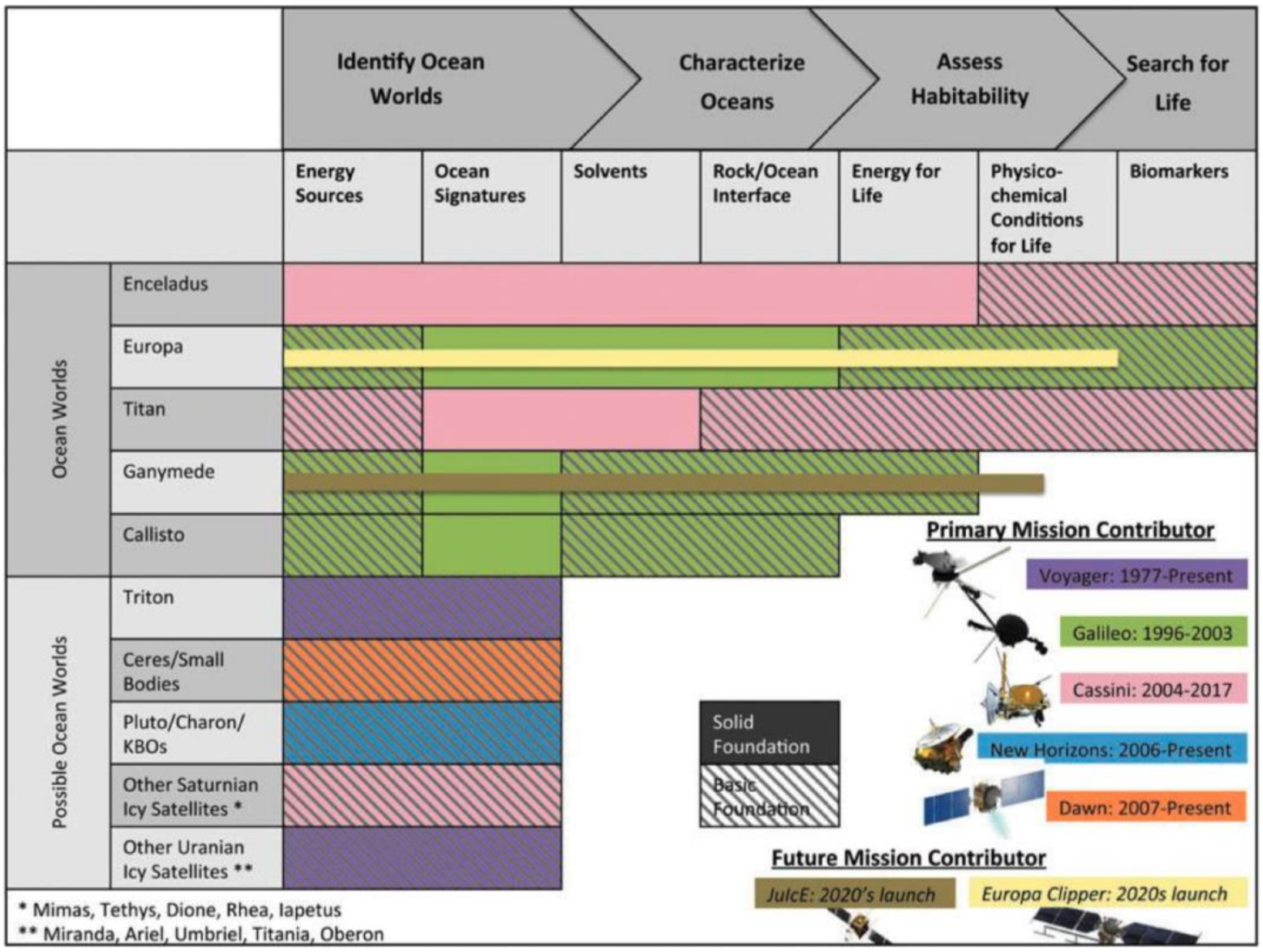

*Figure 4.18: Synthetic representation of the current state of the art in the search for life at confirmed ocean worlds Enceladus, Europa, Titan, Ganymede and Callisto (top lines) and at "possible" ocean worlds (Triton, several dwarf planets and the small satellites of Saturn and Uranus). From Hendrix et al. (2019)*

Building on the findings of Cassini for Titan and Enceladus, JUICE for Ganymede and Europa Clipper for Europa, the next steps can be directly derived from this diagram, as the white areas right of the colored ones. From the most advanced approach to life detection to the most preliminary ones requiring consolidation, the missions to fly concern:

<u>Enceladus</u>: the habitability of this moon has been fully demonstrated by Cassini. There is "green light" for a life detection mission, with the advantages that biosignatures can probably be found directly in the ice particles of the plumes, in addition to the surface. A mission sampling these particles, or even returning them to Earth, seems feasible in a relatively short term.

<u>Europa:</u> its habitability remains to be fully assessed by Europa Clipper. Following this first mission, which will also tell us more on the plume activity of this moon, the final design of a Europa lander mission can be consolidated. Several studies have already gone far into this definition, paving the way to a mission that should be ready to fly by the end of the 30's: the NASA Europa Lander SDT report (Hand et al. 2017) describes a NASA lander mission with sophisticated payload for the detection of biosignatures in the surface and sub-surface. A white paper submitted to ESA in response to the Voyages 2050 call alternatively described a "Joint Europa Mission" (JEM) between NASA and ESA (Blanc et al. 2020, 2021a) combining an ESA orbiter-carrier with the same NASA lander, complemented by a special ESA sub-platform specializing in the analysis of surface samples in liquid phase.

Titan: Understanding the complex world of Titan and its potential habitability is worth several missions, starting with the already selected DragonFly mission. In the next steps, special focus should be on understanding the possible connections between the sub-surface water ocean and the surface lakes and hydrocarbon and weather cycles.

Ganymede: following JUICE, a re-assessment of the potential habitability of this moon should be done before a life detection mission is possibly sent there.

Triton: the next step is to try and detect the presence of an ocean and, if it exists, the exchanges between this ocean and cryovolcanic activity at the surface. This could be done with a "simple" flyby, such as the Trident mission, candidate for selection by NASA in 2021 (Prockter et al. 2019).

#### 3.6.2.4. Missions to giant planet rings

**Science questions and measurement requirements**

Planetary ring systems are some of the most fascinating and beautiful structures in our Solar System. They provide key information on the origins, histories, current states and coupling processes of the giant planet systems of Jupiter, Saturn, Uranus and Neptune. Some may have formed from tidally disrupted moons or objects that wandered too close to the planet, while others may represent the remnants of the material left over from the planetary system's formation. Investigations of ring systems provide insights into the processes that operate in astrophysical disk systems, including exoplanet formation in circumstellar disks. Rings are accessible analogs to general disk processes such as accretion, gap formation, self-gravity wakes, spiral waves, and angular-momentum transfer with embedded masses that can be studied in detail with orbiter missions. Each planetary ring system is unique and distinct, representing a diversity of systems for study and comparison to each other as well as to exoplanet ring systems. Recent reviews of planetary ring systems are included in a comprehensive book edited by Tiscareno and Murray (2018).

Other smaller objects in the outer Solar System support ring systems as well, including the Centaur Chariklo, Kuiper Belt Object Haumea, and possibly Centaur Chiron. No mission studies to these objects have been proposed thus far. The orbits of these bodies are at relatively large heliocentric distances and inclinations, making them difficult to visit. Science questions for these targets include: how do rings form and evolve around small bodies, how stable are these rings, can they be used to constrain the dynamics of their systems, do/did other Solar System objects host ring systems?

The Cassini mission, orbiting Saturn for 13 years, revolutionized our understanding of the entire Saturn system, including its rings, moons, the planet itself, and its magnetosphere (Spilker 2019). The Cassini data set also raised important new questions about the Saturn system as well as planetary rings, which will be answered in the coming decades with new missions to the giant planets. For instance, one important question which Cassini left open, despite the delivery of key new data such as the total mass of the rings, is the one of their origin. While the mass of the rings speaks to their formation in the early solar system, other elements like their low degree of erosion by meteoritic bombardment and low content in non-ice material favor the idea that Saturn's rings may be young (by about 100 Myears): see for instance Crida et al. (2019) and section 4 of Blanc et al. (2021b) for a discussion of this controversy, which remains currently unresolved. Future rings-oriented missions will have to provide critical measurements to properly address the question of the rings origins, and many others: what is the history and evolution of ring systems; what do rings tell us about their planetary systems and exoplanet systems; how do the rings, planet and magnetosphere interact; and how do rings behave at the particle level? Additional science questions include: how do collisional dynamics behave in dense rings, what is the origin of large-scale ring structure, what is the lifetime of each ring system, and what processes determine its evolution? Future missions will address many of these unanswered ring questions.

**Types of Ring missions**

Most of the fundamental interactions in rings occur on spatial scales which are unresolved by previous flyby or orbiter spacecraft. Direct measurement of key ring characteristics such as the coefficient of restitution and particle's velocity dispersion are possible by monitoring individual collisions and would provide critical data on the rings' viscosity and behavior at the particle level. Future planetary ring missions that fly close to the rings will directly study ring particle interactions. Saturn Ring Skimmer and Saturn Ring Observer are examples of such missions. Many open questions still remain about the ring systems of Uranus and Neptune, which have only been visited by the Voyager spacecraft in the 1980s. Future ice giant missions will include ring science objectives and will once again allow detailed measurements of these ring systems that will answer many of the scientific questions outlined in the previous section.

**Saturn Ring Skimmer**

The Saturn Ring Skimmer (SRS) mission (Tiscareno et al. 2020, Decadal White paper) will fly close enough to Saturn's rings to observe individual ring particles and their interactions for the first time, and will provide new insights into the processes operating in astrophysical disks. This mission will provide new insights into the origin, history and evolution of planetary ring systems.

The SRS mission will study not only Saturn's rings, but also Saturn's atmosphere, interior and inner magnetosphere through a series of low altitude flybys over the rings. The spacecraft will fly two orders of magnitude closer to Saturn's rings than the best Cassini ring imaging sequences (Vaquero et al. 2019), although flybys of the rings will be at high velocities. The spacecraft's close proximity to the rings will provide detailed images and spectra of the rings, and also enable in-situ measurements of the material surrounding the rings including the ring atmosphere and tiny, elevated dust particles. The close flyby distance will also provide gravity measurements that will allow more detailed mapping of the mass of the rings. Saturn's atmosphere and deep interior, as well as the inner magnetosphere, will be characterized in more detail as well.

In addition to being scientifically valuable on its own, the SRS mission could be enhanced by the addition of a Saturn atmospheric entry probe. Detailed probe measurements would complement the other Saturn studies possible by this mission. The probe/orbiter mission could also be formulated as a joint NASA/ESA mission, much like the Cassini mission. Flybys of other targets of interest, such as Titan and Enceladus, could be performed as well. Current technologies make this mission feasible during the current decade.

**Saturn Ring Observer**

The Saturn Ring Observer (SRO) mission (Nicholson et al. 2010; Spilker 2003, 2011, 2018), illustrated in Figure 4.19, will hover over Saturn's rings, matching Keplerian velocities with the ring particles directly beneath the spacecraft to monitor the dynamical behavior of ring particles for an extended period of time. SRO will directly answer the question of how individual ring particles behave and interact. The spacecraft can traverse radially across the rings to different ring regions, studying 3-D particle dispersion velocities, individual particle rotations and collision frequencies, and coefficients of restitution as well as the aggregate behaviors such as self-gravity wakes, propellers, edge waves and density waves. Hovering capabilities will provide detailed monitoring of individual ring particles.

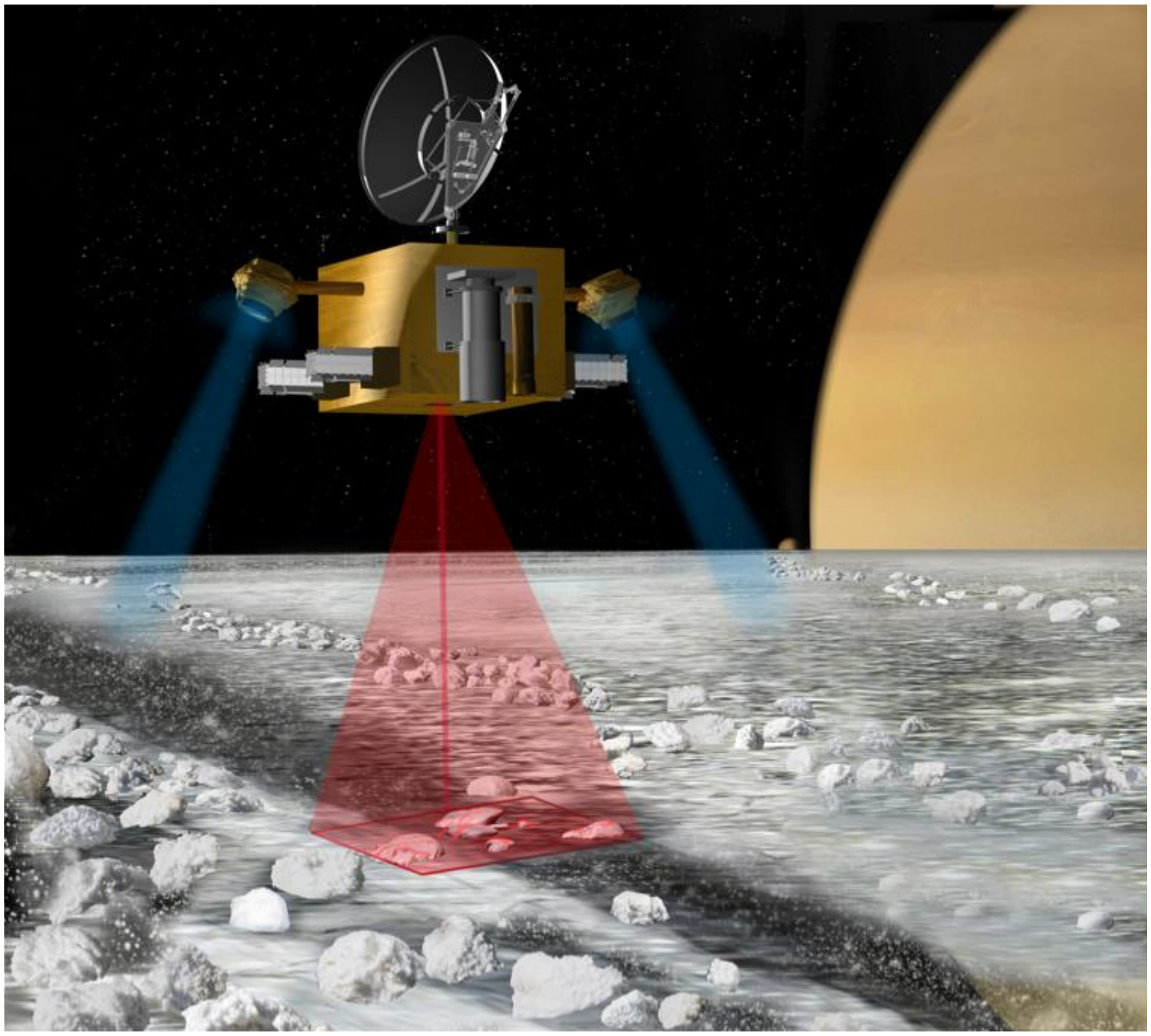

*Figure 4.19: An illustration of the Saturn Ring Observer (SRO) spacecraft hovering over the rings. Blue represents the exhaust plumes from the ion thrusters. Light red illustrates the limits of the scanning lidar, while the dark red line shows the lidar beam at a single instant in time. Lidar observations are used to measure ring particle motions. The actual distance between the rings and spacecraft is typically 2-3 km, much larger than shown here. Reproduced from the Nicholson, P., Tiscareno, M., Spilker, L., 2010. Saturn ring observer study report. Planet. Sci. Decadal Surv. NASA.*

The spacecraft is maintained in a circular orbit vertically displaced above the ring plane by 2-3 km. Constant thrust is provided by a Radioisotope Electric Propulsion (REP) system. The ring studies commence at Saturn's A ring and then move inward across the main rings. Another variant on the SRO concept is a chemical-based mission that uses multiple thrust events during one orbit to maintain a minimum altitude above the rings (Spilker 2003). Mission enhancements, such as those discussed for the SRS mission, are also possible for the SRO mission. Addition of a Saturn probe is a possibility.

**Orbiter studies of ice giant rings**

The ring systems of Uranus and Neptune have only been studied once at close distance by the Voyager flybys of Uranus in 1986 and Neptune in 1989. The Uranian ring system consists of ten narrow rings and at least three broad, diffuse rings. The ε ring is shepherded by two tiny moons, one on each side of the ring. The Neptune ring system is made up of five principal rings, one of which, the Adams ring, contains bright arcs of ring material in which particles are clustered together. A great deal remains to be learned about the ice giant ring systems. A dedicated ice giant orbiter at Uranus and/or Neptune would greatly enhance our understanding of these unique ring systems by providing Cassini-like ring studies.

Planetary ring systems continue to amaze and intrigue us. Studying these unique systems will provide useful information that can be applied to exoplanet ring systems as well.

#### 3.6.2.5. Missions to giant planet magnetospheres

**Science drivers for the exploration of giant planets magnetospheres**

As described in Chapter 3 (Dehant et al. 2021), because of the intense magnetic field of their host planet, Giant Planets' magnetospheres encompass the orbits of their rings and of most of their regular moons. This results in a broad variety of dynamical and electrodynamic interactions between the planet and the other objects in its system, including plasma and charged particle populations, which is mediated by the planetary magnetic field. The Cassini mission at Saturn, Galileo and more recently Juno at Jupiter, which opened a window into the study of these complex interactions. However, their in-depth study and understanding will require new missions with proper platform architecture and instrumentation, for many different reasons. First of all, exploring these magnetospheres in situ is essential to study complex physics of magnetized plasmas (which are very difficult to study in laboratories on Earth, but constitute 99% of the baryonic matter in the Universe). It will also teach us a lot about the interactions of magnetized plasma domains with solid particles (rings), solid surfaces (moon surfaces) and the atmospheres, ionospheres and exospheres of the central planet and its moons (Saur 2021). Finally, beyond the sole perspective of understanding the Solar System, it will also open a window onto the likely complex and still unknown secondary systems of moons, rings and magnetospheres waiting for our discovery around large-mass exoplanets, particularly in the Neptune-like and Jupiter-like families.

As the four giant planets are much bigger than the Earth and have much shorter rotation periods (to first order, a day lasts 10h at Jupiter and Saturn, 18h at Uranus and 16h at Neptune), planetary rotation plays a major role in the global dynamics of those four magnetospheres, much more than the solar wind dynamics or than the reconnection of the planetary magnetic field with the interplanetary magnetic field. The global dynamics of a giant magnetosphere also depends strongly on the tilt angle between the magnetic axis and the spin axis (see Figure 3.21 of chapter 3, Dehant et al. 2021), which is on the order of 11° at Jupiter, null at Saturn, whereas Uranus and Neptune dipole magnetic field axes are largely tilted from their spin axis. This large tilt is at the origin of a very dynamic magnetospheric configuration, with a strong diurnal variation at both planets and a strong seasonal variation at Uranus, whose spin axis almost lies in the ecliptic plane.

Another very unique characteristics of giant planets magnetospheres is the importance of their moon as plasma sources: Io at Jupiter, Enceladus at Saturn provide most of their internal plasma sources. Determining the importance of the regular moons of the ice giants, including Triton, in their plasma populations, will be one of the objectives of future missions to these planets.

In the specific perspective of this book, the exploration of giant planets magnetospheres addresses most of the six key science questions driving the design of future space missions, as summarized in Table 4.4.

| ***Science Question*** | ***Related to Giant Planets (GP) magnetospheres?*** | ***Key measurements to be performed to address this*** |
|---|---|---|
| *Q2: Diversity of Planetary Systems architectures* | *Characterize the global structure of the four GP magnetospheres to capture their diversity. Disentangle the relative effects of planetary rotation, interactions with moons and solar wind dynamics. Determine the relative importance of the different plasmas sources.* | *Magnetometers, particle detectors (electrons/ions), plasma wave measurements onboard orbiters or multi-spacecraft missions* |
| *Q3: Origin of Planetary Systems* | *What is the history of the formation of GP magnetic fields? What explains the diversity of their intensities, tilts, rotation speeds?* | *Magnetometers onboard orbiters (polar orbits)* |
| *Q4: How do Planetary Systems work?* | *Understand atmosphere-ionosphere-magnetosphere, moon-magnetosphere and rings-magnetosphere interactions* | *Magnetometers, particle detectors (electrons/ions), plasma wave measurements onboard orbiters or multi-spacecraft missions* |
| *Q5: Do planetary systems host potential habitats?* | *Effects of moon/charged particles/magnetospheres interactions on moons habitability* | Magnetometers, charged and neutral particle detectors<br>Spectral imaging of moon surfaces in UV, visible and IR |
| *Q6- Where and how to search for life?* | *Role of magnetospheres in transfer of water and other chemical compounds between moons* | |

*Table 4.4: This table describes in which way the exploration of giant planet (GP) magnetospheres addresses five of the six key science questions of the Horizon 2061 exercise, together with the key measurements required to address these questions.*

**Types of missions required to perform the key measurements**

The more explored magnetospheres in the family are the one of Jupiter, followed by Saturn, while Uranus and Neptune were visited by only one Voyager II flyby (see Figure 4.20). Future missions should therefore use Jupiter and Saturn for in-depth studies of their working processes, while the magnetospheres of ice giants still require a first orbital exploration mission to provide a first picture of their configurations and working mechanisms.

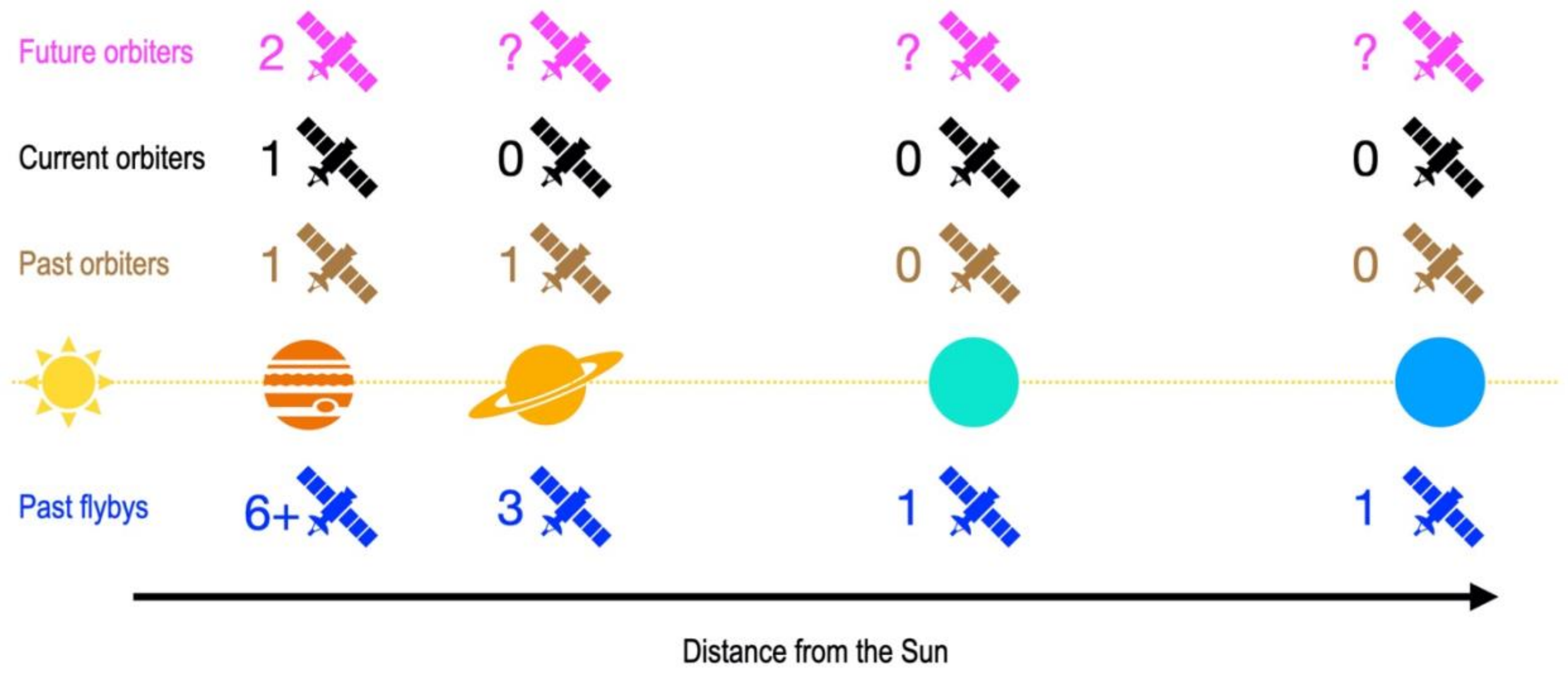

*Figure 4.20: Schematic of planetary giant magnetospheres exploration, along with the respective mean orbital distance of each Giant Planet (Jupiter, Saturn, Uranus and Neptune) to the Sun. Planetary magnetospheres are most often time-dependent and thus require orbiters to be fully understood.*

## Missions in flight or in preparation up to 2040

**JUPITER:** JUICE - JUpiter ICy moons Explorer - is the first large-class mission in ESA's Cosmic Vision 2015-2025 programme, currently in preparation. Planned for launch in 2022 and arrival at Jupiter in 2029, it will spend at least three years making detailed observations of the Jupiter system, including its magnetosphere and three of its largest moons, Ganymede, Callisto and Europa. Its final phase in Ganymede orbit will be a unique opportunity to study the only intrinsic magnetosphere of a moon that has been discovered so far in the Solar System.

**URANUS/NEPTUNE**: realistic scenarios for the future exploration of Ice Giants have been described in section 3.6.2.1 above and Figure 4.14. These combined orbiter-probe missions should include a comprehensive investigation of ice giant magnetospheres. Blanc et al. (2021b) showed that, in order for these missions to address the key questions of magnetosphere science (Table 4.4 above) they should carry a comprehensive particles-and-fields instrument package and include successions of high latitude/high eccentricity/low periapse orbits, medium inclination orbits and near-equatorial orbits for a good coverage of the main planetary fields (magnetic, gravity) and of plasma populations. A series of close targeted moon fly-bys will enable one to characterize moon-magnetospheres interactions. This approach, extensively used by Cassini, Europa Clipper and JUICE, will be well suited for the study of the Uranian moons and can also prove quite effective for Triton (Hosftadter et al. 2019).

#### Missions to be flown between 2040 and 2061

At Jupiter and/or Saturn, a multi-spacecraft mission following the approach used in the Earth's magnetosphere by the Cluster (ESA) and Magnetospheric Multi-Scale (NASA) missions will be required to keep making progresses after Cassini, Juno and JUICE. Inside the magnetosphere, such a multiple spacecraft mission will allow one to use multi-points measurement to distinguish between temporal and spatial effects. It will link the physical processes taking place in critical magnetospheric boundaries (magnetic field reconnection processes, Kelvin-Helmoltz instabilities) to processes coupling them to the planets upper atmosphere (field-aligned currents, aurora, radio emissions). It should also aim at providing a deeper insight into the role of moons in the dynamics, energy transfer processes and plasma sources of giant planet magnetospheres. At Jupiter, radiation belts and magnetotail dynamics could be explored through a Van Allen Probe/THEMIS-like mission. With one spacecraft inside the magnetosphere and one in the upstream solar wind on the dayside, the role of the solar wind in the driving of these magnetospheres could be accurately separated from the role of internal drivers (planetary rotation and moons).

### 3.6.3. Technology challenges for future missions

Key generic technology developments will be required to enhance or enable future outer planet missions. In addition, several of the mission types just described require specific technology developments.

Future missions to the Ice Giants (Uranus and Neptune) will be particularly demanding on:

- Power, and development of higher efficiency radioisotope power systems (RPSs). The long flight times and long durations of these missions require light-weight, efficient power sources to provide higher specific power for a given RPS mass. Assuming solar power Is not feasible at Uranus or Neptune, the use of nuclear power solutions (Radioisotope Power Sources) seems mandatory.

- Telemetry: how to return back to Earth the significant data volumes required by magnetospheric science and spectro-imaging?
- Orbital choices/longevity of spacecraft and instruments: magnetospheric studies require very large orbits, with periapsides close to the planet and apoapsides as far as possible, for solar wind monitoring in the dayside, and magnetotail investigations in the night side, and as many orbits as possible for a longer temporal coverage.

Multi-point missions to Jupiter or Saturn for magnetosphere studies will require the operation of several platforms, and possibly the development of Cube/Nanosat(s) that can be resilient in the Jovian magnetosphere environment, to be dropped off by a main orbiter. If these missions include an « upstream monitor » capable of providing solar wind parameters upstream of the planet, a special release and orbital scenario will have to be designed to inject it, for instance, in orbit about the L1 Lagrangian point of the giant or icy giant planets. This « upstream » spacecraft could be used by the Heliospheric community in order to monitor the propagation of the solar wind in the outer Solar System (trans-disciplinary science).

The advent of small, miniaturized spacecraft, often called CubeSats, can provide the unique opportunity to directly sample ring particles by sending a small daughter spacecraft to rendezvous with a ring particle, collect a sample and return it to the primary spacecraft such as SRO for detailed imaging of its surface and for compositional analysis. Other CubeSats could attach themselves to ring particles, and radio back data about collisions and the behavior of the ring particles over the several-day lifetime of the CubeSat. CubeSats would enable a new set of on-site ring observations.

Additional technology development is also needed to ensure the Saturn Ring Observer mission. Higher efficiency RPSs are needed along with more efficient electric propulsion systems to support the hover spacecraft's mass. The efficiencies of the current REP systems are not sufficient to support a capable spacecraft in hover orbit. Another SRO technology development would include remote operations to sense and maneuver to avoid collisions with any hazardous ring particles that have been deflected above the ring plane. One-way communication from Saturn to Earth is over one hour so the spacecraft would need to detect and react to any hazards in a much shorter time.

Lander missions to the regular moons of giant planets will require EDL (Entry, Descent, Landing) systems capable of landing in full autonomy on sometimes chaotic terrain, or even EDLA (Entry, Descent, Landing, Ascent) systems in the case of sample return missions. Once on the moon surface, the landers will have to perform their assigned science operations in extreme conditions of low temperature, and also of very harsh radiation environment in the case of Europa. Astrobiology-oriented missions will have to operate specific suites of science investigations based on sophisticated strategies to search for bio-signatures at the surfaces and sub-surfaces of ocean moons: see for instance Hand et al. (2017); Blanc et al. (2020, 2021a) and Prieto-Ballesteros et al. (2019).

Finally, magnetospheric studies and interaction with the solar wind require specific instrumentation. All future missions aiming to explore Giant Planets magnetospheres require in-situ measurements of:

i. The distribution functions of all constituents of the plasma
ii. The DC/AC magnetic and electric fields
iii. The distribution functions of energetic particles

All quantities must be measured with high spatial, temporal and directional information.
Future mission need to provide instrumentation enabling higher performance with smaller size, lower power and lower cost : miniaturization of the list of instruments given above is a key aspect of future exploration. Radiation tolerance is also non-negligible, particularly so for the exploration of the inner Jovian magnetosphere, where the Io torus, its main plasma source, stands in the region of harshest radiation environment.

### 3.6.4. New infrastructures and services needed

To fully realize the scientific potential of future giant planet missions, enhanced collaborations between ESA, NASA and other space agencies would provide the expertise and resources to send a highly capable spacecraft and comprehensive science payload to study those targets. The flight times to reach the outer planets are on the order of a decade or more, so sending diverse, comprehensive, capable and complementary scientific payloads and advanced spacecraft would provide the greatest science return for a given mission. A collaboration between « Planetology » and « Heliophysics » divisions of space agencies could also be an opportunity to use the cruise phase as an outer Solar System solar wind exploration mission.

Ideally, this enhanced collaboration should begin in the concept study phase, and include all aspects of the mission from initial concept development, spacecraft and instrument design, selection and build, launch, joint in-flight operations, through scientific analysis, publication and data archiving. Multinational efforts have the potential to maximize the scientific return and sometimes to enable the most challenging missions, as will be shown in chapter 7.

## 3.7. From the Trans-Neptunian Solar System to the interstellar medium

### 3.7.1. A new frontier of space exploration.

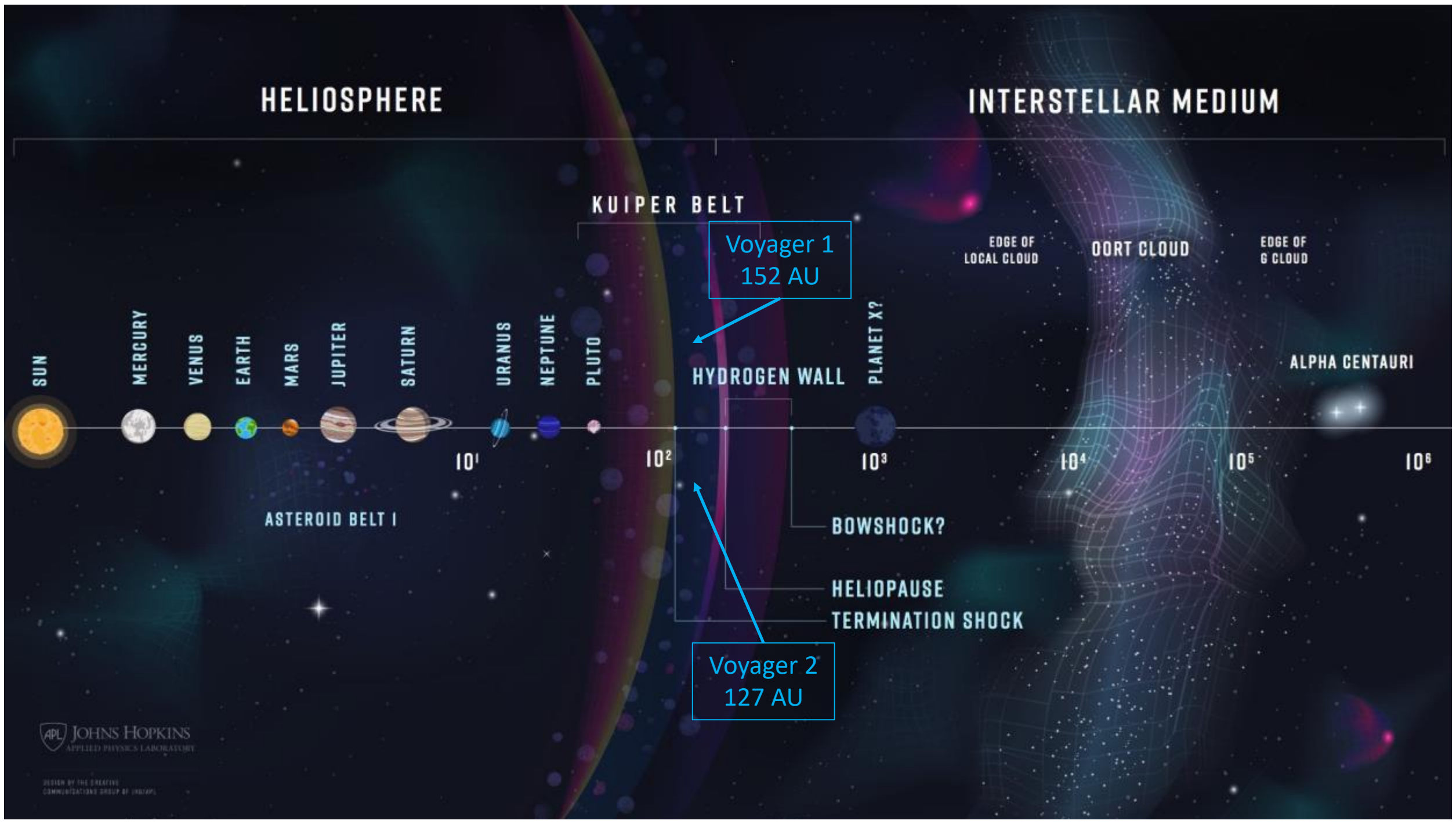

*Figure 4.21: Representation of Solar System objects, families of objects and heliospheric boundaries encompassing the Interstellar Medium and the nearest star, Alpha Centauri, plotted in AU on a logarithmic scale. Distances to the Sun of Voyager 1 and 2 for June 2021 are indicated. Courtesy Ralph Mc Nutt, Johns Hopkins University.*

A logarithmic scale of sun-centered distances is needed (Figure 4.21) to place the Solar System in the broader context of its local Galactic environment. On this scale, the population of Trans-Neptunian Objects (TNOS) extends from 50 AU to the heliopause region and beyond. For dynamical reasons, a spacecraft visiting a TNO will cross the heliopause and its different boundary layers (termination shock, hydrogen wall, external bow shock if any, before flying into the "Very Local Interstellar Medium" (VLISM)), right into one of the local interstellar clouds surrounding the Solar System (see chapter 3, section 5.5 and Figure 3.29). This complex medium, never explored in situ yet, extends all the way to the Oort cloud, the farthest among the populations of small bodies, and to the nearby stars. The closest of them Proxima Centauri, known to host an exoplanet in its habitable zone, appears to be a logical destination in a very long-term, perhaps multi-centennial perspective. Exploring the Trans-Neptunian Solar System and heliospheric boundaries to reach out into the VLSM is one of the "new frontiers" of planetary exploration. As for the other "provinces" previously described, the next steps can build on the heritage of space missions flown before 2020 and of a very large body of telescopic observations, with a first period (approximately 2021-2040) using currently available technologies, and the following bi-decadal period flying more ambitious missions benefiting from more advanced technologies. Table 4.5 summarizes this perspective, which will be developed in the following sub-sections for each of the destinations addressed by the lines of the table: Trans-Neptunian objects, heliospheric boundaries and the VLISM.

As the table shows, the dynamical characteristics of future missions exploring Heliospheric boundaries will take them into the VLISM. While building on different observational heritage (Voyager, Cassini and IBEX for heliospheric boundaries, telescopic observations for the VLISM), the scientific objectives for the two regions will addressed by the same type of missions: long-range, long-duration missions reaching out to the VLISM.

| **Time period** | **2020 heritage** | **2021-2040** | **2041-2061** |
|---|---|---|---|
| Trans-Neptunian Solar System | New Horizons<br>Telescopic observations | Explore the diversity of TNO's and dwarf planets (including multiple systems) with flybys | Orbiter and/or sample collection from selected objects and multiple systems |
| Heliospheric boundaries | Voyager, Cassini, IBEX | • First mission crossing all heliospheric boundaries and cruising in the VLISM<br>• First global observations of the Solar System and its debris disk from outside | Multiple missions in different directions:<br>• Exploration of heliospheric boundaries and resolution of its 3-D shape<br>• Long-term, long range exploration of VLISM in the directions of the different local clouds |
| Interstellar Medium (VLISM) | Telescopic observations | | |

*Table 4.5: Main mission objectives for future missions to the Trans-Neptunian Solar System, to heliospheric boundaries and to the VLISM : current heritage; mission objectives for the periods 2021-2040 (using current technologies) and 2041-2061 (as new propulsion and power delivery technologies become available.*

### 3.7.2. Exploring the Trans-Neptunian Solar System

Our knowledge of the very broad family of Trans-Neptunian Objects (TNOs), and among them of the dwarf planets orbiting in this distant region, is based on an impressive body of telescopic observations which followed the discovery of the first TNO in 1992. The New Horizons mission opened the era of the space observation of these objects with its spectacular fly-bys of the Pluto-Charon system in 2015, followed in 2019 by the fly-by of 486958 Arrokoth, a surprising bi-lobe shaped object nicknamed “Ultima Thule” by the New Horizons team (see chapter 3, figure 3.16). The body of observations and models available (see Prialnik et al. 2019) gives an idea of the extreme diversity of this family in size, color and albedo, shapes, with many grouped in multiple systems like the Pluto-Charon system, a binary system with four smaller moons orbiting it.

The main objectives of the exploration of this family is thus first to explore this diversity, and to characterize representative objects in each of the groups that can be identified this first inventory.

**3.7.2.1 2021-2040**

During this first period, one should take advantage of the current technologies to fly by as many different objects as possible, if possible sampling the different groups of objects, to characterize them in the same way as New Horizon did. The missions should be optimized to visit several objects. Given the fast trajectories used for these visits, missions aiming at crossing the heliopause and exploring the ISM will be the best choice.

**3.7.2.2 2041-2061**

Building on a better characterization of some representative objects, and of the family as a whole, the next period should fly missions with significantly larger Delta V capacities, likely based on new non-chemical propulsion technologies, that could orbit one of the most interesting systems (the Pluto-Charon system is likely one of the best choices) and/or collect samples during its fly-by.

### 3.7.3. Exploring heliospheric boundaries and the VLISM

McNutt et al. (2019) gave detailed descriptions of the next steps after Voyager in the exploration of heliospheric boundaries and the VLISM. Additional material can also be found in the Taikong magazine report #20. This section only summarizes their main conclusions.

The heliosphere, the extension in space of the Sun's magnetic field and expanding atmosphere, the solar wind, is the Sun's astrosphere and our shelter in the Galaxy. Heliospheric boundaries are opaque to a fraction of the components of the Interstellar medium, reducing by about 75% the flux of Galactic Cosmic Rays propagating inside the Solar System and impacting its planets. In this sense, the interaction of the heliosphere with the VLISM plays a key role in the habitability of the Solar System and of its planets. By exploring the heliospheric boundaries and better understanding their workings and interactions, one will better the role of astrospheres as shelters for the development of habitable planets around other stars.

Missions flying beyond the outermost boundaries of the heliosphere into the VLISM will explore and characterize for the first time the complex structure and dynamics of the interstellar medium, with its mix of neutral gas and plasma populations of different temperatures, dust and cosmic rays, regulated by the poorly known interstellar magnetic field. They will carry a science payload that can measure all these components in situ and look back at the Solar System to observe it for the first time "from outside" as a single object, and possibly detect in the infra-red its disk of dust particles and small bodies.

**3.7.3.1 2021-2040**

The next logical step is to build on the heritage of the first in-situ exploration of heliospheric boundaries by the Voyagers and on the remote sensing measurements of dust and charged particles by Cassini and IBEX to design a mission that will be able to cross and characterize all boundaries of the heliosphere, from the inner shock and complex magnetic field pile-up region extending beyond it, to the heliopause itself and a putative outer shock, and will materialize Humankind's first steps into the Interstellar Medium. McNutt et al. (2019) describe the science objectives, science payload, optional mission profiles and technology challenges of this "Interstellar Probe" mission, under study for NASA as part of the Heliophysics Decadal Survey. Though technically challenging, this mission should be ready for flight in the few years to come.
This mission should be able to reach the heliopause in a time of the order of ten years at most. For a heliopause located around 120 AU from the Sun, it must therefore fly at a speed on the order of 12 AU/year. Several options have been considered by the Interstellar Probe study to achieve this objective, all of them using first a Jupiter gravity assist: (1) passive JGA and direct escape, (2) active JGA, and finally (3) JGA followed by low-altitude fly-by of the Sun at a few solar radii, the most efficient but also the most challenging one.
Figure 4.22 shows the configuration of the spacecraft composite (panel (b), based on a New Horizon-type bus, that should perform this daring maneuver. The spacecraft is protected from the solar radiation by a heavy abrasive shield, and must carry an engine (below the spacecraft in the figure) that will burn close to the perihelion to inject the spacecraft to a Solar System escape trajectory. Panel (a) shows the asymptotic escape velocity achieved for different types of engines (color curves) and different perihelions from 3 to 5 solar radii. As the panel shows, the current optimal solution uses a Castor 30XL engine, and corresponds to a perihelion of about 4 Rs. Flying closer to the Sun at 5, 4 or 3 $R_S$ (successive configurations from left to right on panel (b)) does not provide a significant gain in this parameter range, as the shield mass has to be increased with decreasing perihelion).

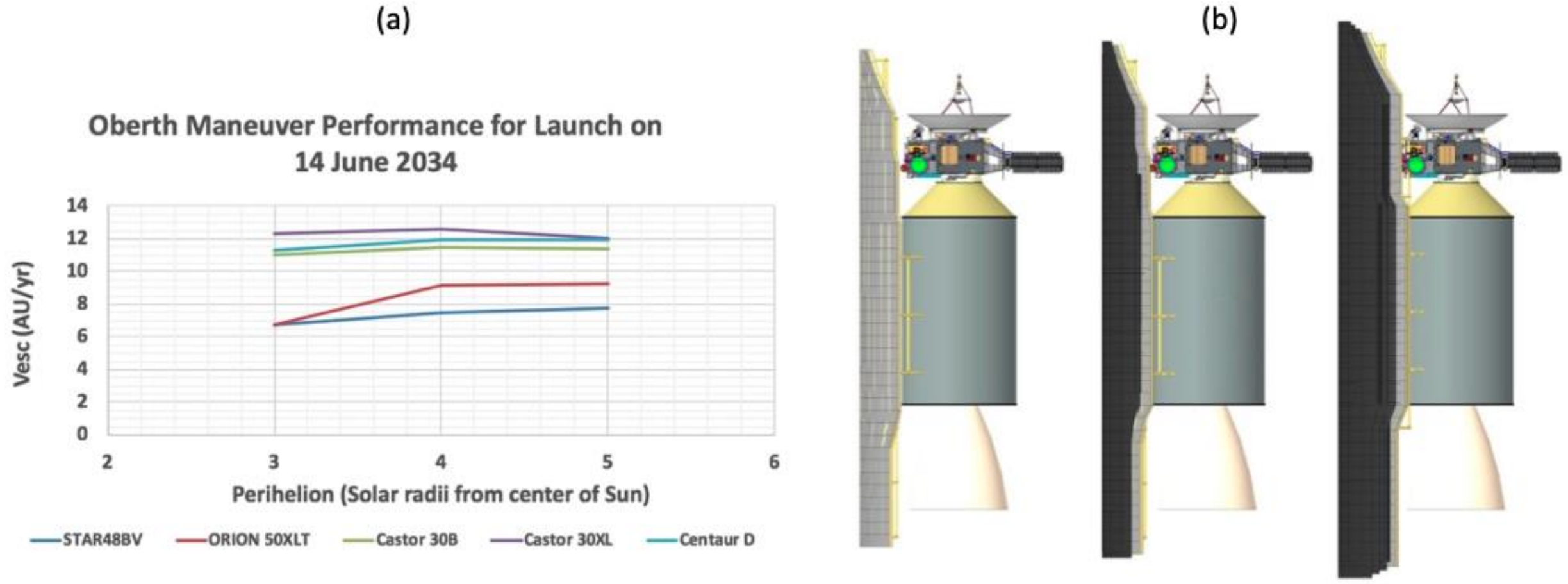

*Figure 4.22: Escape velocity achieved by an interstellar probe of the New Horizon class (panel a) for different perihelion distances and different types of engines. The corresponding configurations of the compositive spacecraft are shown from left to right in panel (b) for perihelions at 5, 4 and 3 $R_S$. From McNutt et al. (2019).*

(a) Probe I to nose of heliopause

(b) Probe II to tail of heliopause

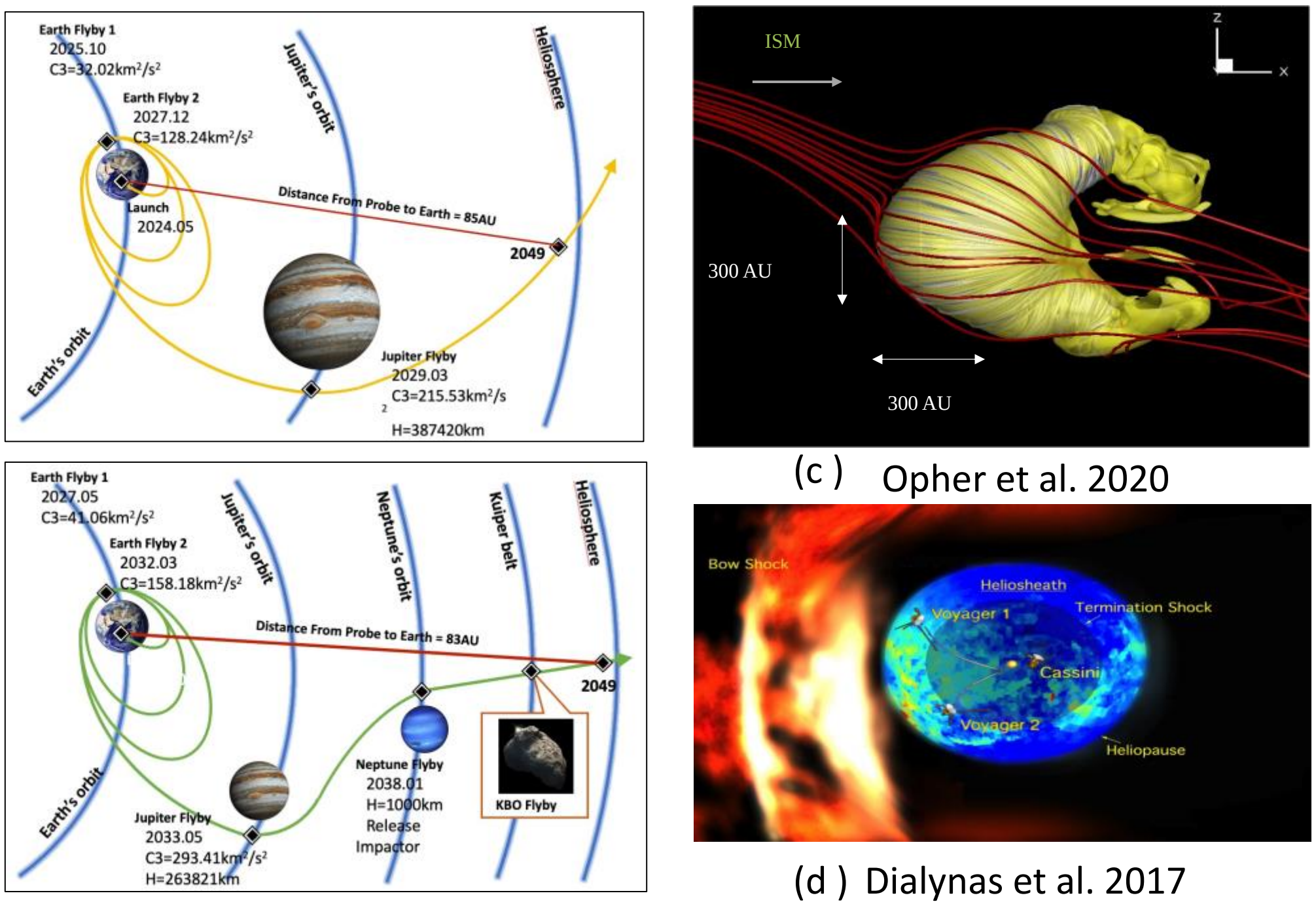

(c) Opher et al. 2020

(d) Dialynas et al. 2017

*Figure 4.23: The 3-D shape of the heliosphere and its boundary, the heliopause, remain basically unknown today; different models using different assumptions on the physical parameters of the outer heliosphere and of the surrounding very local interstellar medium (panels c and d) predict radically different shapes, from a "croissant with two tails" shape (panel (c ) from Opher et al. 2020) to an egg shape (panel (d) from Dialynas et al. 2017). The only possible strategy to distinguish between these different shapes is to send different probes to cross the heliopause in different directions, as proposed by the Chinese study of an "Interstellar Heliospheric Probe": using Jupiter fly-by's 6 years apart (panels (a) and (b)), it will be possible to cross the heliopause in two opposite directions, for instance towards its nose and towards its tail, thus providing a preliminary indication of the asymmetry of the heliosphere along the direction of arrival of the interstellar wind. Ideally, a third probe flying out of the ecliptic would be needed to really sense the 3-D structure of the heliosphere.*

### 3.7.3.2 2041-2061

Following the first steps back to the heliopause after Voyager and into the VLISM, the next wave of missions should be used to explore the extreme variations of characteristics of the heliopause and of the adjacent interstellar clouds with the direction of flight. The shape of the heliopause itself is currently unknown, and can be only inferred from theoretical models. Figure 4.23 (c, d) shows the diversity of heliospheric configurations predicted by different models, depending on their modeling approach and their assumptions on the parameters of the VLISM. Does the heliopause have an egg shape, as predicted by Dialynas et al. (2017) or does it extend two magnetic tail lobes into it, as predicted by Opher et al. (2020). Only in situ measurements made in very different directions, at least one along the direction of arrival of the interstellar wind (the "nose" of the heliosphere), another one in the opposite direction, can discriminate between these models. This is the exploration strategy proposed in the concept of "Interstellar Heliospheric Probe" currently studied in China (see Taikong # 20, 2020); by sending two probes at a few years interval and using Jupiter and Neptune gravity assist, this mission will directly measure the asymmetry of the heliosphere along the axis of the direction of arrival of the interstellar wind. Following this example, the period 2041-2061 should be used to send interstellar probes in a diversity of directions to unravel the complex geometry of heliosphere-VLISM interface regions and, beyond it, the diversity of interstellar clouds surrounding the Solar System.

### 3.7.4. Technology and methodology challenges of trans-Neptunian exploration

With missions flying deeper into the interstellar medium, spaceflight will start expanding into a totally new domain of parameter space: mission ranges measured in hundreds of AU; mission durations covering several decades and several generations of scientists and engineers, in a distant space where solar power is simply irrelevant. While the first steps of this major adventure of exploration of the environment of our planetary system can and should be accomplished as soon as possible based on an optimal adaptation of current technologies, totally new approaches to power, propulsion, telecommunications, equipment and software maintenance, mission and Human resources management will need to be developed to enable a systematic exploration by Humankind of its close Galactic neighborhood. This is a good price to pay for this new episode of Humankind's exploration of the cosmos.

# 4. Conclusions: from future missions to infrastructure and technology requirements

## 4.1. Introduction

This chapter started from the six science questions guiding the Horizon 2061 exercise to identify the generic mission architectures needed to address these questions when flying to the different destinations of the Solar System. Building on the scientific knowledge gained from past missions, a new generation of missions, shown as the "blue missions" in Table 4.6, will have to be flown. These "blue missions" first include those already in planning or in preparation by space agencies, which will fly typically between now and 2040. Following them, a new generation of more technically challenging missions should be flown in the 2041 to 2061 time frame, which we will summarize first. These missions will require new technology developments and/or upgraded or new dedicated infrastructures, to be developed during the two coming decades, which will be summarized in conclusion to this chapter.

| | | Earth-Moon system | Mars | Venus | Mercury | Planet | Moons & Rings | Planet | Moons & Rings | Asteroids | Comets | Trojans | KBO | HP Boundaries | VLISM | Oort cloud |
|---|---|---|---|---|---|---|---|---|---|---|---|---|---|---|---|---|
| Human outpost | | Artemis Lunar Village | Mars deep drilling | | | | | | | | | | | | | |
| Utilization of in situ resources | | PROSPECT | Mars 2020 | | | | | | | | | | | | | |
| Sample Return | | Apollo, Luna , Chang'e | MSR, MMX | Atm./surface Sample return | | | Enceladus, Europa? | | | Hayabusa | Comet Sample return | | | | | |
| In situ exploration | Network | ILRS | Geophys. network | | | | | | | | | | | | | |
| | Mobile | Luna, Chang'e | MER, MSL, Mars 2020 | Balloons | | | | | | MASCOT | | | | | | |
| | Station | Luna, Chang'e | Phoenix, Insight | Vega | | | Moon landers | | | | Philae | Orbiter + lander | | | | |
| | Penetrator | LCROSS | | | | | ? | | | ARM | Deep Impact | | | | | |
| | Atm. probe | | | Vega | | Galileo | Huygens | … with probe | | | | | | | | |
| Orbital Observation | Small satellites | | ? | | | | | | | | | | | | | |
| | Orbiter | LRO, Chandrayaan, SMART-1, Selene, Chang'e, etc. | Mex, MRO, Odyssey, Maven, TGO, Mangalyaan | Magellan, Vex, Akatsuki | Messenger, Bepi-Colombo | Galileo, Cassini, Juno, JUICE, EC | Callisto<br>JUICE | Ice giant systems missions … | Triton orbiter | Dawn, NEAR, Psyche | Rosetta | Orbiter + lander | Pluto-Charon Orbiter | | | |
| Fly-by | | | | | | | | | | | Giotto, Stardust, Deep Impact Comet interceptor | Lucy | Multiple probes<br>1st Interstellar Heliospheric Probe<br>New Horizon | Voyager | | |
| Meteorites and cosmic dust collections | | | | ? | | | | | | | | | | | | |
| Telescope observations from Earth and its orbit | | | | | | | | | | | | | | | | |
| **MISSIONS ADDRESSING THE KEY SCIENCE TO FLY: BY 2040 (green) BY 2061 (light blue)** | | **Earth-Moon system** | **Mars** | **Venus** | **Mercury** | **Planet** | **Moons & Rings** | **Planet** | **Moons & Rings** | **Asteroids** | **Comets** | **Trojans** | **KBO** | **HP Boundaries** | **VLISM** | **Oort cloud** |
| | | **Terrestrial planets** | | | | **Gas giant systems** | | **Ice giant systems** | | **Small bodies** | | | | **Heliopause and beyond** | | |

*Table 4.6: "Mission types vs destinations » matrix showing the green missions (missions already flown) and the "blue missions" (missions in planning, in preparation or to be flown in the 2041-2061 time frame) that will need to be flown to address the six key science questions guiding the Horizon 2061 exercise.*

# 4.2. A new generation of missions beyond the current programmatic horizon.

The general trend of the distribution of "blue" missions in this table parallels the one found for "green" missions in the introduction to this chapter, but pushes the boundaries to further destinations and more complex missions.

For the closest accessible destinations (the Moon and Mars) planetary exploration should reach the highest level of complexity by 2061, with the establishment of human outposts at both places whose sustainability will rely on a broad spectrum of In Situ Resource Utilization (ISRU) technologies.

One step further in the Solar System, one finds a clear dominance of sample return missions that will reach increasingly challenging destinations, or will concern an increasing diversity of sampling sites: this will be the major goal of Mars exploration for the coming decades. Samples should also be returned from comets and Jupiter trojans, as well as from Venus despite the exceptional technical difficulties of achieving it. Among all giant planet moons, one can also expect that samples will be returned more easily from Enceladus, a body of major astrobiological interest, at least via fly-by's through its plumes.

Farther away, in situ exploration and multi-platform orbital surveys should progressively prevail over orbital exploration at gas giants and their systems: atmospheric probes, moon landers, multi-point studies of their magnetospheres, ring hoppers and skimmers will serve more focused studies or their origins, workings or habitability. The next steps in the exploration of ice giants should be accomplished by pluri-disciplinary orbiter missions at each of the two systems. Combined with the delivery of atmospheric probes to their atmospheres, these missions will provide insight into the formation scenarios and workings of the most abundant class of planets in our galactic neighborhood (see chapter 2, Rauer et al. 2021) and pave the way to more focused missions to characterize the habitability of the most promising of their moons, starting very likely with a Triton orbiter.

Exploration of the Trans-Neptunian Solar System will likely remain dominated by probes combining fly-bys of individual TNOs with long-range travel across the heliospheric boundaries and into the very local interstellar medium. By sending probes in different directions, it will be possible to explore at the same time the diversity of TNO's, the three-dimensional geometry of the heliosphere and of its boundaries, and the heterogeneity of the VLISM around it.

Our idea of the diversity of secondary systems in the Solar System would remain incomplete, however, without an orbiter mission to one of the multi-object TNO systems, such as Pluto-Charon.

## 4.3. New technology requirements for a new wave of missions.

This new generation of "blue missions" will generate a broad spectrum of requirements on technologies and infrastructures. Starting with technology requirements, one can identify them by just examining the same set of blue missions, starting this time from the most distant destinations. Missions to the Trans-Neptunian Solar System and to the interstellar medium will mostly remain long-duration cruise missions. For them, far away from the sun and Earth, power, propulsion, data downlink capacities will remain the main bottleneck. Their requirements will push forward the corresponding technologies towards non-solar energy sources (necessarily nuclear), long-duration and high specific impulse propulsion systems (likely electric) and a very high degree of on-board data processing and compression before downlink. Progress made on these generic technologies will also become available for closer destinations, even considering that new generations of solar cells will likely push outwards the boundaries of solar power over the same time frame.

Among the objects populating the still poorly explored Trans-Neptunian Solar System, visiting those transiting for the first time closer in, like pristine comets and interstellar objects, will require mission scenarios "on alert", following the example of Comet Interceptor to be launched by ESA in 2028 (Snodgrass and Jones 2019).

Giant planets and their moons, particularly icy ones, and Venus will offer great challenges of a different kind because of their extreme environments. Their exploration will require innovative multi-platform mission architectures in which interplanetary carriers and orbiters will progressively be specialized to the function of delivering science platforms and sometimes transferring back samples to Earth. The diversity of science platforms will be tailored to the specific environments in which they will have to operate: atmospheric entry probes, landers operating in extreme temperature, pressure and sometimes radiation conditions, mobile elements moving not only at the surface of their target, but also flying in their atmospheres or jumping from one exploration site to another (following the example of NASA's Dragonfly mission to Titan, Lorenz et al. 2018), networks or clusters of platforms providing multi-point measurements wherever the scientific objectives will require. Just like evolution and selection pushed living beings on Earth to adapt to an increasing diversity of niches, the science platforms of the future will have to adapt to the even larger diversity of planetary environments. Given their long communication time with Earth and the necessity of optimizing the scientific return of their missions, they will require an ever increasing degree of on-board autonomy and the use of advanced Artificial Intelligence tools to perform their missions.

The scientific instruments carried and operated on these platforms will reflect the increasing diversity of scientific disciplines and measurement techniques required by planetary sciences. In this moving context, sample return caches, geosciences investigations (geophysics and geochemistry) and astrobiology experiments dedicated to the detection and characterization of biosignatures will occupy a larger and larger fraction of payloads.

The continuity of operations between carriers-orbiters and in situ science platforms will have to rely on advanced EDLA (Entry, Descent, Landing and Ascent) systems equally well adapted to the environments where they will deliver science platforms and from where they will have to carry back samples to orbit.

Finally, for the Moon and later Mars, where human outposts could progressively be established, opening expanded possibilities for their scientific investigation, a full new panoply of technologies enabling a sustainable human presence will have to emerge and be proven before being used for routine operations: in situ resource utilization (ISRU), advanced environment and life support systems, in-space and on-site assembly and manufacturing processes.

Finally, for all destinations, optimization of mass, power and telecom budgets will push towards an increased miniaturization of existing technologies, enabling additional in situ measurements with compact mission architectures.

To conclude this section, Table 4.7 lists horizontally the specific technology requirements associated with the different types of mission, listed vertically in the table. Chapter 5 (Grande et al, 2022) will present a projection of future technology developments in response to these requirements.

| MISSIONS ADDRESSING THE KEY SCIENCE TO FLY BY 2061 | | Large telescope assemby | Formation flying spectrometers | Miniaturization | LF space Interfero-metry | Deep drilling | ISRU | Ascent vehicle | ARV | Power; RTGs | Extreme temperature sample collection, cryogeny | Radiation hardening | heatshield | Communications, data rate | Multi-point measurements | Bio detectors | propulsion | Thin solar arrays |
|---|---|---|---|---|---|---|---|---|---|---|---|---|---|---|---|---|---|---|
| Human outpost (Moon, Mars) | | | | | | | ■ | ■ | ■ | ■ | | | | ■ | | | ■ | |
| Venus Sample Return | | | | | | | | ■ | ■ | | ■ | | ■ | | | | | |
| Mars Sample Return, deep drilling | | | | | | ■ | | ■ | ■ | ■ | | | ■ | | | ■ | | |
| Comet / Trojan sample return | | | | | | ■ | | | | ■ | ■ | | ■ | ■ | | ■ | ■ | ■ |
| In situ exploration | Network | | ■ | ■ | | | | | | ■ | | ■ | | ■ | ■ | | | |
| | Mobile | | | ■ | | | | | | ■ | | ■ | | ■ | ■ | ■ | | |
| | Station | | | ■ | | | | | | ■ | | ■ | | ■ | ■ | ■ | | |
| | Penetrator | | | ■ | | | | | | ■ | | ■ | ■ | ■ | ■ | ■ | | |
| | Atm. probe | | | | | | | | | ■ | | ■ | ■ | ■ | ■ | ■ | | |
| Orbital observation | Swarm | | ■ | ■ | | | | | | ■ | | ■ | | ■ | ■ | | | |
| | Small sat. | | ■ | ■ | | | | | | ■ | | ■ | | ■ | ■ | | | |
| | Orbiter | | | | | | | | | ■ | | ■ | | ■ | | | | |
| Fly-by (KBO, Oort Cloud, interstellar probes) | | | | ■ | | | | | | ■ | | ■ | | ■ | | | ■ | ■ |
| Giant Observatories | | ■ | ■ | | | | | | | | | | | | | | | |

*Table 4.7: A concise summary of some of the most important critical technologies that will be required to fly the "blue missions" identified in table 4.6 to their destinations.*

## 4.4. Advanced infrastructures for future planetary missions.

The "blue missions" of Table 4.6 will also need the support of new generations of infrastructures and services to be flown, operate at their destination, return data, and to maximize their scientific return, as emphasized throughout Section 3. These critical support elements generally serve not a single mission, but a broad diversity of missions and destinations. They can be classified into eight themes briefly reviewed here, following the logical path of the generation of new scientific knowledge by a planetary mission or a telescope observation: from the acquisition of the data, through their delivery to the scientific users, to the production of new scientific results, this path will end up with a special focus on the necessary human resources.

<u>Infrastructures for scientific observations (telescopes and observatories):</u> Section 2 reviewed the critical role that Earth-based telescopes, observing in a variety of wavelengths from radio to X rays, will continue to play in planetary sciences. Access to the new emerging generation of giant telescopes, ground-based or in space, will be extremely important for the planetary science community.

<u>Infrastructures for planetary mission operations:</u> from the launch facilities through the communication and navigation support to mission operation centers, they will have to offer services of increasing quality and duration to the demanding "blue missions" contributing to pillar 2.

<u>Infrastructures for long-term human exploration:</u> heavy launchers, orbital stations (including the Lunar gateway) and Moon or Mars outposts will support the long-term presence of astronauts conducting scientific experiments.

Monitoring of space weather & other space-related hazards: observation services monitoring and, when possible, predicting space weather events will be a necessity for long-duration human flights, just like weather forecast today has to support airborne navigation.

Infrastructures for sample collection, curation and analysis: sample return from a broad range of solar system objects will be increasingly important in the decades to come. To implement the processing chain linking the collection of samples to their return to Earth, their curation and their analysis under heavy planetary protection constraints, world-class facilities for the preservation, curation, analysis and dissemination of samples will be needed.

Data systems and virtual observatories: continued progress in planetary research will rely even more in the decades to come on a simultaneous access to a diversity of sources of information and analysis: space and ground-based observation data, results of sample analyses, laboratory experiments, numerical simulations and models, intensive computing. The enormous volume of data produced independently by these diverse tools will need to be easily accessible and manageable by the scientific end-users. The role of advanced data centers and virtual observatories will be more and more critical in the production of new scientific knowledge.

Human resources and socio-economic services: finally, operating the diversity of facilities mentioned above will require the training of a new generation of young scientists, engineers, mission specialists and astronauts. It is this new generation which will conduct scientific experiments on the future lunar outposts, or design new ways of operating planetary missions. Long-term multi-decadal missions beyond the heliopause, which will constitute Humankind's first steps outside of its own planetary system, will require new approaches to the management of missions and to the transmission of knowledge between generations: this major challenge on human resources will add to the purely technical challenges of interstellar journeys.

These themes in the future development of infrastructures and services are extensively explored in Chapter 6 (Foing et al., 2022).

*Acknowledgements: The research on the Saturn Rings Observer mission concept described in this book was carried out at the Jet Propul- sion Laboratory, California Institute of Technology, under a contract with the National Aeronautics and Space Administration to support the SS2012 Planetary Science Decadal Survey. The authors acknowledge fundings from the National Aeronautics and Space Administration, from the European Space Agency, from the French Space Agency, CNES and from the International Space Science Institute.*